\documentclass[aps,prd,preprint,superscriptaddress,showpacs,floatfix,preprintnumbers,nofootinbib]{revtex4-1}
\usepackage[utf8]{inputenc}

\usepackage{cancel}

\usepackage[
    letterpaper, 
    left = 2.5cm, 
    right = 2.5cm, 
    top = 2.4cm, 
    bottom = 2.4cm, 
    headsep = 0.5cm, 
    footskip = 1.5cm]{geometry}

\usepackage[dvipsnames]{xcolor}      
\definecolor{lcolor}{rgb}{0.5,0,0}
\definecolor{citcolor}{rgb}{0,0.0,1}

\usepackage[breaklinks,colorlinks,urlcolor=blue,citecolor=citcolor,linkcolor=lcolor,linktoc=all]{hyperref}
\usepackage{color}
\usepackage{graphicx}	
\graphicspath{{./figures/}}
\usepackage[utf8]{inputenc}
\usepackage{amsmath} 
\usepackage{amssymb}
\usepackage[ragged]{footmisc}

\makeatletter
\g@addto@macro\bfseries{\boldmath}
\makeatother

\usepackage{tikz}
\usepackage[customcolors]{hf-tikz}
\usepackage{mciteplus}

\usetikzlibrary{arrows,cd,shapes,decorations.pathmorphing,decorations.markings,shadings}
\tikzset{
  big arrow/.style={
    decoration={markings,mark=at position 1 with {\arrow[scale=4,#1]{>}}},
    postaction={decorate},
    shorten >=0.4pt},
  big arrow/.default=blue}

\def\picSc{0.8} 
\def\bigBlobSc{0.75}

\def\lineW{0.5} 

\newcommand\gluonLine[4]{
	\draw[decorate, decoration={snake,amplitude=.4mm,segment length=2mm,post length=0mm}, line width=\lineW mm, Black!80] (#1,#2) -- (#3,#4);
}

\newcommand\doubleGluonLine[4]{
	\draw[decorate, style=double, decoration={snake,amplitude=.4mm,segment length=2mm,post length=0mm}, line width=\lineW mm, Black!80] (#1,#2) -- (#3,#4);
}

\newcommand\quarkLine[5]{
	\tikzset{->-/.style={decoration={
  markings,
  mark=at position ##1 with {\arrow{>}}},postaction={decorate}}}
	\draw[->-=#5, line width=\lineW mm, Black!80]  (#1,#2) -- (#3,#4);
} 
\newcommand\ghostLine[5]{
	\tikzset{->-/.style={decoration={
  markings,
  mark=at position ##1 with {\arrow{>}}},postaction={decorate}}}
	\draw[->-=#5, loosely dotted, line width=\lineW mm, Black!80]  (#1,#2) -- (#3,#4);
} 

\newcommand\blobNode[4]{\node[circle,line width=\lineW mm,Black!70,draw,fill=Gray!30] at (#1,#2)[scale=#3] {#4};}

\newcommand{\eq}{Eq.~}
\newcommand{\eqs}{Eqs.~}
\newcommand{\Sec}{Sec.~}
\newcommand{\fig}{Fig.~}

\newcommand{\app}{Appendix~}
\newcommand{\apps}{Appendices~}
\newcommand{\tab}{Tab.~}
\newcommand{\tabs}{Tabs.~}
\newcommand{\nn}{\nonumber \\ }
\newcommand{\nr}[1]{(\ref{#1})}
\renewcommand{\Ref}{Ref.~}
\newcommand{\Refs}{Refs.~}
\newcommand{\Lh}{\Lambda_{\text{h}}}
\newcommand{\mE}{m_{\text{E}}}

\newcommand{\muQ}{\mu_q}
\newcommand{\D}[1]{D^{#1}}
\newcommand{\DZ}{\Delta_0}
\newcommand{\Dz}[1]{\DZ^{#1}}
\newcommand{\DN}[1]{\Delta_{#1}}
\newcommand{\Dn}[2]{\Delta_{#1}^{#2}}
\newcommand{\DDN}[1]{D_{#1}}
\newcommand{\DDn}[2]{D_{#1}^{#2}}
\newcommand{\PI}[1]{\Pi^{#1}}
\newcommand{\Pii}[2]{\Pi_{#1}(\hat{#2})}
\newcommand{\nc}{N_c}
\newcommand{\da}{d_A}
\newcommand{\intkp}{\int_{KP}}
\newcommand{\intkpr}{\int_{KPR}}

\newcommand{\intkpMod}{\int_{\widetilde{KP}}}
\newcommand{\intkprMod}{\int_{\widetilde{KPR}}}
\newcommand{\prjoT}{\mathcal{P}_{\text{T}}}
\newcommand{\prjoL}{\mathcal{P}_{\text{L}}}
\newcommand{\prjof}{\mathcal{P}_{D}}

\newcommand{\ProjIX}[3]{\mathcal{P}^{#1}_{\text{#2}}(\hat{#3})}
\newcommand{\Proji}[2]{\mathcal{P}_{#1}(\hat{#2})}

\newcommand{\kt}{{\vec{k}}}
\newcommand{\pt}{{\vec{p}}}
\newcommand{\rt}{{\vec{r}}}
\newcommand{\ut}{{\vec{u}}}
\newcommand{\vt}{{\vec{v}}}
\newcommand{\st}{{\vec{s}}}
\newcommand{\ttt}{{\vec{t}}}

\newcommand{\yt}{{\vec{y}}}

\renewcommand{\vec}{\mathbf}
\newcommand{\Gam}[1]{\Gamma^{#1}}
\newcommand{\Gamz}[1]{\Gamma_0^{#1}}
\newcommand{\dGam}[1]{\delta\Gamma^{#1}}
\newcommand{\II}[1]{I_{\text{#1}}}
\newcommand{\IIUD}[2]{I^{\text{#1}}_{\text{#2}}}
\newcommand{\uvtext}[1]{[#1]^{\text{UV}}}

\renewcommand{\epsilon}{\varepsilon}

\newcommand{\ud}{\mathrm{d}}
\newcommand{\upd}{\mathrm{d}}
\newcommand{\dNum}[1]{\!\operatorname{d}^{#1}\!}
\newcommand{\dInt}{\ud}

\renewcommand{\Im}{\operatorname{Im}}
\newcommand{\gamE}{\gamma_{\text{E}}}
\newcommand{\tr}[1]{\operatorname{Tr}\left[ #1 \right] }
\newcommand{\sx}{s_{\chi}}
\newcommand{\cx}{c_{\chi}}
\newcommand{\tx}{\tan \chi}
\newcommand{\ctx}{\cot \chi}
\newcommand{\Ave}[2]{\left\langle #2 \right\rangle_{#1}}
\newcommand{\AveT}[1]{\left\langle #1 \right\rangle_{3}}
\newcommand{\Span}{\operatorname{span}}
\renewcommand{\Re}{\operatorname{Re}}
\usepackage[acronym]{glossaries}
\newacronym{LO}{LO}{leading-order}
\newcommand{\LO}{\gls*{LO}}
\newacronym{QM}{QM}{quark matter}
\newcommand{\CQM}{cold \gls*{QM}}
\newacronym{QCD}{QCD}{quantum chromodynamics}
\newcommand{\QCD}{\gls*{QCD}}
\newcommand{\NLO}[1]{N$^{#1}$LO}
\newacronym{HTL}{HTL}{Hard-Thermal-Loop}
\newcommand{\HTL}{\gls*{HTL}}
\newacronym{UV}{UV}{ultraviolet}
\newcommand{\UV}{\gls*{UV}}
\newacronym{IR}{IR}{infrared}

\newcommand{\pDU}[2]{p_{#1}^{#2}}
\newcommand{\apDU}[2]{\alpha_s^{#1} p_{#1}^{#2}}
\newcommand{\pPowIrReg}[3]{p_{#1,#2}^{#3}}
\newcommand{\aPowpIrReg}[3]{\alpha_s^{#1} p_{#1,#2}^{#3}}
\newcommand{\aOpDU}[2]{\alpha_s \pDU{#1}{#2}}

\begin{document}

\title{Cold quark matter at N3LO: soft contributions}

\preprint{HIP-2021-10/TH}
\author{Tyler Gorda}
\affiliation{Technische Universit\"{a}t Darmstadt, Department of Physics, D–64289 Darmstadt, Germany}
\affiliation{Helmholtz Research Academy for FAIR, D–64289 Darmstadt, Germany}
\author{Aleksi Kurkela}
\affiliation{Faculty of Science and Technology, University of Stavanger, Stavanger, Norway}
\author{Risto Paatelainen}
\affiliation{Helsinki Institute of Physics and Department of Physics, University of Helsinki, Finland}
\author{Saga S\"appi}
\affiliation{European Centre for Theoretical Studies in Nuclear Physics and Related Areas (ECT*) and Fondazione Bruno Kessler, Strada delle Tabarelle 286, I-38123, Villazzano (TN), Italy}
\affiliation{Helsinki Institute of Physics and Department of Physics, University of Helsinki, Finland}
\author{Aleksi Vuorinen}
\affiliation{Helsinki Institute of Physics and Department of Physics, University of Helsinki, Finland}

\begin{abstract}
    High-order perturbative calculations for thermodynamic quantities in QCD are complicated by the physics of dynamical screening that affects the soft, long-wavelength modes of the system. Here, we provide details for the evaluation of this soft contribution to the next-to-next-to-next-to-leading order (\NLO{3}) pressure of high-density, zero-temperature quark matter (QM), complementing our accompanying paper in \Ref\cite{Gorda:2021znl}. Our calculation requires the determination of the pressure of the hard-thermal-loop (HTL) effective theory to full two-loop order at zero temperature, which we go through in considerable detail. In addition to this, we comprehensively discuss the structure of the weak-coupling expansion of the QM pressure, and lay out a roadmap towards the evaluation of the contributions missing from a full \NLO{3} result for this quantity. 
\end{abstract}

\maketitle

\tableofcontents

\newpage
\section{Introduction}
\label{sec:intro}

Determining the equation of state (EOS) of quantum-chromodynamic matter in extreme conditions using perturbation theory is a longstanding challenge almost as old as \QCD\ itself (see e.g.~\cite{Ghiglieri:2020dpq} for a review). In the case of high-temperature quark-gluon plasma (QGP), the calculation has reached a partial next-to-next-to-next-to-leading order (\NLO{3}) level \cite{Kajantie:2002wa,Kajantie:2003ax,DiRenzo:2006nh}. At such high orders, a complication in perturbative calculations arises from the emergence of collective phenomena at long wavelengths, most importantly the physics of dynamical in-medium screening. To address this, all-loop-order resummations must be performed in order to reach a fixed order in the strong coupling constant $\alpha_s$. 

At high temperatures $T$, reaching the partial \NLO{3} accuracy was made possible on one hand by technical advances in the evaluation of multi-loop sum-integrals \cite{Arnold:1994ps,Zhai:1995ac} and on the other hand by the seminal works of Kajantie et al.\ \cite{Kajantie:2002wa,Kajantie:2003ax,Hietanen:2004ew,Hietanen:2006rc,DiRenzo:2006nh}, where a resummation of 
soft screened modes of momentum scales $\alpha_s^{1/2} T$ and $\alpha_s T$ was performed using the dimensionally reduced effective theories electrostatic \QCD\ (EQCD) and magnetostatic \QCD\ (MQCD) \cite{Kajantie:1997tt,Braaten:1995cm,Braaten:1995jr}. These calculations left only the contribution of the hard momentum scale $\pi T$ missing from the full \NLO{3}  EOS of hot QGP, which constitutes a conceptually simple but technically very demanding challenge. 

Screening phenomena closely analogous to those encountered at high temperatures appear also in the context of dense and \CQM\ \cite{Manuel:1995td,Gerhold:2004tb}, where phenomenological motivation stems from model-independent studies of the neutron-star matter EoS \cite{Annala:2017llu,Annala:2019puf}. Here, the last fully completed order in perturbation theory dates back to the seminal papers of Freedman and McLerran \cite{Freedman:1976dm,Freedman:1976ub}, who determined the EOS to \NLO{2} accuracy. At this level, the calculation becomes sensitive to the physics of screening, which these authors addressed through an all-loop-order diagrammatic resummation. The framework of dimensional reduction is unavailable at low temperatures, and challenges related to extending this resummation to higher orders have so far prevented bringing the EOS of \CQM\ to the same level of perturbative accuracy as its nonzero-$T$ counterpart, although some progress in this direction has recently been achieved in \cite{Kurkela:2016was,Gorda:2018gpy}. In the present paper, complementing an accompanying letter \cite{Gorda:2021znl}, we finally perform this resummation using the \HTL\ effective theory, determining the soft contributions to the EOS up to and including the \NLO{3} order. While \Ref\cite{Gorda:2021znl} concentrates on an in-depth analysis of the result, here we provide extensive details of the technical aspects of the calculation, and in addition discuss the computations needed to determine the last contributions missing from a full \NLO{3} result for the EOS of \CQM.

The physical picture behind perturbative calculations at high densities is as follows. In a medium characterized by a large quark chemical potential $\muQ$ and zero temperature, \CQM\ contains a filled Fermi sea of quarks from zero momentum up to the scale $\muQ$.\footnote{In this section, we shall describe the situation for a single quark flavor, for simplicity.} The free Fermi pressure of this system of quarks forms the \LO\ description of the pressure $p$ of \CQM\ and scales as $\muQ^{4}$ in the case of massless quarks.\footnote{Note that at high density, quarks on the Fermi surface undergo pairing through attractive channels of gluonic interactions, leading to a different ground state \cite{Alford:2007xm}. However, these effects do not enter at any finite order in the weak-coupling expansion. At sufficiently high densities, the pairing gap $\Delta$, which only depends on $\mu$ in a mild way, becomes small in comparison to the chemical potentials, and the pairing contributions to the pressure become suppressed by a factor $\Delta^2/\mu^2$.} While there are no on-shell gluons in the medium, off shell gluons are present because the quarks are color charged. Interactions between the quarks and gluons in \QCD\ lead to corrections to this \LO\ pressure as a function of the strong coupling constant $\alpha_s$. 

\begin{figure}[t]
    \includegraphics[width=0.5\textwidth]{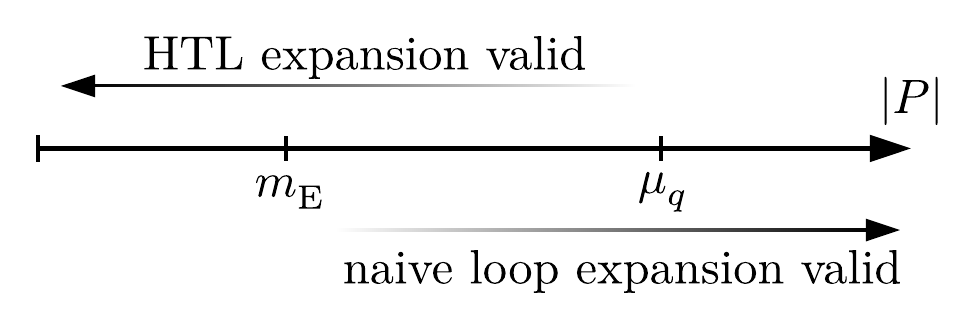}
    \caption{Illustration of the ranges of gluonic momenta $P$ where different approximations can be made. 
    }
\label{fig:expand_axis}
\end{figure}

Because of this Fermi sea of quarks, the propagation of both quarks and gluons through \CQM\ becomes modified. Low-momentum quarks are Pauli blocked and cannot propagate, as those states are filled by the medium. Thus, these low-momentum quarks do not contribute to higher-order loop corrections to $p$, leaving the scale $\mu_q$ (dubbed ``hard'') as the only relevant scale for quarks.
For gluons, the picture is more complicated and involves two different approximations that can be used in different regions of momentum space: the naive loop expansion and the \HTL\ expansion (see \fig\ref{fig:expand_axis}). Hard, short-wavelength modes can be treated similarly to the quarks in a naive loop expansion, while the soft, long-wavelength gluons become qualitatively modified by the medium and require resummations of arbitrary numbers of one-quark-loop insertions within calculations. These modifications lead to, e.g., nonanalyticities $\ln \alpha_s$ in the weak-coupling expansion in the pressure of \CQM.

The rest of this Introduction is organized as follows. In \Sec\ref{sec:two_expansions}, we introduce the naive loop expansion and the \HTL\ expansion and motivate their respective regions of validity. After this, in \Sec\ref{sec:power_count_soft} we explain how to power-count the contributions of the resummed soft gluons. In \Sec\ref{sec:computing_the_pressure}, we then discuss the analytic structure of the different contributions to the pressure of \CQM, proceeding from \LO\ to \NLO{3}. Finally, in \Sec\ref{sec:what_we_compute} we explain what precisely is computed in the present article, and walk the reader through the overall structure of the paper.

\subsection{Two expansions for gluons in cold quark matter}
\label{sec:two_expansions}

Whether gluonic propagation is qualitatively modified by scattering from hard quarks in loop corrections depends on the magnitude of the propagating gluonic momentum \cite{Blaizot:2001nr}. This can be seen most clearly from the dispersion relation of the gluonic modes with momenta $P$, which is schematically of the form
\begin{equation}
    P^{2} + \Pi(P) = 0.    
\label{eq:gluon_dispersion}
\end{equation}
Here we will work consistently in a Euclidean framework, and $\Pi(P)$ is a generic component of the Euclidean gluon polarization tensor. This tensor is parametrically of the order of the square of the in-medium effective mass scale $\mE$, related to the one-loop Debye mass. For a single massless quark in $d = 3$ spatial dimensions it has the value $\mE^2 = (2 / \pi ) \alpha_s \muQ^2$.\footnote{This effective mass scale is related to the asymptotic HTL mass \cite{Andersen:1999sf} $m_\infty$ by $\mE^{2} = 2 m_\infty^{2}$. In the case of multiple quark flavors $f$, this effective mass scale becomes $\mE^2 = (2\alpha_s / \pi) \sum_f \mu_f^2$.} If the free part of \eq\eqref{eq:gluon_dispersion} parametrically exceeds the interaction part, the interactions can be treated as perturbations to the free propagation of  gluons and be dealt with using a naive perturbative (loop) expansion. We can see that this occurs for $|P| \gg \mE$.  

\begin{figure}[bt]
    \centering
    \raisebox{-0.42\height}{\includegraphics[width=0.42\textwidth]{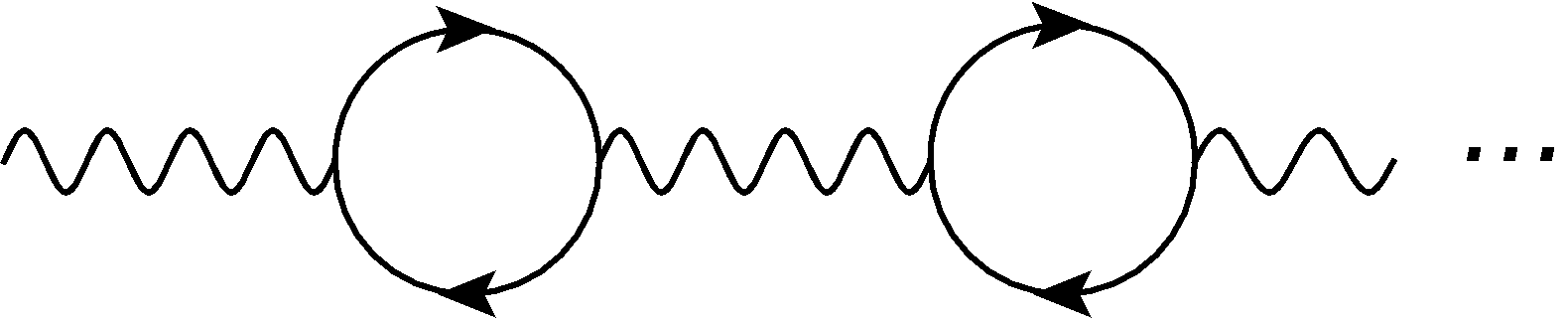}}
    \hspace{+1.5cm}
    \raisebox{-0.42\height}{\includegraphics[width=0.15\textwidth]{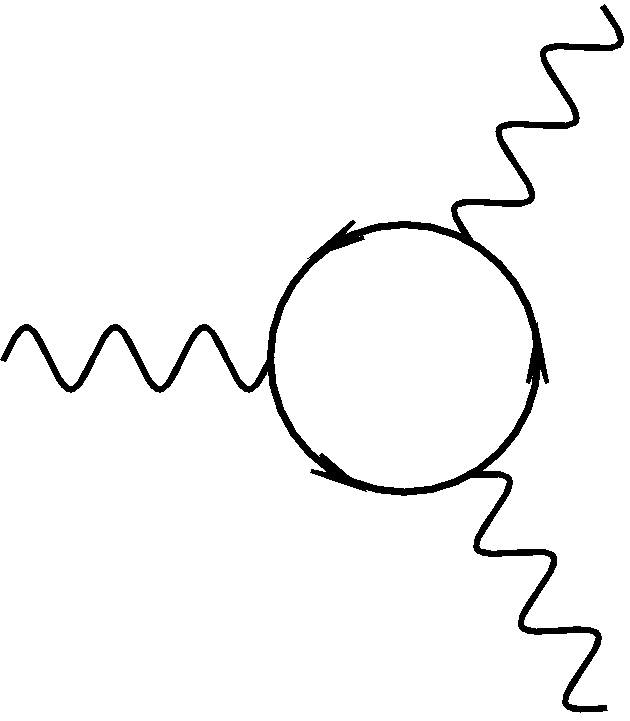}}
    \vspace{0.5cm}
    \caption{Two contributions that enter at the same order as the corresponding bare terms for soft gluonic momenta.}
\label{fig:gluon_resum}
\end{figure}

If, on the other hand, the gluon has momentum $|P| \lesssim \mE$ (dubbed ``soft''), then its propagation is qualitatively modified. In particular, generic low-momentum gluonic excitations require a nonzero excitation energy proportional to $\mE$.\footnote{Note that for some special directions of the Euclidean four-vector $P$, these excitations may still be massless. At high temperature, unscreened magnetic gluons lead to the generation of a further ultrasoft mass scale, but this is not the case in \CQM, as soft gluons are not Bose enhanced (see below).} This behavior arises because the gluonic self energy $\Pi^{\mu \nu}(P)$ has a nonzero $|P| \to 0$ limit
\begin{equation}
\label{eq:htl_pi_lim}
    \lim_{|P| \to 0}\Pi^{\mu \nu}(P) \equiv \Pi^{\mu \nu}_{\text{HTL}}(\hat{P}) \neq 0
\end{equation}
where we have suppressed the color indices and defined a unit four-vector $\hat{P} = P / |P|$ in the direction of $P$, a notation we shall use prominently in this work. Here, we have also identified the \HTL\ self energy, which is the low-momentum limit of the full self energy. It is important to note that in \CQM, only the quark loops contribute to this HTL self energy: hard gluon and ghost loops are not directly populated by the medium (i.e.,\ there are no on-shell gluons present) and thus do not differ from their vacuum values (i.e.~depend on $\muQ$) at \LO. It then follows that only the single quark loops must be resummed into the gluonic propagators, as all other corrections can be naively expanded around that limit. This resummation is depicted in \fig\ref{fig:gluon_resum}.

The full kinematics of the one-loop quark contribution to the gluon polarization tensor results in a $\Pi^{\mu \nu}(P)$ that depends not only on $\hat{P}$, but also on the magnitude of the gluonic momentum $P$.  However, if this magnitude is parametrically less than the hard scale, $|P| \ll \muQ$, one may systematically expand the polarization tensor in powers of the ratio of the external gluon momentum and the internal quark loop momentum, $|P|/|Q|$, which constitutes part of the \HTL\ framework as described by Braaten and Pisarski \cite{Braaten:1989mz}. In addition to this modified propagation,  the interaction between soft gluons also becomes modified in \CQM, as 
$n$ soft gluons interacting through a quark loop with momentum $|Q| \gtrsim \muQ$ enters at the same order as the bare coupling between $n$ soft gluons (if it exists). For example, for soft gluons the interaction shown in the right panel of \fig\ref{fig:gluon_resum} enters at the same order as the bare three-gluon coupling. However, both quark and ghost fields remain unresummed in Euclidean space --- the quark propagators are protected in the infrared by the nonzero chemical potential, while the ghosts are known not to develop a thermal mass \cite{Laine:2016hma}.

In summary, there are two different approximations that can be made for gluons in \CQM, depending on the magnitude of their momenta $P$: if $|P| \ll \muQ$, the HTL expansion becomes valid, and if $|P| \gg \mE$, the naive loop expansion becomes valid. This is shown pictorially in \fig\ref{fig:expand_axis}. As demonstrated in \Ref\cite{Gorda:2018gpy}, the integration region $\mE \ll |P| \ll \muQ$, where both approximations are valid (dubbed ``semisoft'' in \Ref\cite{Gorda:2018gpy}) leads to a logarithm of the coupling:
\begin{equation}
    \int_{\mE}^{\muQ} \frac{\upd^{4} P}{P^{2} + \Pi(P)} 
    \to \int_{\mE}^{\muQ} \Pi_{\text{HTL}}(\hat{P}) \frac{\upd^{4} P}{(P^{2})^2} 
    \sim \Ave{\hat{P}}{\Pi_{\text{HTL}}(\hat{P})} \ln \left( \frac{\mE}{\muQ} \right) 
    \sim \mE^2 \ln \alpha_s.
\end{equation}
Here, the notation $\langle \cdot \rangle_{\hat{P}}$ indicates an average over all four-dimensional Euclidean angles. Since such a logarithm can arise from any integral over a resummed gluonic momentum, at higher orders, where multiple resummed gluons may contribute, we should expect to find contributions to the pressure containing factors of $\ln^{n} \alpha_s$, where $n$ is the number of resummed gluonic momenta in a given resummed diagram. At \NLO{3}, this leading logarithm was computed already in \Ref\cite{Gorda:2018gpy}.

We now address the question of how one should power count such resummed, soft gluons, to see where they first contribute.

\subsection{Power counting the soft contributions}
\label{sec:power_count_soft}

As per the discussion in \Ref\cite{Gorda:2018gpy}, the soft gluons occuring in loop corrections and that require resummation are phase-space supressed by the integration measure 
\begin{equation}
    \int^{\mE} \upd^{4} P \to \mE^{4} \sim \alpha_s^2 \muQ^4.
\end{equation}
Thus, these contributions do not enter the weak-coupling expansion of the pressure until \NLO{2}.  Moreover, unlike the case at high temperatures \cite{Ghiglieri:2020dpq}, gluons occurring in loop corrections are not populated by the medium, and thus their occupation numbers are not Bose enhanced. This has the important implication that interactions between multiple soft gluons are perturbative in a loop expansion within the long-wavelength HTL theory. To see this, consider adding a soft gluon with loop momentum $P$ to a soft gluon line of momentum $K$. Doing this through three-point vertices leads to (defining $R \equiv P - K$ as the other internal loop momentum)
\begin{equation}
   \raisebox{-0.35\height}{\includegraphics[height=1.8cm]{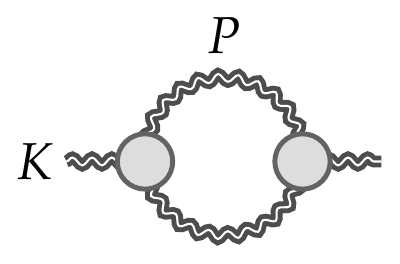}}  \sim  \alpha_s \int_{|P| \sim \mE} \frac{\Gamma(K,P,R)^2 \, \upd^{4} P}{[P^{2} + \Pi(P)][R^2+\Pi(R)]} = O(\alpha_s \mE^2),
\end{equation}
where the two effective vertices are represented by $\Gamma(K,P,R)^2$ in the numerator. For soft momenta, these vertices scale in the same way as the bare vertices, leading to ${\Gamma(K,P,R)^2 \sim \mE^2}$. Similarly, for soft $P, K, R$, the denominator scales as $\mE^4$, leading to the final power counting shown above. We thus see that this is a perturbative correction to the self energy $\Pi(K) \sim \mE^2$. The case of adding a gluon through a four-point vertex is identical, as for the purposes of power counting a single 4g HTL vertex behaves similarly to two 3g HTL vertices. We therefore see that these soft corrections are indeed perturbative. A full account of the form of these effective vertices, and our notation for the double lines and blob vertices for the HTL diagrams are introduced in detail in Appendix \ref{sec:HTLappendix}.

%
%
%
%

The fact that interactions between soft gluons are perturbative means that we can systematically improve the soft sector using a loop expansion within the HTL theory.
This situation is qualitatively different from that encountered in high-temperature QGP, where the presence of ``ultrasoft'' gluonic momenta of order $\alpha_s T$  famously leads to the Linde problem and the emergence of fundamentally nonperturbative contributions to the pressure at $O(\alpha_s^3)$ (see e.g.\ \Ref\cite{Ghiglieri:2020dpq} for a discussion of this subtle topic).

Let us now make one brief remark about regularization. The cutoff description implicitly used so far in these discussions (and in \Ref\cite{Gorda:2018gpy}) is very convenient for identifying the physical sources of the logarithms. However, in detailed computations (especially at higher orders) such an approach has drawbacks. In particular, once there are multiple soft gluons, performing the entire computation with a cutoff is very cumbersome. Thus, at this point we make the choice to use dimensional regularization to regulate not only the hard UV divergences arising in the full theory but also the intermediate ones arising from different kinematic regions 
(which we will introduce momentarily), once those regions are separated. This will provide a much more streamlined framework for self-consistently determining all the different contributions to the pressure at higher orders. 

\subsection{Computing the pressure of cold quark matter}
\label{sec:computing_the_pressure}

Following the logic from above, we deduce the following structure for the pressure of \CQM, valid up to and including the \NLO{3} terms:
{\allowdisplaybreaks
\begin{align}
    p  = p_\text{FD} 
    +\aOpDU{1}{h}  & +\apDU{2}{h}  + \apDU{3}{h} \nonumber \\*
    & +\apDU{2}{s}  +\apDU{3}{s}  \nonumber  \\*
    &  \quad \quad \quad\, + \apDU{3}{m} . \label{eq:schematics}
\end{align}
}%
Here, $p_\text{FD}$ is the pressure of a free Fermi gas of quarks, while the other terms arise from interaction corrections among or across modes of different types. Terms on first line arise from hard modes and can be computed through a naive loop expansion in full QCD. Terms on the second line arise from soft modes and their interactions, and can be determined within the HTL theory. Finally, the remaining term on the third line arises from interactions between the soft and hard modes and requires a partial HTL resummation. 

Due to the ambiguous semisoft momentum range $\mE \ll P \ll \muQ$, the splitting between the different kinematic regions is not unique.  This ambiguity leads to ultraviolet (UV) divergences within the $\pDU{i}{s}$ that cancel against corresponding infrared (IR) divergences within the $\pDU{i}{h}$ (and mixed UV-IR divergences in $\pDU{3}{m}$ at \NLO{3}). This cancellation will be further remarked on briefly below. The ambiguity also makes these coefficients dependent on a factorization scale $\Lh$, which arises from the dimensionally regularized integration measure in our case, and which will be canceled when summing over the different kinematic contributions at a given order. In terms of the factorization scale, the divergences also lead to expressions of the form $\ln( \Lh / \mE) $ and $\ln(\muQ/\Lh)$ from the UV limit of the soft sector and the IR limit of the hard sector, respectively. As the $\Lh$ dependence cancels in the sum over all kinematic contributions, these logarithms will generate precisely the $\ln( \muQ/ \mE) \sim \ln \alpha_s$ terms discussed in \Sec\ref{sec:two_expansions}. It is important to note here that the factorization scale $\Lh$ is not a momentum cut-off between the soft and hard sectors of the theory, and thus need not lie between the scales $\mE$ and $\muQ$. This becomes relevant when analyzing the behavior of the result, and is further discussed in \Ref\cite{Gorda:2021znl}.

\begin{figure}
    \centering
    \includegraphics[height = 2.6 cm ]{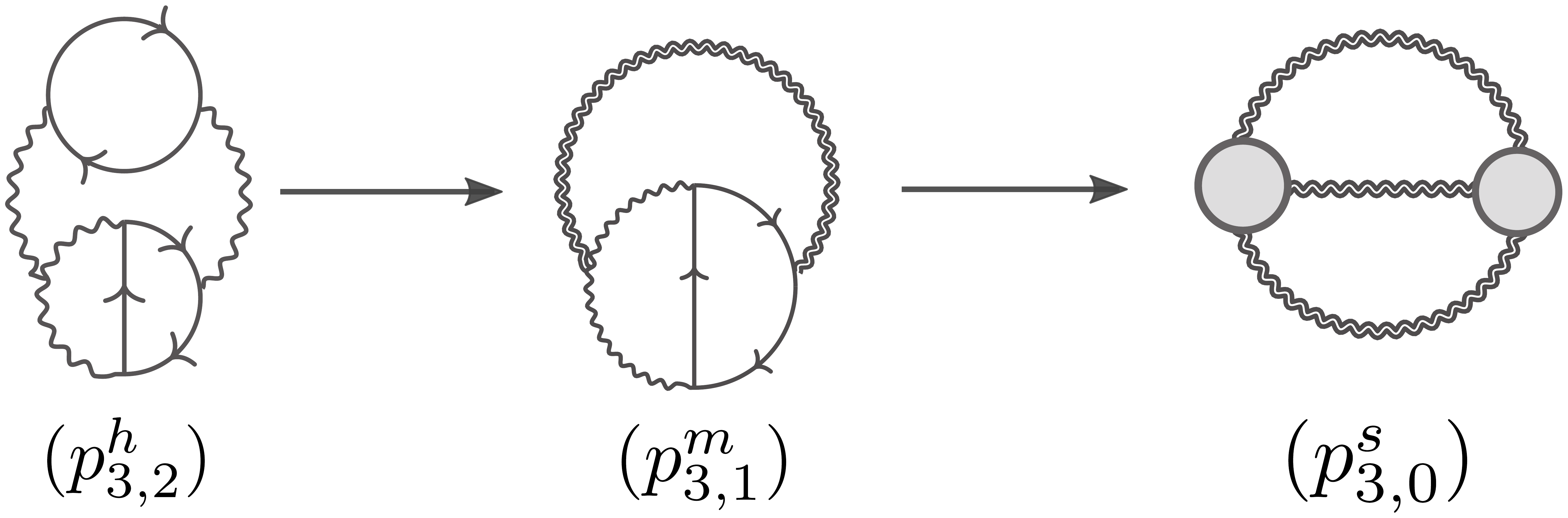}
    \caption{Example of progressively HTL-resumming a diagram, leading to a mixed inter\-mediate diagram and then finally to a fully soft 2-loop HTL diagram. The second index on the $\pPowIrReg{i}{k}{n}$ denotes the number of remaining hard gluonic (or ghost) loop momenta in a diagram.}
    \label{fig:zorro}
\end{figure}

To discuss the structure of the terms in \eq\eqref{eq:schematics} in more detail, we find it useful to further classify the diagrams contributing to the different coefficients $\pDU{i}{n}$. This further classification is based on the HTL limits of the different diagrams which are obtained by taking soft kinematics of the all the gluon lines. Diagrammatically this is reached by (i) resumming all gluon lines and (ii) contracting all quark loops into points and absorbing them into propagators and HTL vertices (see a sample illustration of this process of ``HTL resummation'' in \fig\ref{fig:zorro}). Concretely, we classify the contributions by the loop order of the resulting HTL diagram after taking fully soft kinematics.\footnote{Note that this classification is well-defined at zero temperature, where HTL corrections arise purely from quarks. At high $T$, this classification is not unique because gluons can also be absorbed into propagators and vertices.} 
In general, for $0<j<i$, an $(i+1)$-loop hard diagram is classified as a part of $\pDU{i,j}{h}$ if its fully soft limit is a $j$-loop HTL diagram. The remaining $(i+1)$-loop hard diagrams, which lead to $i$-loop HTL diagrams, are classified as $\pDU{i,0}{h}$.

This leads to the decomposition of the \NLO{2} and \NLO{3} terms 
\begin{equation}
    \begin{split}
    \pDU{2}{h} &= \pDU{2,1}{h} + \pDU{2,0}{h}, \\
    \pDU{2}{s} &= \pDU{2,0}{s} ,
    \end{split}
    \label{eq:n2lo_breakdown}
\end{equation}
    and
\begin{equation}
    \begin{split}
    \pDU{3}{h} &= \pDU{3,2}{h} + \pDU{3,1}{h}+ \pDU{3,0}{h}, \\
    \pDU{3}{m} &= \pDU{3,1}{m} + \pDU{3,0}{m},\\
    \pDU{3}{s} &= \pDU{3,0}{s},
    \end{split}
    \label{eq:n3lo_breakdown}
\end{equation}
where the second subscript denotes the number of hard gluon (or ghost) loop momenta, except for the above-mentioned exceptional $\pDU{i,0}{h}$ terms. We show the relation between these contributions and the fully resummed HTL diagrams in \fig\ref{fig:visual_organization}.
Organizing the terms in this fashion guarantees that the sum of all contributions in a given column is independent of the factorization scale as well as associated divergences.\footnote{We show how this works for the two-loop HTL column (with soft, mixed, and hard contributions) via a simple worked example in \app\ref{sec:four_regs_sum}.}

Next, we discuss in detail each of the terms appearing in \eqs\eqref{eq:schematics}--\eqref{eq:n3lo_breakdown}.

\subsubsection{\texorpdfstring{Classification of diagrams up to \NLO{2}}{Classification of diagrams up to NNLO}}

\begin{figure}[t]
    \centering
    \raisebox{-0.40\height}{\includegraphics[width=0.25\textwidth]{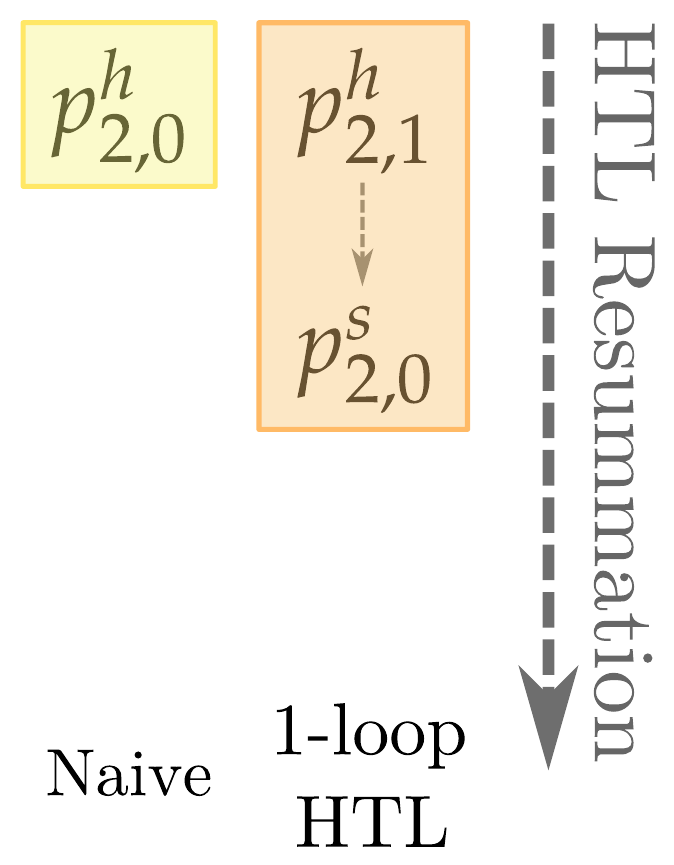}}
    \hspace{1.5cm}
    \raisebox{-0.40\height}{\includegraphics[width=0.35\textwidth]{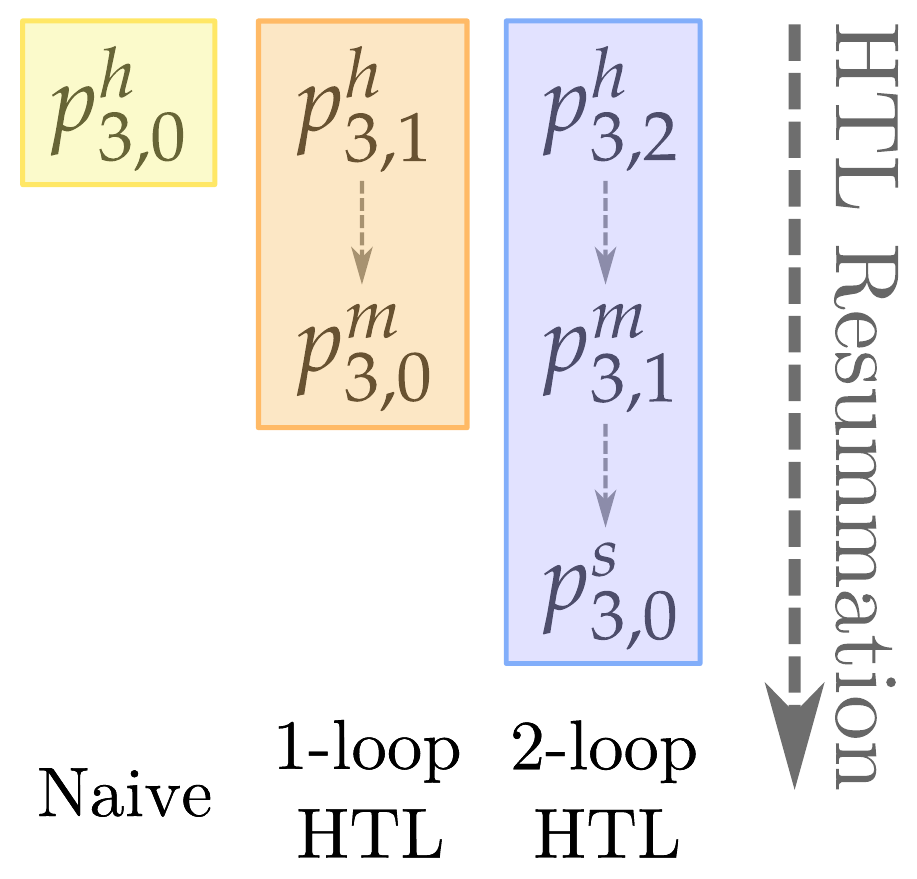}}
    \vspace{0.2cm}
    \caption{\label{fig:visual_organization}(Color online) The periodic table of diagrams, showing the relation between the different subclasses of contributions at \NLO{2} (left) and \NLO{3} (right). Moving downward within a single column corresponds to HTL resumming the harder contributions; this procedure also decreases the number of hard gluon (or ghost) loop momenta by one at each step, rendering the last term in each column IR safe.}
\end{figure}

Up to NLO, the contributions to the pressure of \CQM\ are simple and do not require resummations:

\begin{itemize}

    \item LO: This is simply the free Fermi pressure, arising from a single diagram with full quark kinematics;
    \begin{equation}
        p_\text{FD}= \raisebox{-0.42\height}{\includegraphics[height=1.2cm]{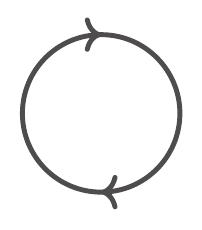}}.
    \end{equation}
    \item NLO: In this contribution, only a hard gluon contributes, since a soft gluon is phase-space suppressed. The quark loop requires the full kinematics.
    \begin{equation}
    \aOpDU{1}{h} =  \raisebox{-0.42\height}{\includegraphics[height=1.2cm]{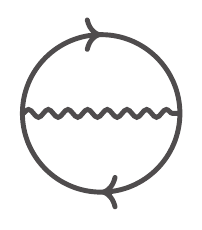}}.
    \end{equation}
\end{itemize}

At \NLO{2}, there are contributions from both hard and soft gluons, corresponding to the coefficients $\pDU{2}{h}$ and $\pDU{2}{s}$ in \eq\eqref{eq:schematics}, respectively. As the diagrams in $\pDU{2}{s}$ are fully resummed, the diagrams are IR safe and there is only one subclass of diagrams contributing to
\begin{equation}
\apDU{2}{s}= \aPowpIrReg{2}{0}{s} =  \raisebox{-0.42\height}{\includegraphics[height=1.2cm]{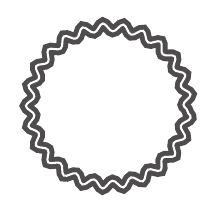}}.
\end{equation}
On the other hand, $\pDU{2}{h}$ can be further subdivided into two subclasses 
\begin{equation}
    \aPowpIrReg{2}{1}{h} =\raisebox{-0.42\height}{\includegraphics[height=1.2cm]{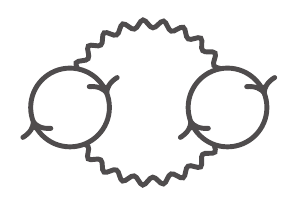}}
\end{equation}
and
\begin{equation}
\aPowpIrReg{2}{0}{h} =     \raisebox{-0.42\height}{\includegraphics[height=1.2cm]{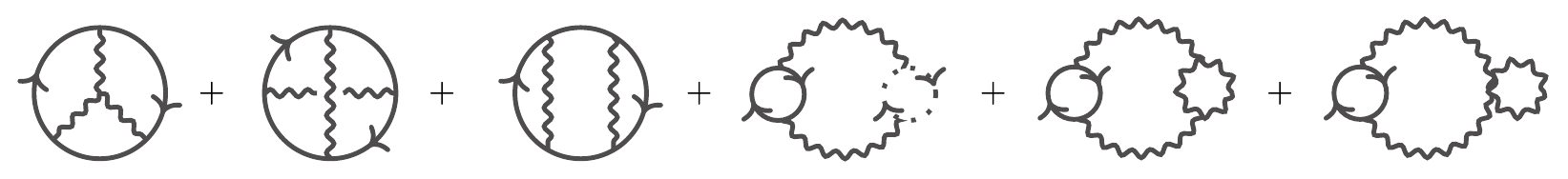}},
\end{equation}
grouped by their IR properties. Here, the hard $\pPowIrReg{2}{1}{h}$ and soft $\pPowIrReg{2}{0}{s}$ become by construction identical for semisoft gluon kinematics, and the diagram in $\pDU{2,0}{s}$ can in fact be generated from that in $\pDU{2,1}{h}$ by HTL resumming the gluon loop in the hard diagram. One might worry that this leads to a double counting of contributions, but this is not the case since at $T=0$ the semisoft region of $\pPowIrReg{2}{1}{h}$ or $\pPowIrReg{2}{0}{s}$ gives rise to scalefree integrals that vanish in dimensional regularization: in the first case, the quark loops can be replaced with HTL self-energy insertions, in the second, the resummed gluon propagator is re-expanded; in both cases, the integration over the gluonic loop momentum is then scalefree in the semisoft region.\footnote{The situation is qualitatively different at nonzero $T$, where one indeed needs to subtract a naive, expanded HTL contribution from the corresponding graphs (see e.g.~\Refs\cite{Ipp:2006ij,Kurkela:2016was}). Working at $T=0$ similarly simplifies this issue at the \NLO{3} level, helping us avoid the double counting of contributions at that order.}  The $\pPowIrReg{2}{0}{h}$ contribution is on the other hand IR safe and diagrammatically distinct. This is precisely the structure illustrated in the left panel of \fig\ref{fig:visual_organization}.

The \NLO{2} contributions to the pressure of \CQM\ were first determined by Freedman and McLerran in \Refs\cite{Freedman:1976dm,Freedman:1976ub} without the use of the HTL theory. In the modern language of dimensional regularization, the logarithmic contributions to the pressure arise solely from the subclasses $\pPowIrReg{2}{0}{s}$ and $\pPowIrReg{2}{1}{h}$ above, through the $(1/\epsilon) \times \epsilon$ terms therein, with $\epsilon\equiv (4-D)/2$ and $D$ standing for the spacetime dimensionality. 
Additionally, one finds $1/\epsilon$ terms cancelling between these $\pPowIrReg{2}{0}{s}$ and $\pPowIrReg{2}{1}{h}$ subclasses. In the case of $\pPowIrReg{2}{0}{s}$, such a term arises from a UV divergence, while in $\pPowIrReg{2}{1}{h}$, it arises from an IR divergence. 

\subsubsection{\texorpdfstring{Classification of diagrams at \NLO{3}}{Classification of diagrams up to NNNLO}}

\begin{figure}[t]
    \centering
        \includegraphics[width=0.48\textwidth]{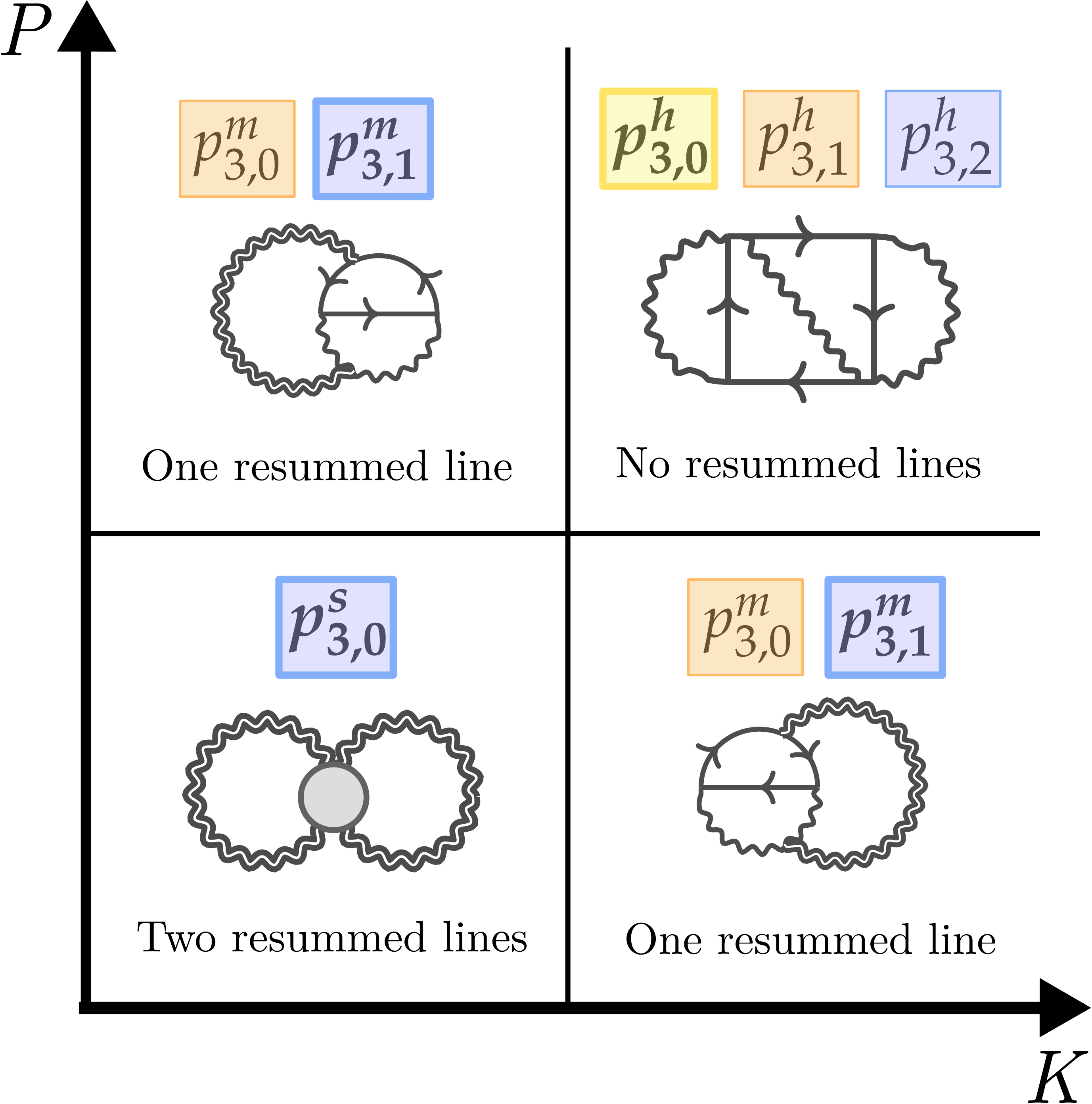}
    \caption{\label{fig:summaryfig} (Color online) A visual summary of the six subclasses of \NLO{3} contributions,  classified in terms of the number of resummed lines in each class. The class used to show a representative diagram is written in boldface, and color coding of the contributions shows the correspondence with \fig\ref{fig:visual_organization}.} 
 \end{figure}

The organization of the \NLO{3} contributions was displayed already in \eq\eqref{eq:n3lo_breakdown} and the right panel of \fig\ref{fig:visual_organization} and is further visually summarized in \fig\ref{fig:summaryfig}.

Similar to the \NLO{2} calculation, at \NLO{3} there are multiple classes of contributions, i.e.~those which arise from either two, one, or zero soft gluons, corresponding to the coefficients $\pDU{3}{s}$, $\pDU{3}{m}$, and $\pDU{3}{h}$ in \eq\eqref{eq:schematics}, respectively. 
In this section, we give a more detailed account of all the subclasses making up these contributions, proceeding according to the HTL diagrams they are related to (i.e., the colored columns in \fig\ref{fig:visual_organization}). 

The soft contribution is again fully resummed and IR safe and therefore forms only one subclass, namely
\begin{equation}
    \aPowpIrReg{3}{0}{s} = 
    \raisebox{-0.42\height}{\includegraphics[height=1.2cm]{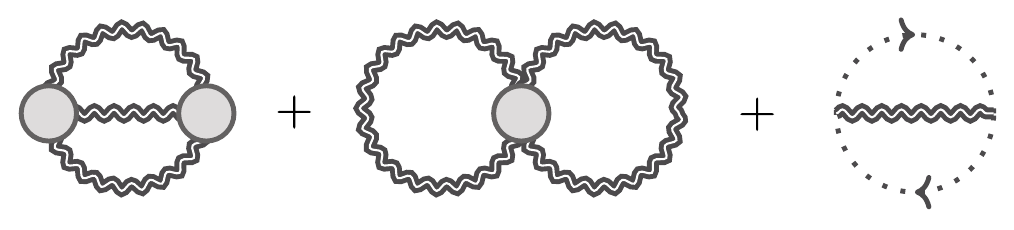}}.
\end{equation}
These diagrams are intimately related to those in the mixed contribution\footnote{Here and in the following diagrams, we assume an implied summation over the fermion and ghost directions in each loop, to reduce the number of diagrams shown. One can find the full list of contributions, with the correct symmetry factors, in \Ref\cite{Kajantie:2001hv}.}
\begin{align}
    \aPowpIrReg{3}{1}{m} = 
    &\raisebox{-0.42\height}{\includegraphics[height=1.2cm]{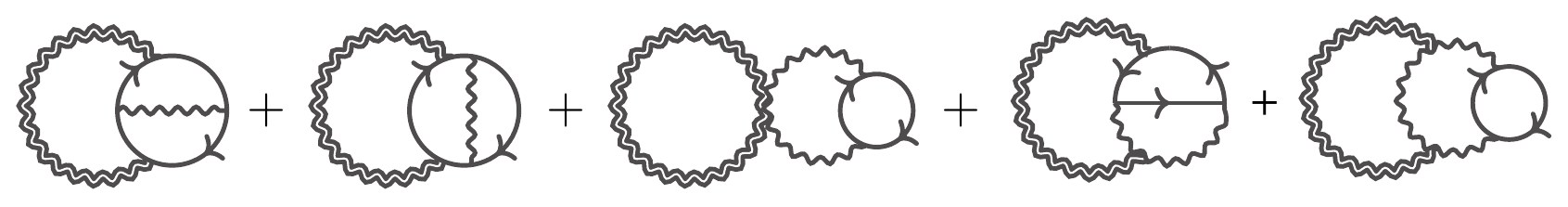}} \nonumber \\
    &+\raisebox{-0.42\height}{\includegraphics[height=1.2cm]{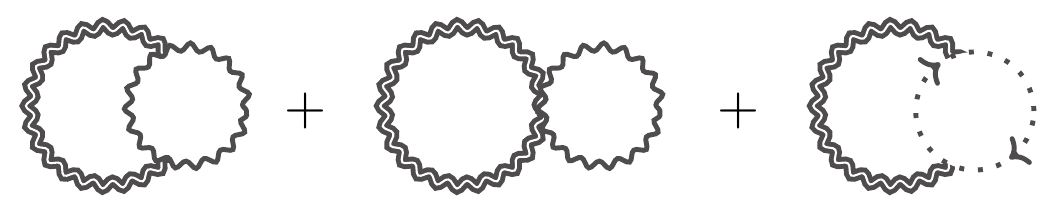}}.
\end{align}
The diagrams in $\pPowIrReg{3}{1}{m}$ become identical to those in $\pPowIrReg{3}{0}{s}$ in the semisoft region, while the diagrams of $\pPowIrReg{3}{0}{s}$ can be generated from those in $\pPowIrReg{3}{1}{m}$ by HTL resumming the one unresummed gluon line. Furthermore, the kinematics of the resummed gluon line in this mixed contribution is soft, and hence one may expand in the small gluonic momentum. In this sense, these hard corrections within the mixed $\pPowIrReg{3}{1}{m}$ contributions can be thought of as corrections to the HTL self energy.

Similarly, the diagrams in 
\begin{align}
    \aPowpIrReg{3}{2}{h} = 
    &\raisebox{-0.42\height}{\includegraphics[height=1.2cm]{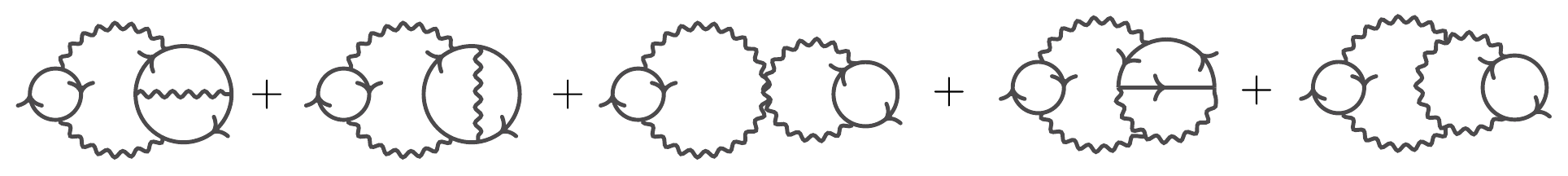}} \nonumber \\
    &+\raisebox{-0.42\height}{\includegraphics[height=1.2cm]{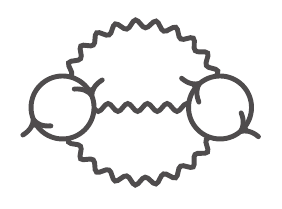}}+\raisebox{-0.42\height}{\includegraphics[height=1.2cm]{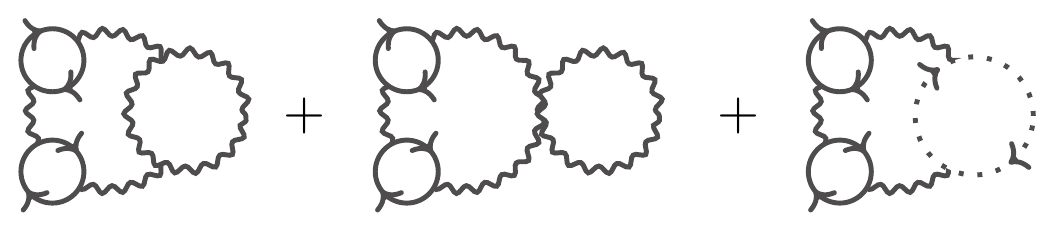}}
\end{align}
are intimately related to those in $\pPowIrReg{3}{1}{m}$ above. Upon HTL resumming one of the gluon lines, one obtains the diagrams of $\pPowIrReg{3}{1}{m}$. 

Now we turn to the second column in the right panel of \fig\ref{fig:visual_organization} related to the one-loop HTL diagrams. The one-loop diagram contributes at  \NLO{2} (namely, in $\pPowIrReg{2}{0}{s}$), so to contribute at \NLO{3} it has to be dressed with a hard quark line.\footnote{The hard gluon or ghost contributions already appear in $\pPowIrReg{3}{1}{m}$, and are related instead to the two-loop HTL diagrams.} This comes about naturally when one of the momenta running in the HTL self energies becomes hard, giving rise to the diagram
\begin{equation}
    \aPowpIrReg{3}{0}{m} = \raisebox{-0.42\height}{\includegraphics[height=1.2cm]{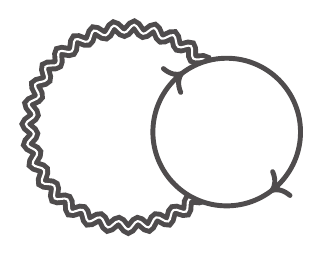}},
\end{equation}
where we can again expand in the soft gluon momentum. However, since the leading-order term in the small-momentum expansion gives back the lower-order one-loop HTL result, we must instead use the NLO soft kinematics \cite{Manuel:2016wqs,Carignano:2017ovz,Carignano:2019ofj} to obtain the \NLO{3} contribution.

The diagram in $\pPowIrReg{3}{0}{m}$ is on the other hand related to the graph\footnote{It is worth noting that this diagram represents the leading large-$N_f$ behavior of the pressure, which has been determined in Ref.~\cite{Gynther:2009qf} in the high-$T$ limit.}
\begin{align}
    \aPowpIrReg{3}{1}{h} = 
    &\raisebox{-0.42\height}{\includegraphics[height=1.2cm]{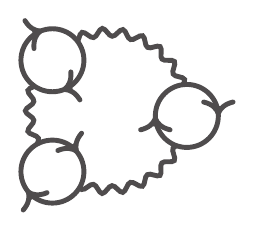}}.
\end{align}
For semisoft kinematics, the diagram in $\pPowIrReg{3}{0}{m}$ becomes identical to the one in $\pPowIrReg{3}{1}{h}$.

Finally, there remains one further subclass where no resummations are necessary, namely $\pPowIrReg{3}{0}{h}$ which contains the remaining IR-safe four-loop diagrams containing a single quark loop, here with the full quark kinematics. It reads
{
\allowdisplaybreaks
\begin{align}
    \aPowpIrReg{3}{0}{h}  = 
    &\raisebox{-0.42\height }{\includegraphics[height=1.3cm]{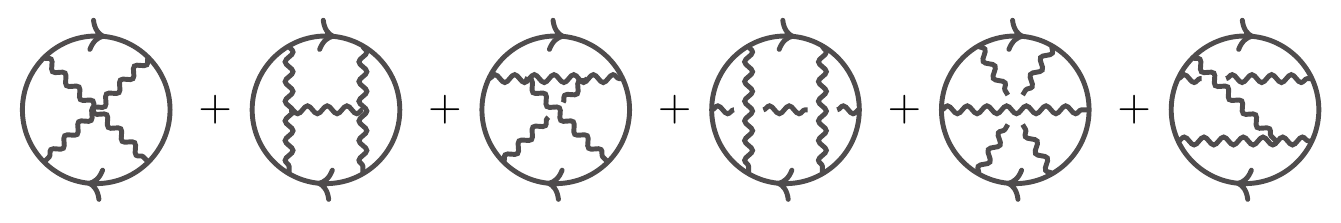}} \nn
    &\raisebox{-0.42\height }{\includegraphics[height=1.3cm]{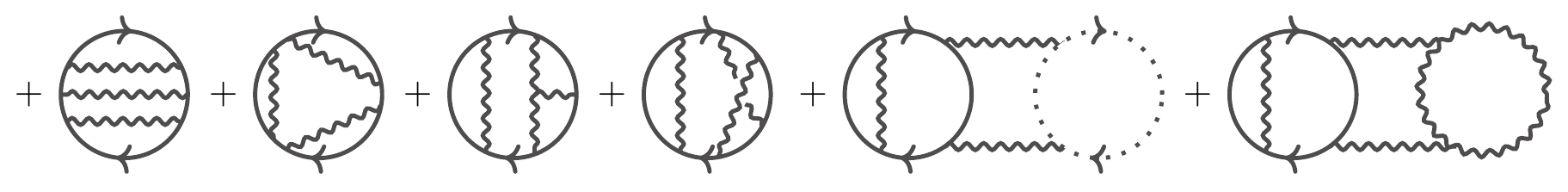}} \nn
    &\raisebox{-0.42\height }{\includegraphics[height=1.3cm]{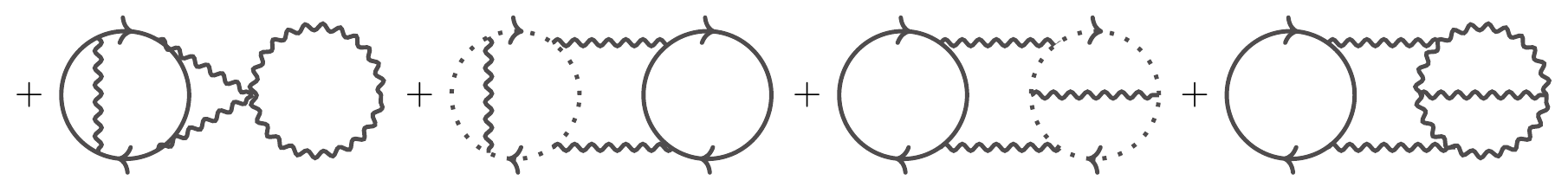}} \nn
    &\raisebox{-0.42\height }{\includegraphics[height=1.3cm]{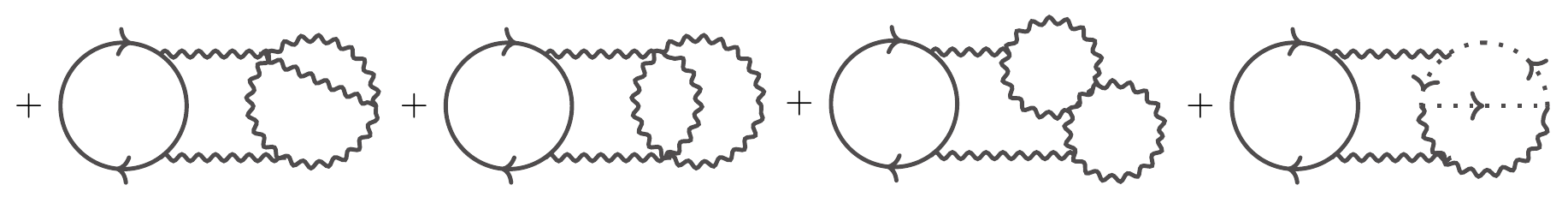}} \nn
    &\raisebox{-0.42\height }{\includegraphics[height=1.3cm]{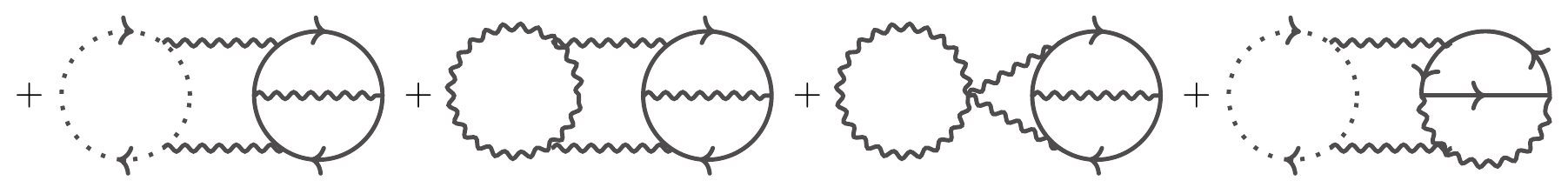}} \nn
    &\raisebox{-0.42\height }{\includegraphics[height=1.3cm]{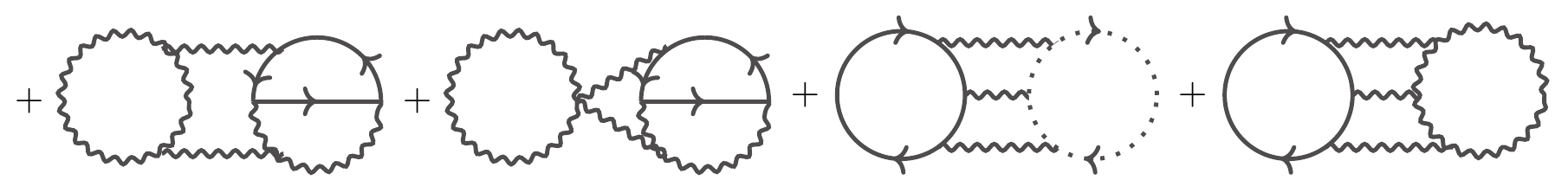}} \nn
    &\raisebox{-0.42\height }{\includegraphics[height=1.3cm]{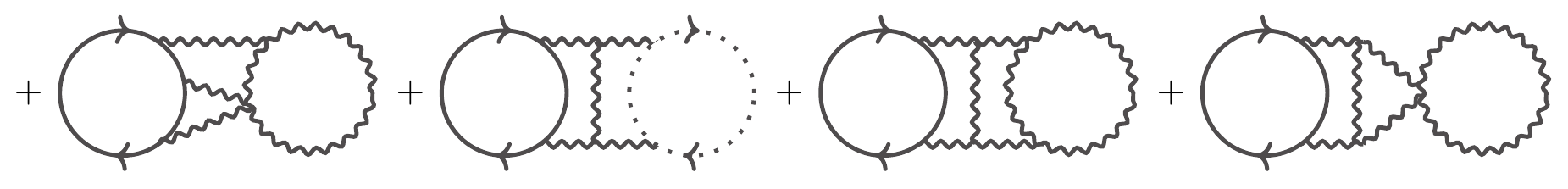}} \nn
    &\raisebox{-0.42\height }{\includegraphics[height=1.3cm]{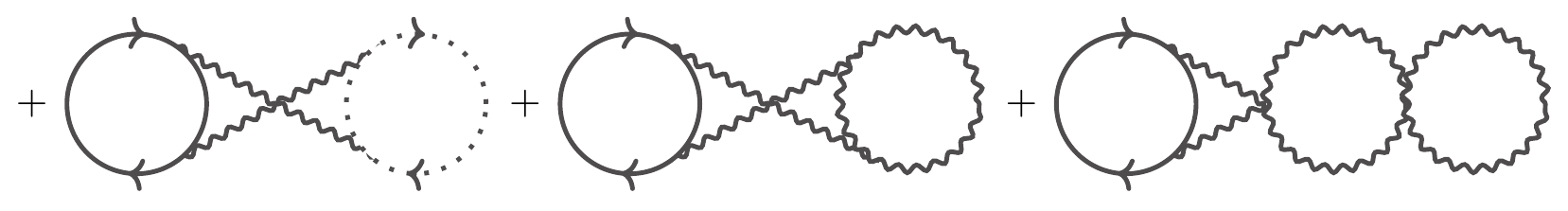}} \nn
    &\raisebox{-0.42\height }{\includegraphics[height=1.5cm]{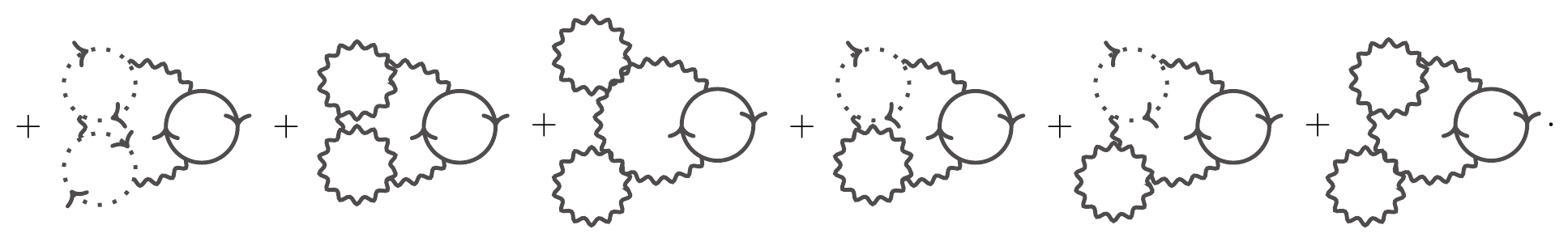}} 
\end{align}
}%
The set of all four-loop vacuum diagrams in QCD has previously been written down using a different organization in e.g.\ \Ref\cite{Kajantie:2001hv}. Comparing the two, we see that each diagram is correctly reproduced in our organization, and we have checked agreement with the symmetry factors as well.

Out of the different contributions to the \NLO{3} pressure, all carry dependence on the factorization scale $\Lh$, while all but the soft contributions, determined in this work, depend additionally on the renormalization scale $\bar{\Lambda}$. The latter of these scales is related to regulating UV divergences in the hard sector, and leads to the true scale dependence of the physical pressure. The dependence on $\Lh$, related to UV divergences in the soft contributions and IR divergences in the hard and mixed contributions, will on the other hand cancel upon summing all the different parts of the pressure together. For technical reasons, similar calculations carried out for the high-temperature pressure using the dimensionally reduced effective theory EQCD often set the scale parameters of the full and effective theories equal, but this is by no means mandatory, as it is equally possible to keep two scale parameters in the calculation, letting one regulate IR and the other UV divergences.

\subsection{What we compute in this work}
\label{sec:what_we_compute}

In this paper, we determine the contribution $\pDU{3}{s}$ to the cold-QM pressure, defined in \eq\eqref{eq:schematics}, which is equal to the fully soft subclass $\pPowIrReg{3}{0}{s}$ at \NLO{3}. While it does not constitute a complete new order in the weak-coupling expansion of the pressure, it amounts to a complete kinematic contribution that has furthermore been speculated to play a crucial role in the slow convergence of the quantity \cite{Blaizot:2000fc}. In addition, as we will show below, we can recover the known $O(\alpha_s^3 \ln^2 \alpha_s)$ contribution to the pressure from this region alone. 

In the computation, we use dimensional regularization and work in the limit of vanishing quark masses, which amounts to evaluating the full two-loop HTL pressure at zero temperature, without expanding in the in-medium effective mass scale. We note that there has been previous research on higher-order HTL thermodynamics \cite{Blaizot:2000fc,Andersen:2010wu,Andersen:2011sf,Mogliacci:2013mca, Haque:2014rua}. However, these works all expand the HTL diagrams in powers of the effective mass, and so do not perform the full resummation that we need.

The general structure of the paper is as follows. In \Sec\ref{sec:organizing}, we introduce the setup, conventions, and machinery used in the calculation of $\pDU{3}{s}$. We explain the power counting of all the contributions in detail and present notations that allow us to easily extract the UV-sensitive integral contributions. In \Sec\ref{sec:compute}, we then explain our steps for evaluating these integral expressions, and display results for the different contributions along with many details, especially for the UV-sensitive terms. Finally, in \Sec\ref{sec:pResult}, we present our final result for the pressure in the soft region. 

We also discuss cross-checks of our computation, remark upon the sizes of different contributions, and provide a small outlook for the remainder of the full \NLO{3} pressure. Following the main text is a large collection of appendices summarizing the Euclidean-space HTL framework used throughout this work, as well as the additional machinery that we have developed to tackle the computation. We have collected all of this into the appendices to aid future researchers who wish to use the Euclidean-space HTL framework in their work.

\section{Organizing the computation}
\label{sec:organizing}
\subsection{Starting expression and convention}

\begin{figure}
    \centering
        \includegraphics[width=0.5\textwidth]{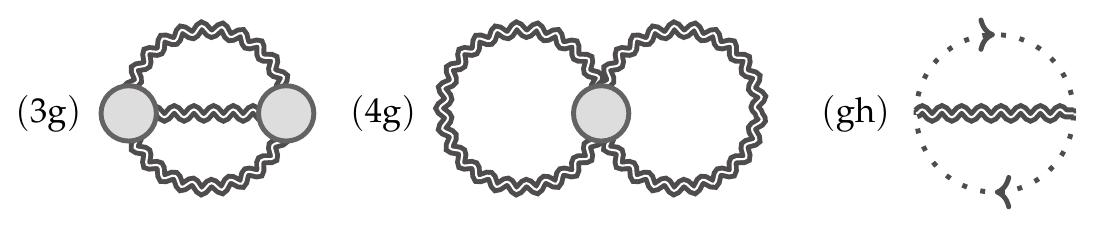}
    \caption{\label{fig:2_loop_graphs}The three different two-loop HTL diagrams contributing to the pressure at \NLO{3}, constituting the $\pDU{3}{s}$ contribution.
    }
\end{figure}

The expression corresponding to the fully soft contribution to the cold-QM pressure that we will evaluate in this paper is [cf.\ \eq\eqref{eq:schematics}]
\begin{equation}
    \apDU{3}{s}
    = 
    g^{2} N_c d_A \bigl[\II{3g} + \II{4g} +  \II{gh}  \bigr], 
\label{eq:2-loop_HTL}
\end{equation}
with $\II{3g}$, $\II{4g}$, $\II{gh}$ labeling the diagrams repeated in \fig\ref{fig:2_loop_graphs}. Here, $g = \sqrt{4 \pi \alpha_s}$ is the QCD gauge coupling, $N_c$ is the number of quark colors, and $d_A = N_c^2 - 1$ is the dimensionality of the adjoint representation of the SU($N_c$) gauge group, or the number of gluons. We will perform this computation in dimensional regularization in $D = d+1$ spacetime dimensions.

Eq.~(\ref{eq:2-loop_HTL}) is the expression for the two-loop HTL pressure examined also in \Ref\cite{Andersen:2002ey}, but in the zero temperature limit and at nonzero density. 
To evaluate this expression, we find it easier manipulate it using techniques that will be discussed below, rather than using the expressions in \Ref\cite{Andersen:2002ey} as a starting point.

First, we write down explicit expressions for the resummed two-loop graphs under study. Using the Feynman rules of \app\ref{sec:feynman_rules}, we readily obtain
{\allowdisplaybreaks
\begin{align}
\label{eq:G3g}
    \II{3g} &= \frac{1}{12} \intkpr \Gamma^{\mu \nu \rho}(K,P,R) \Gamma^{\mu' \nu' \rho'}(K,P,R) \D{\mu \mu'}(K) \D{\nu \nu'}(P) \D{\rho \rho'}(R), \\
\label{eq:G4g}
    \II{4g} &= -\frac{1}{8} \intkp \Gamma^{\mu \nu \rho \sigma}(K,P,-K,-P) \D{\mu \rho}(K) \D{\nu \sigma}(P), \\
\label{eq:Ggh}
    \II{gh} &= \frac{1}{2} \intkpr \frac{K^{\mu} P^{\nu}}{K^2 P^2} \D{\mu \nu}(R). 
\end{align}
}%
These expressions form the starting point of our diagrammatic analysis. We have here introduced the following notations, which are discussed in more detail in the appendices:

\paragraph*{Metric and vector conventions:} We work in Euclidean 
space
with metric $\delta^{\mu\nu}$. We write the components of our four-vectors 
as
$K^{\mu} = (K^0,k^{i})$, 
where $K^0 = K_0$,  $k^i = k_i$ for $i=1,2,3$.   The scalar product between two Euclidean four-vectors $K$ and $P$ is given by 
$K \cdot P = K_0P_0 + \kt \cdot \pt$,
and we use the notation $|\vec{k}|^{2} = \kt \cdot \kt$ for the magnitudes of the spatial part of the momentum.
We will also repeatedly use the notation
\begin{equation}
\label{eq:unitVectors}
    \hat{K} = \frac{K}{|K|}, \quad
    \hat{\vec{k}} = \frac{\vec{k}}{|\vec{k}|}, 
\end{equation}
for the unit vectors in the direction of $K$ and $\vec{k}$, respectively, the former already defined around \eq\eqref{eq:htl_pi_lim}.

\paragraph*{Integration measures:} The integration measures are defined in $D = d+1$ space-time dimensions. The symmetric integral $\intkpr$ we use frequently is defined as
\begin{equation}
\label{eq:intmeasurePKR}
    \intkpr \equiv 
    \left (\frac{e^{\gamE}\Lh^2}{4\pi}\right )^{4-D}
    \int \frac{\upd^D K}{(2\pi)^D} \int \frac{\upd^D P}{(2\pi)^D}  \int \frac{\upd^D R}{(2\pi)^D}(2\pi)^D \delta^{(D)}(K+P+R),
\end{equation}
where $\Lh$ is a factorization scale and the factor $(e^{\gamE}/4\pi)^{(4-D)}$, with $\gamE$ the Euler--Mascheroni constant, is introduced so that one absorbs the \UV-divergent part with a universal constant. The integral $\intkp$ in \eq\eqref{eq:G4g} follows from $\intkpr$ upon doing the trivial integration over $R$.

\paragraph*{Propagator:} The HTL-resummed gluon propagator $\D{\mu \nu}$ is defined in the covariant gauge as
\begin{equation}
\label{eq:fullpropa}
    \D{\mu\nu}(K) \equiv \ProjIX{\mu \nu}{T}{K} G_T(K) + \ProjIX{\mu \nu}{L}{K} G_L(K)  + \xi \frac{K^{\mu}K^{\nu}}{(K^2)^2},
\end{equation} 
where the parameter $\xi$ fixes the gauge and 
\begin{equation}
\label{eq:gfuncdef}
G_I(K) \equiv \frac{1}{K^2 + \Pi_{I}(K)}, \quad I \in \{ \text{T}, \text{L} \} .
\end{equation}
In this equation, and from this point on, we drop the HTL label on the HTL self energies, for brevity. It turns out that the computation is most efficient to perform in the $\xi = 1$ gauge, which we shall use throughout the rest of the text. In the class of covariant $R_\xi$-gauges, the gauge parameter $\xi$ only appears in the propagator, and we have explicitly checked that the expression in \eq\eqref{eq:2-loop_HTL} is gauge-independent in the sense of being independent of the $\xi$-parameter, with the conclusion being supported by \Ref\cite{Andersen:2002ey}. 


The standard projectors $\prjoT^{\mu \nu}(\hat{K})$ and $\prjoL^{\mu \nu}(\hat{K})$ used in \eq\eqref{eq:fullpropa} are defined in \Sec\ref{sec:HTLappendix}, and the reader is advised to consult the appendix when necessary. We do repeat here the definition of the symmetric and transverse HTL self-energy tensor
\begin{equation}
    \Pi^{\mu\nu}(K) = \ProjIX{\mu\nu}{T}{K} \Pi_\text{T}(K) + \ProjIX{\mu\nu}{L}{K} \Pi_\text{L}(K)
\end{equation}
where the coefficient functions are given by 
\begin{equation}
\label{eq:piTandpiL}
\begin{split}
\Pi^{\mu\mu}(K) & = (d-1)\Pi_\text{T}(K) + \Pi_\text{L}(K),\\
\Pi^{00}(K) & =  \frac{\vert \kt\vert^2}{K^2} \Pi_\text{L}(K).
\end{split}
\end{equation}
The trace of the one-loop HTL self energy is defined to be $\mE^2$, so that $\mE^2 \equiv \Pi^{\mu \mu}(K)$. An alternative explicit definition of $\Pi^{\mu\nu}(K)$ is given in the aforementioned appendix.

\paragraph*{Vertices:} The effective three- (3g) and four-gluon (4g) vertices are obtained by adding the HTL loop to the bare vertex. 
We write these quantities as
\begin{equation}
    \begin{split}
        \Gamma^{\mu\nu\rho}(P,Q,R) & \equiv \Gamma_{0}^{\mu\nu\rho}(P,Q,R) + \delta\Gamma^{\mu\nu\rho}(P,Q,R), \\
        \Gamma^{\mu\nu\rho\sigma}(P,Q,R,S) & \equiv \Gamma_{0}^{\mu\nu\rho\sigma}(P,Q,R,S) + \delta\Gamma^{\mu\nu\rho\sigma}(P,Q,R,S),
    \end{split}
\end{equation}
where the subscript ``0'' is always understood as referring to the bare quantity. The HTL vertices $\delta \Gamma$ are only defined when the sum of all of their arguments is zero, and they have the property that they are totally symmetric in their indices and traceless in any pair of indices. Furthermore, the 3g HTL vertex is even (odd) under even (odd) permutations of the momenta $P$, $Q$, and $R$, while the 4g HTL vertex is even under all permutations of the momenta $P$, $Q$, $R$, and $S$. We note here that like the HTL self energies, the HTL vertices are proportional to $\mE^{2}$, and satisfy the generalized Ward identities
\begin{equation}
\label{eq:WardVertices}
    \begin{split}
        P^{\mu} \delta\Gamma^{\mu\nu\rho}(P,Q,R) & = \Pi^{\nu\rho}(R) - \Pi^{\nu\rho}(Q), \\
        P^{\mu} \delta\Gamma^{\mu\nu\rho\sigma}(P,Q,R,S) & = \delta\Gamma^{\nu\rho\sigma}(Q,R,S+P) - \delta \Gamma^{\nu\rho\sigma}(Q+P,R,S).
    \end{split}
\end{equation}
We also note that the HTL vertices are independent of the gauge parameter $\xi$, even without fixing the gauge. The explicit integral expressions for the $d$-dimensional HTL vertices are given in \app\ref{sec:HTLvertices}.

\subsection{Isolating the UV-sensitive terms}

Due to the two-loop structure of the quantity under consideration and the fact that the HTL self energies are independent of the magnitude of the gluonic momenta, we can deduce that the general structure of our final result for $\apDU{3}{s}$ will be 
\begin{equation}
    \apDU{3}{s}
    = 
    \frac{g^{2} N_c d_A \mE^{4}}{(2\pi)^{6}}  
    \left( \frac{\mE}{\Lh} \right)^{-4\epsilon}
    \biggl[ \frac{p_{-2}}{(2 \epsilon)^{2}} + \frac{p_{-1}}{2 \epsilon} + p_{0} \biggr].
\label{eq:2loopSeriesSimple}
\end{equation}
When this is expanded out fully in $\epsilon$, our final result for the soft contributions to the pressure of \CQM\ becomes
\begin{equation}
    \apDU{3}{s} 
    = 
    \frac{g^{2}\! N_c d_A \mE^{4}}{(2\pi)^{6}}  
    \biggl\{
    \frac{p_{-2}}{(2 \epsilon)^{2}} + \frac{p_{-1} \!-\! 2 p_{-2} \ln\!\big(\frac{\mE}{\Lh}\big)}{2 \epsilon} + \Bigl[ p_{0} - 2 p_{-1} \ln\!\Big(\frac{\mE}{\Lh}\Big) + 2 p_{-2} \ln^{2}\!\Big(\frac{\mE}{\Lh}\Big) \Bigr]\!
    \biggr\}
\label{eq:2loopSeriesExpanded}
\end{equation}
and contains both double and single-logarithmic terms of the ratio $\mE / \Lh$.
Note that, since we have resummed the soft sector, the expression is IR safe. Thus, the terms in \eq\eqref{eq:2loopSeriesSimple} that enter with negative powers of $\epsilon$ arise from the UV. 
We are thus led to the following conclusion: in order to isolate the $p_{-2}$ and $p_{-1}$ parts of the pressure, we must isolate the UV behavior of the integrals \eqs\eqref{eq:G3g}--\eqref{eq:Ggh}. To do this, we will first power count to isolate the power-law divergences in the UV (which do not contribute at all in dimensional regularization). This in turn will lead us to  introduce a notation to isolate the terms identified by the power counting.

\subsubsection{Power counting} 
In the UV, the bare and HTL parts of the vertices and propagators no longer enter at the same order in $\mE$. Therefore, by expanding the vertices and unfolding the propagators, we will be able to isolate the UV-sensitive terms. Let us first begin by expanding out the vertices into their bare $(0)$ and HTL $(\mathrm{H})$ parts, and splitting $\II{3g}$ and $\II{4g}$ into the following pieces:
\begin{equation}
    \begin{split}
        \label{eq:G4g3gsplit}
        \II{3g} & = \IIUD{(0,0)}{3g} + 2 \IIUD{(0,H)}{3g} + \IIUD{(H,H)}{3g}, \\
        \II{4g} & = \IIUD{(0)}{4g} + \IIUD{(H)}{4g}, 
    \end{split} 
\end{equation}
with
\begin{equation}
    \begin{split}
        \label{eq:3gsplit}
        \IIUD{(0,0)}{3g} &= \frac{1}{12} \intkpr \Gamz{\mu \nu \rho}(K,P,R) \Gamz{\mu' \nu' \rho'}(K,P,R) \D{\mu \mu'}(K) \D{\nu \nu'}(P) \D{\rho \rho'}(R), \\
        \IIUD{(0,H)}{3g} &= \frac{1}{12} \intkpr \Gamz{\mu \nu \rho}(K,P,R) \dGam{\mu' \nu' \rho'}(K,P,R) \D{\mu \mu'}(K) \D{\nu \nu'}(P) \D{\rho \rho'}(R), \\
        \IIUD{(H,H)}{3g} &= \frac{1}{12} \intkpr \dGam{\mu \nu \rho}(K,P,R) \dGam{\mu' \nu' \rho'}(K,P,R) \D{\mu \mu'}(K) \D{\nu \nu'}(P) \D{\rho \rho'}(R),
    \end{split}
\end{equation}
and
\begin{equation}
    \begin{split}
        \label{eq:4gsplit}
        \IIUD{(0)}{4g} & = -\frac{1}{8} \intkp \Gamz{\mu \nu \rho \sigma}(K,P,-K,-P) \D{\mu \rho}(K) \D{\nu \sigma}(P), \\
        \IIUD{(H)}{4g} & = -\frac{1}{8} \intkp \dGam{\mu \nu \rho \sigma}(K,P,-K,-P) \D{\mu \rho}(K) \D{\nu \sigma}(P).
    \end{split}
\end{equation}
The factor of two in the second line of \eq\eqref{eq:G4g3gsplit} follows by symmetry.  For the purposes of power counting, we need  the scalings for each of the vertices in the region $K, P, R \sim \Lambda$ with $\Lambda \gg \mE$. This can be determined from the explicit expressions in \apps\ref{sec:feynman_rules} and \ref{sec:HTLvertices}.  Since our leading term in the $\epsilon$ expansion \eqref{eq:2loopSeriesSimple} is $O(\epsilon^{-2})$, subleading contributions will also contribute to the divergent terms. 
For completeness, we include in \tab\ref{tab:vertexPowCount} the vertex scalings in all of the possible momentum regions.
\begin{table}
\begin{ruledtabular}
    \begin{tabular}{llll}
        Vertex & $K,P,R \sim \mE$ & $K\sim \mE;\, P,R \sim \Lambda$  & $K, P, R \sim \Lambda$ \\
        \hline
        $\Gamz{\mu\nu\rho}(K,P,R)$  & $\sim \mE$ & $\sim \Lambda + \mE$ & $\sim \Lambda$  \\
        $\dGam{\mu\nu\rho}(K,P,R)$ & $\sim \mE$ & $\sim \mE^{2}\Lambda^{-1} + \mE^{3} \Lambda^{-2} + \cdots$  &   $\sim \mE^{2}\Lambda^{-1}$  \\
        $\Gamz{\mu\nu\rho\sigma}(K, P, -K, -P)$ & $\sim\Lambda^{0}$ & $\sim \Lambda^{0}$ &   $\sim \Lambda^{0}$  \\
        $\dGam{\mu\nu\rho\sigma}(K, P, -K, -P)$ & $\sim \Lambda^{0}$ & $\sim \mE^{2}\Lambda^{-2} + \mE^{4} \Lambda^{-4} + \cdots$ & $\sim \mE^{2}\Lambda^{-2}$    
    \end{tabular}
\end{ruledtabular}
\caption{\label{tab:vertexPowCount} Scales of the bare and resummed vertices in different momentum regions. Note that the second column also represents the other cases where the three momenta $K, P, R$ are simply permuted.}
\end{table}

Now, unlike the vertices, which split into a simple sum of two terms of different orders in the UV, the propagators in \eq\eqref{eq:fullpropa} expand into an infinite number of terms there, namely,
\begin{equation}
    \label{eq:propExpand}
        \D{\mu \nu}(K) \simeq \frac{1}{K^{2}} \delta^{\mu \nu} - \frac{\Pi(K)^{\mu \nu}}{(K^{2})^2} + \frac{[\Pi(K)^{2}]^{\mu \nu}}{(K^{2})^{3}} - \cdots 
\end{equation}
where we have used a compact notation for the power of $\Pi^{\mu \nu}(K)$,
\begin{equation}
    [\Pi(K)^{n}]^{\mu \nu} = \overbrace{\Pi^{\mu \alpha_1}(K) \Pi^{\alpha_1\alpha_2}(K) \cdots \Pi^{\alpha_{n-1}\nu}(K)}^{n \, \Pi\text{s}}
\end{equation}
and have used the simple form of the bare gluonic propagator in $\xi = 1$ gauge in the leading term, namely 
\begin{equation}
\label{eq:bareProp}
    \DZ^{\mu\nu}(K) \equiv \frac{1}{K^2}\delta^{\mu\nu}.
\end{equation}
Taking a momentum $K$ of the order of some large $\Lambda \gg \mE$, the expansion in \eq\eqref{eq:propExpand} goes as
\begin{equation}
\label{eq:propExpandPowCountHard}
       \D{\mu \nu}(K) \sim \Lambda^{-2} + \mE^{2} \Lambda^{-4} + \mE^{4} \Lambda^{-6} + \cdots,
\end{equation}
which shows that we have an infinite number of contributions, with the higher terms becoming less important.  On the other hand, when the momentum flowing through the vertex becomes soft $K \sim \mE$, the propagator simply scales as 
\begin{equation}
\label{eq:propExpandPowCountSoft}
    \D{\mu \nu}(K) \sim \mE^{-2},
\end{equation}
so that no expansion is possible.

Let us now use these scalings to power count one piece of \eq\eqref{eq:3gsplit}. When $K, P, R \sim \Lambda$, the term with two bare vertices scales as 
\begin{equation}
\begin{split}
\label{eq:3gSchematicHard}
    \IIUD{(0,0)}{3g} \sim&\, \Lambda^{4} \Lambda^{4} 
    \bigl( \Lambda \bigr) 
    \bigl( \Lambda \bigr) 
    \biggl( \frac{1}{\Lambda^{2}} + \frac{\mE^{2}}{\Lambda^{4}} + \frac{\mE^{4}}{\Lambda^{6}}\cdots \biggr) 
    \biggl( \frac{1}{\Lambda^{2}} + \frac{\mE^{2}}{\Lambda^{4}} + \cdots \biggr) 
    \biggl( \frac{1}{\Lambda^{2}} + \frac{\mE^{2}}{\Lambda^{4}} + \cdots \biggr) \\
    \sim&\, \Lambda^{4} + \Lambda^{2} \mE^{2} + \mE^{4} + \cdots,
\end{split}
\end{equation}
where the two leading $\Lambda^{4}$ terms come from the integration measure,\footnote{Note that we are power counting in $D = 4$.} the next two $\Lambda$ terms are from the bare vertices, and the final three series are from the propagators. We thus see that $\IIUD{(0,0)}{3g}$ contains two power-law divergent terms in this hard--hard UV region, which we would like to peel away. On the other hand, in the hard--soft region $K \sim \mE; P, R \sim \Lambda$, we have
\begin{equation}
\begin{split}
\label{eq:3gSchematicSoft}
    \IIUD{(0,0)}{3g} \sim&\, 
    \mE^{4} 
    \Lambda^{4} 
    \bigl( \Lambda + \mE )(\Lambda + \mE \bigr) 
    \biggl( \frac{1}{\mE^{2}} \biggr) 
    \biggl( \frac{1}{\Lambda^{2}} + \frac{\mE^{2}}{\Lambda^{4}} + \cdots \biggr) 
    \biggl( \frac{1}{\Lambda^{2}} + \frac{\mE^{2}}{\Lambda^{4}} + \cdots \biggr)  \\
    \sim&\, \Lambda^{2} \mE^{2} + \Lambda \mE^{3} + \mE^{4} + \cdots,
\end{split}
\end{equation}
which shows that there is also a power-law divergence here, even though one of the propagators is still resummed. Observe that letting one of the lines become soft shifts the leading term in the expansion of the integral to a higher power in $\mE$. Note also that in this region, there are subleading contributions from the vertices, in addition to the propagators. Finally, there is the soft--soft region. However, this region does not probe the UV, and so will not lead to any divergences at all.

Inspecting the above equations, we see that these power-law divergent terms arise from the first few leading terms in the expansion of the propagators, which motivates us to introduce a new notation in the following section.

\subsubsection{Peeling away the bare propagators} 
We have called the leading term in \eq\eqref{eq:propExpand} (the bare propagator) $\DN{0}^{\mu \nu}(K)$; let us extend this notation to label the other terms in the expansion as well:
\begin{equation}
    \begin{split}
        \DN{n}^{\mu \nu}(K) \equiv&\, (-1)^{n} [\overbrace{\DZ(K)\cdot\Pi(K)\cdot\DZ(K)\cdots}^{n\,{\Pi}s}\DZ(K)]^{\mu \nu}, \qquad n\geq 0, \\
        ={}& (-1)^{n} \frac{[\Pi(K)^{n}]^{\mu \nu}}{(K^{2})^{n+1}}.
    \end{split}
\end{equation}
Here we make the identification $[\Pi(K)^0]^{\mu \nu} = \delta^{\mu \nu}$ to match the leading term.  This allows us to write the expansion in \eq\eqref{eq:propExpand} as
\begin{equation}
    \D{\mu\nu}(K) \simeq\, \DN{0}^{\mu\nu}(K) + \DN{1}^{\mu\nu}(K) + \DN{2}^{\mu\nu}(K) + \cdots.
\end{equation}
We can now introduce the following notation for the resummed propagator with the $n$ leading terms removed:
\begin{equation}
    \DDN{n}^{\mu\nu}(K) \equiv\, \D{\mu\nu}(K) - \sum_{k = 0}^{n - 1} \DN{k}^{\mu \nu}(K), \qquad n \ge 1.
\end{equation}
Consequently, the $\DDN{n}^{\mu\nu}(K)$ are still resummed expressions, while the $\DN{n}^{\mu\nu}(K)$ are not. Note that both $\DN{n}^{\mu\nu}(K)$ and $\DDN{n}^{\mu\nu}(K)$ are $D$-dimensionally transverse for every $n$, and that the following relations hold for any $n\ge 1$:
{
    \allowdisplaybreaks
\begin{align}
    \DDN{n}^{\mu \nu}(K) ={}&\, (-1)^{n} [\overbrace{\DZ(K)\cdot\Pi(K)\cdot\DZ(K)\cdots}^{n\,{\Pi}s} D(K)]^{\mu\nu},  \nn
        ={}& (-1)^{n} \frac{[\Pi(K)^{n}]^{\mu \alpha}}{(K^{2})^{n}} \D{\alpha \nu}(K), \label{eq:DDN_explicit}\\
    \D{\mu\nu}(K) ={}&\, \DN{0}^{\mu\nu}(K) + \DN{1}^{\mu\nu}(K) + \cdots + \DN{n-1}^{\mu\nu}(K) + \DDN{n}^{\mu\nu}(K) \label{eq:DDN_sum},\\
    \DDN{n-1}^{\mu\nu}(K) ={}&\, \DN{n-1}^{\mu\nu}(K) + \DDN{n}^{\mu\nu}(K), \label{eq:DDN_recursion} \\
    \DN{n}^{\mu\nu}(K) \sim{}& \mE^{2n} \Lambda^{-2(n+1)} \quad\text{in UV,} \label{eq:DNpowCount} \\
    \DDN{n}^{\mu\nu}(K) \sim{}& \mE^{2n} \Lambda^{-2(n+1)} \quad\text{in UV.} \label{eq:DDNpowCount}
\end{align}
}%
Notice also the full propagator at the end of both lines of \eq\eqref{eq:DDN_explicit}, and the fact that \eq\eqref{eq:DDN_sum} is not a partial sum of an infinite series, but is exact. 

\subsubsection{Applying the new notation} 

Using this new notation, we can make the schematic expressions in \eqs\eqref{eq:3gSchematicHard}--\eqref{eq:3gSchematicSoft} more explicit. To this end, we rewrite $\IIUD{(0,0)}{3g}$ into the following form:
\begin{align}
    \IIUD{(0,0)}{3g} =& 
    \frac{1}{12} \intkpr \!\!\!
    \Gam{\mu \nu \rho}_{0,KPR} 
    \Gam{\mu' \nu' \rho'}_{0,KPR} 
    \bigl[ \DZ(K) + \DDN{1}(K) \bigr]^{\mu \mu'} 
    \bigl[ \DZ(P) + \DDN{1}(P) \bigr]^{\nu \nu'} 
    \bigl[ \DZ(R) + \DDN{1}(R) \bigr]^{\rho \rho'} \nn
    =& \frac{1}{12} \intkpr 
    \Gam{\mu \nu \rho}_{0,KPR} 
    \Gam{\mu' \nu' \rho'}_{0,KPR} 
    \bigl[ 
        \Dz{\mu \mu'}(K) 
        \Dz{\nu \nu'}(P) 
        \Dz{\rho \rho'}(R) 
        + 3 \Dz{\mu \mu'}(K) 
        \Dz{\nu \nu'}(P) 
        \Dn{1}{\rho \rho'}(R) \nn
        &\phantom{\intkpr 
        \Gam{\mu \nu \rho}_{0,KPR} 
        \Gam{\mu' \nu' \rho'}_{0,KPR} \bigl[ }
        + 3 \Dz{\mu \mu'}(K) 
        \Dz{\nu \nu'}(P) 
        \DDn{2}{\rho \rho'}(R)
        + 3 \DDn{1}{\mu \mu'}(K) 
        \DDn{1}{\nu \nu'}(P)
        \Dz{\rho \rho'}(R)  \nn
        &\phantom{\intkpr 
        \Gam{\mu \nu \rho}_{0,KPR} 
        \Gam{\mu' \nu' \rho'}_{0,KPR} \bigl[ }
        + 
        \DDn{1}{\mu \mu'}(K) 
        \DDn{1}{\nu \nu'}(P)
        \DDn{1}{\rho \rho'}(R)
    \bigl].
\end{align}%
We stress that this equation is not an expansion, but holds exactly.\footnote{Note that, from now on, we will use this more compact expression for the vertices.} Observe that the two terms on the second line of the above equation contain only $\Delta$s, which have no mass scale. Therefore, these two terms are power-law divergent and thus vanish in dimensional regularization. These terms correspond precisely to the two power-law divergent terms in \eq\eqref{eq:3gSchematicHard}. The remaining power-law divergence in \eq\eqref{eq:3gSchematicSoft} is located in one particular part of the term containing $\DDN{2}$ on the third line of the above equation, as can be seen by counting the bare propagators. This divergence will not become explicit until we insert the explicit expression for the vertices and perform the contractions.

In addition to identifying the terms that contain only $\Delta$s and are trivially power-law divergent in the UV, this notation also allows us to identify precisely the terms which are logarithmically UV-sensitive. We can do this as follows. From the scaling of the $\DN{n}$ and $\DDN{n}$ in \eqs\eqref{eq:DNpowCount}--\eqref{eq:DDNpowCount}, we see that we can determine the scaling of a product of $\DN{n}$s and $\DDN{n}$s by simply summing the subscripts: if the sum is $N$ then the product goes as $\mE^{2N} \Lambda^{-2(N+1)}$ in the UV. However, we know that the logarithmically UV-sensitive terms (ones that lead to the $O(\epsilon^{-2})$ and $O(\epsilon^{-1})$ terms in the expansion) must scale as $\mE^{4}$ times a dimensionless integral in the UV, as such a structure allows it to contribute through the whole UV tail of the integral. 
This gives us a powerful prescription for identifying the logarithmally UV-sensitive terms: we can simply sum the subscripts on the propagator terms (and include one $\mE^{2}$ per HTL vertex if necessary) and see whether the result is $\mE^{4}$.

We now carry out the above procedure for all of the terms in \eq\eqref{eq:G4g3gsplit} as well as for $\II{gh}$ to peel away all the power-law UV divergences
and to identify the logarithmically UV-sensitive terms. We then find the following: for the parts of $\II{3g}$,
{
    \allowdisplaybreaks
\begin{align}
    \IIUD{(0,0)}{3g} =& 
    \frac{1}{12} \intkpr\!\!\!
    \Gam{\mu \nu \rho}_{0,KPR} 
    \Gam{\mu' \nu' \rho'}_{0,KPR} 
    \bigl[ 
        3 \Dz{\mu \mu'}(K) 
        \Dz{\nu \nu'}(P) 
        \DDn{2}{\rho \rho'}(R)
        + 3 \DDn{1}{\mu \mu'}(K) 
        \DDn{1}{\nu \nu'}(P)
        \Dz{\rho \rho'}(R)
    \bigl] 
    \nn
    &+\frac{1}{12} \intkpr\!\!\! 
    \Gam{\mu \nu \rho}_{0,KPR} 
    \Gam{\mu' \nu' \rho'}_{0,KPR} 
    \bigl[ 
        \DDn{1}{\mu \mu'}(K) 
        \DDn{1}{\nu \nu'}(P)
        \DDn{1}{\rho \rho'}(R)
    \bigl]  \nn
    \equiv& 
    \Bigl[\IIUD{(0,0)}{3g}\Bigr]^{\text{UV}}_{002} + 
    \Bigl[\IIUD{(0,0)}{3g}\Bigr]^{\text{UV}}_{110} + 
    \Bigl[\IIUD{(0,0)}{3g}\Bigr]_{111}, \label{eq:I3G00def} \\[2ex]
    \IIUD{(0,H)}{3g} =&
    \frac{1}{12} \intkpr\!\!\! 
        \dGam{\mu' \nu' \rho'}_{KPR} 
        \Gam{\mu \nu \rho}_{0,KPR} 
        \bigl[ 
            3 
            \Dz{\mu \mu'}(K)
            \Dz{\nu \nu'}(P)
            \DDn{1}{\rho \rho'}(R) 
        \bigr]
    \nn
    &+\frac{1}{12}\intkpr\!\!\! 
    \dGam{\mu' \nu' \rho'}_{KPR} 
    \Gam{\mu \nu \rho}_{0,KPR} 
    \bigl[ 
        3 
        \DDn{1}{\mu \mu'}(K)
        \DDn{1}{\nu \nu'}(P)
        \Dz{\rho \rho'}(R) 
        +
        \DDn{1}{\mu \mu'}(K) 
        \DDn{1}{\nu \nu'}(P) 
        \DDn{1}{\rho \rho'}(R)
    \bigr] \nonumber \\*
    \equiv& 
    \Bigl[\IIUD{(0,H)}{3g}\Bigr]^{\text{UV}}_{001} + 
    \Bigl[\IIUD{(0,H)}{3g}\Bigr]_{110} + 
    \Bigl[\IIUD{(0,H)}{3g}\Bigr]_{111}, \label{eq:I3G0Hdef} \\[2ex]
    \IIUD{(H,H)}{3g} =& 
    \frac{1}{12}\intkpr\!\!\!
    \dGam{\mu \nu \rho}_{KPR}
    \dGam{\mu' \nu' \rho'}_{KPR}
    \bigl[
        D^{\mu \mu'}(K)
        D^{\nu \nu'}(P)
        D^{\rho \rho'}(R)
    \bigr] \nonumber \\*
    \equiv&
    \Bigl[\IIUD{(H,H)}{3g}\Bigr]^{\text{UV}} ; 
\end{align}
}
for the parts of $\II{4g}$,
{
\allowdisplaybreaks
\begin{align}
\label{eq:split4g}
    \IIUD{(0)}{4g} =& 
    -\frac{1}{8}\intkp \!\!\! 
    \Gamz{\mu \nu \rho \sigma} \DDn{1}{\mu \rho}(K) \DDn{1}{\nu \sigma}(P) \nn
    \equiv& 
    \Bigl[\IIUD{(0)}{4g}\Bigr]^{\text{UV}}_{11}, \\[2ex]
    \IIUD{(H)}{4g} =& 
    -\frac{1}{8}\intkp \!\!\!
    \dGam{\mu \nu \rho \sigma}
        \DDn{1}{\mu \rho}(K) \DDn{1}{\nu \sigma}(P) \nn
        \equiv& \Bigl[\IIUD{(H)}{4g}\Bigr]_{11}; \label{eq:I4GHdef}
\end{align}
}%
and for $\II{gh}$,
\begin{align}
\label{eq:splitgh}
    \II{gh} =&
    \frac{1}{2}\intkpr 
    \frac{K^{\mu} P^{\nu}}{K^{2} P^{2}} \DDn{2}{\mu \nu}(R)  \nn
    \equiv&
    \left[\II{gh}\right]^{\text{UV}}_{2}.
\end{align}
Here, we have introduced the notation $[I]_{ijk}$ to mean the integral $I$ with the propagators replaced by $D(K) \mapsto \DDN{i}(K)$, $D(P) \mapsto \DDN{j}(P)$, and $D(R) \mapsto \DDN{k}(R)$ if $i, j, k > 0$. In the case of $i, j, k = 0$, the replacement is instead $D(K) \mapsto \DN{0}(K)$, $D(P) \mapsto \DN{0}(P)$, or $D(R) \mapsto \DN{0}(R)$. (In the case of fewer subscripts, the notation is the same, but replacing only those propagators appearing in $I$.) We have additionally added a superscript ``UV'' to those terms which are logarithmically UV sensitive, and hence contribute to the $O(\epsilon^{-2})$ or $O(\epsilon^{-1})$ terms of the final answer. From this point on, we shall simply refer to them as the ``UV terms''. Consequently, those terms which do not have the UV labeling are UV finite and thus can be computed in $d = 3$ spatial dimensions. Most of these non-UV-sensitive terms cannot be analytically simplified much more, and so for the rest of this organizational section we shall not manipulate them. Their contributions are listed in \Sec\ref{sec:Ifinite}.

\subsubsection{Performing the contractions} 
In most of the UV terms, the Lorentz contractions are relatively straightforward and lead to some simplifications. The general procedure for evaluating all of them, except 
those in
$\bigl[\IIUD{(H,H)}{3g}\bigr]^{\text{UV}}$, which we will separately consider in \Sec\ref{sec:IF}, is as follows:
\begin{enumerate}
    \item Substitute in the form of the bare propagator $\DN{0}$ in \eq\eqref{eq:bareProp} to the expressions, which contracts some of the indices of the vertices together.
    \item Use both (i) the expression for the bare vertex and (ii) the generalized Ward identities in \eq\eqref{eq:WardVertices} to eliminate all vertices from the expressions. This allows us to avoid the explicit expressions for the HTL vertices.
    \item Use the symmetry of the integration measure $\intkpr$ to permute $K, P, R$ as necessary to further simplify the expressions.
\end{enumerate}

Let us quickly illustrate this procedure in two cases, one with only bare vertices $\bigl[\IIUD{(0,0)}{3g}\bigr]^{\text{UV}}_{002}$, and one with a 3g HTL vertex $\bigl[\IIUD{(0,H)}{3g}\bigr]^{\text{UV}}_{001}$. In the first case, we have
{\allowdisplaybreaks
\begin{align}
    \Bigl[\IIUD{(0,0)}{3g}\Bigr]^{\text{UV}}_{002} ={}&\,
    \frac{1}{4}
    \intkpr 
    \Gam{\mu \nu \rho}_{0,KPR} 
    \Gam{\mu' \nu' \rho'}_{0,KPR} 
        \Dz{\mu \mu'}(K) 
        \Dz{\nu \nu'}(P) 
        \DDn{2}{\rho \rho'}(R) \nn
    ={}&\, \frac{1}{4} \intkpr 
    \frac{1}{K^{2}P^{2}}
    \Gam{\mu \nu \rho}_{0,KPR} 
    \Gam{\mu \nu \rho'}_{0,KPR} 
        \DDn{2}{\rho \rho'}(R)  \nn
    ={}&\, 
    \frac{1}{4} \intkpr
    \frac{1}{K^{2}P^{2}}
    \Bigl\{ (4d - 2) [K\cdot \DDN{2}(R) \cdot K] + (4 R^{2} + P^{2} + K^{2})  \tr{\DDN{2}(R)} \Bigr\}  \nn
    ={}&\, \frac{1}{4} \intkpr
    \frac{1}{P^{2}R^{2}}
    \Bigl\{ (4d - 2) [P\cdot \DDN{2}(K) \cdot P] + (4 K^{2} + 2 P^{2})  \tr{\DDN{2}(K)} \Bigr\} \nonumber \\[1ex]
    ={}&\, \intkpr
    \frac{(4d - 2)}{4P^{2}R^{2}}
    [P\cdot \DDN{2}(K) \cdot P] 
    + \intkpr \frac{K^{2}}{P^{2} R^{2}}  \tr{\DDN{2}(K)} + [\text{UV pow. div.}],
\end{align}
}%
where from the first to the second line, we used step (1) above; from the second to the third line, we used step (2); and from the third to the fourth line, we used step (3) of our procedure.  Finally, in the last line, we separated out a UV power divergence, which vanishes in dimensional regularization. This is precisely the power-divergent term identified in \eq\eqref{eq:3gSchematicSoft}. Here, we have introduced a Lorentz-index-free notation, where we represent contractions as dots or traces depending on the structure:
\begin{equation}
    \begin{split}
        P \cdot \DDN{2}(K) \cdot P ={}&  P^{\mu} P^{\nu} \DDN{2}^{\mu \nu}(K), \\
        \tr{\DDN{2}(K)} ={}&  \DDN{2}^{\mu\mu}(K).
    \end{split}
\end{equation}
We will use this notation extensively through the rest of the work. 

In the second case, we have
{
    \allowdisplaybreaks
\begin{align}
    \Bigl[\IIUD{(0,H)}{3g}\Big]^{\text{UV}}_{001} ={}& 
		\frac{1}{4} \intkpr 
        \dGam{\mu' \nu' \rho'}_{KPR} 
        \Gam{\mu \nu \rho}_{0,KPR} 
		\Dz{\mu \mu'}(K)
		\Dz{\nu \nu'}(P)
		\DDn{1}{\rho \rho'}(R) \nn
	={}&
		\frac{1}{4} \intkpr 
		\frac{1}{K^{2} P^{2}}
        \dGam{\mu \nu \rho'}_{KPR} 
        \Gam{\mu \nu \rho}_{0,KPR} 
		\DDn{1}{\rho \rho'}(R), \nn
	={}&
		\frac{1}{4} \intkpr 
		\frac{1}{K^{2} P^{2}}
		\DDn{1}{\rho \rho'}(R)
        \left[ \Pi^{\rho \rho'}(K) + \Pi^{\rho \rho'}(P) - 2 \Pi^{\rho \rho'}(R) \right] \nn
	={}&
		\frac{1}{2} \intkpr 
		\frac{1}{K^{2} P^{2}}
        \Bigl\{
            \tr{\DDN{1}(R) \Pi(K)} - \tr{\DDN{1}(R) \Pi(R)}
        \Bigr\} \nn
	={}&
		\frac{1}{2} \intkpr 
		\frac{1}{P^{2} R^{2}}
        \Bigl\{
            \tr{\DDN{1}(K) \Pi(P)} - \tr{\DDN{1}(K) \Pi(K)}
        \Bigr\},
\end{align}
}%
where we used the same steps in the same order. Here, the generalized Ward identity was applied in the contraction of the bare 3g vertex and the 3g HTL vertex. Additionally, we also made use of the symmetries of the 3g HTL vertex discussed in \app\ref{sec:HTLvertices}.

Now let us further massage 
the above expression 
by using (i) the fact that 
\begin{equation}
    [\DDN{n-1}(K) \cdot \Pi(K)]^{\mu \nu} = -K^{2} \DDn{n}{\mu \nu}(K),
    \label{eq:DDN_recursion2}
\end{equation}
for any $n$ (this follows from \eq\eqref{eq:DDN_explicit}), and (ii) the fact that, since $P^{2} \Dz{\mu \nu}(P) = \delta^{\mu \nu}$,
\begin{align}
    [\DDN{1}(K) \cdot \Pi(P)]^{\mu \nu} ={}& [\DDN{1}(K) \cdot \Pi(P) \cdot P^{2} \DZ(P)]^{\mu \nu}\nn
    ={}& P^{2} \Bigl\{ \DDN{1}(K) \Pi(P)[D(P) - \DDN{1}(P)] \Bigr\}^{\mu \nu} \nn
    ={}& -(P^{2})^{2} \Bigl\{ \DDN{1}(K) [\DDN{1}(P) - \DDN{2}(P)] \Bigr\}^{\mu \nu},
\end{align}
where in the last line we used the recursion relation in 
\eq\eqref{eq:DDN_recursion2} immediately above. Thus, plugging in the above and rearranging, we find
\begin{align}
    \Bigl[ \IIUD{(0,H)}{3g} \Bigr]^{\text{UV}}_{001}
	={}&
        \frac{1}{2}
		\intkpr 
        \Bigl\{
            -\frac{P^{2}}{R^{2}} \tr{\DDN{1}(K) \DDN{1}(P)}  
            + \frac{K^{2}}{P^{2}R^{2}} \tr{\DDN{2}(K)} 
            + \frac{P^{2}}{R^{2}} \tr{\DDN{1}(K) \DDN{2}(P)}
        \Bigr\}.
\end{align}
Note that the last term in this expression  goes as $\mE^{6}$ in the UV, and so can only contribute to the constant term.

Our final results for all the UV terms after the above procedure are as follows: for the parts of $\II{3g}$,
    \begin{align}
            \Bigl[\IIUD{(0,0)}{3g}\Bigr]^{\text{UV}}_{002} ={}& 
                \intkpr
                \frac{(4d - 2)}{4P^{2}R^{2}}
                [P\cdot \DDN{2}(K) \cdot P] 
                + \intkpr \frac{K^{2}}{P^{2} R^{2}}  \tr{\DDN{2}(K)}, \\
            \Bigl[\IIUD{(0,0)}{3g}\Bigr]^{\text{UV}}_{110} ={}&
                \intkpr \Bigl\{ \Bigl[ \frac{K^{2}}{R^{2}} - \frac{1}{4} \Bigr] \tr{\DDN{1}(K) \DDN{1}(P)} + \frac{2}{R^{2}} [P \cdot \DDN{1}(K) \cdot P] \tr{\DDN{1}(P)} \Bigr\}, \\
            \Bigl[\IIUD{(0,H)}{3g}\Bigr]^{\text{UV}}_{001} ={}&
                \frac{1}{2}
                \intkpr 
                \Bigl\{
                   - \frac{P^{2}}{R^{2}} \tr{\DDN{1}(K) \DDN{1}(P)}  
                   + \frac{K^{2}}{P^{2}R^{2}} \tr{\DDN{2}(K)} 
                   + \frac{P^{2}}{R^{2}} \tr{\DDN{1}(K) \DDN{2}(P)}
                \Bigr\},  \\
            \Bigl[\IIUD{(H,H)}{3g}\Bigr]^{\text{UV}} ={}&
                \frac{1}{12}
                \intkpr
                \dGam{\mu \nu \rho}_{KPR}
                \dGam{\mu' \nu' \rho'}_{KPR}
                \bigl[
                    D^{\mu \mu'}(K)
                    D^{\nu \nu'}(P)
                    D^{\rho \rho'}(R)
                \bigr] ;
        \end{align}
for the parts of $\II{4g}$,
        \begin{equation}
            \Bigl[\IIUD{(0)}{4g}\Bigr]^{\text{UV}}_{11} =
            -\frac{1}{4}
            \intkp
            \Bigl\{ \tr{\DDN{1}(K)}\tr{\DDN{1}(P)} - \tr{\DDN{1}(K) \DDN{1}(P)} \Bigr\};
        \end{equation}
and for $\II{gh}$,
        \begin{equation}
            \left[\II{gh}\right]^{\text{UV}}_{2} = - \frac{1}{2} \intkpr \frac{1}{P^{2} R^{2}} [P \cdot \DDN{2}(K) \cdot P].
        \end{equation}
The sum of the above terms is quite compact, as there are a few cancellations. The total reads:
\begin{align}
\label{eq:ItotUV}
    \bigl[\II{tot}\bigr]^{\text{UV}} ={}& \intkpr \biggl\{
        \frac{2P^{2}}{R^{2}}
        [\hat{P} \cdot \DDN{1}(K) \cdot \hat{P}] 
        \tr{\DDN{1}(P)} 
        -\frac{1}{4} 
        \tr{\DDN{1}(K)} 
        \tr{\DDN{1}(P)} 
        \nn
        &\quad+
        \frac{2 K^{2}}{P^{2} R^{2}}
        \tr{\DDN{2}(K)} 
        +
        \frac{d-1}{R^{2}}
        \bigl[
            \hat{P} \cdot \DDN{2}(K) \cdot \hat{P}
            \bigl]
        +
        \frac{P^2}{R^{2}}
        \tr{\DDN{1}(K) \DDN{2}(P)}
        \nn
        &\quad
        +
        \frac{1}{12} \dGam{\mu \nu \rho}_{KPR}\dGam{\mu' \nu' \rho'}_{KPR}
        D^{\mu \mu'}(K)
        D^{\nu \nu'}(P)
        D^{\rho \rho'}(R)
        \biggr\}.
\end{align}
Here, we have factored out all magnitudes from the dot products, using our unit-vector notation introduced in \eq\eqref{eq:unitVectors}.  Note the very important point that the HTL self energy $\Pi^{\mu\nu}(K)$ inside of the propagators depends only on $\hat{K}$, and not 
on 
the magnitude $|K|$. In light of this, we will from now on write $\Pi^{\mu\nu}(\hat{K})$ to make this clear. 

At this point, we are ready to tackle performing the integrations of all the terms isolated and scalarized above.

\section{The computation}
\label{sec:compute}

After the manipulations performed in the previous section, evaluating $\pDU{3}{s}$ has been reduced to computing $[\II{tot}]^{\text{UV}}$ in \eq\eqref{eq:ItotUV} and the sum of the non-UV terms in \eqs\eqref{eq:I3G00def}--\eqref{eq:splitgh}. Our general procedure for evaluating these integrals can be summarized as follows:
\begin{enumerate}
    \item Rescale all the magnitudes of momentum variables by $|K| \mapsto \mE |K|$ etc.\ to pull out the $\mE$ dependence and make the integrands dimensionless.
    \item Perform the trivial $R$ integral to set $R \mapsto -(K+P)$, and change variables in the remaining $K$ and $P$ integrals to write them as integrals over the magnitude of the four-vectors $|K|$ and $|P|$, and the remaining angles.
    \item Further transform from the magnitudes of the momenta $(|K|,|P|)$ to Euclidean polar $(X,\chi)$ coordinates, given by 
\begin{equation}
\label{eq:Xchange}
    |K| = X \sin \chi  \equiv X \sx, \quad |P| = X \cos \chi  \equiv X \cx, \;\text{with} \;\,  \chi \in [0, \pi / 2], \quad X \in [0, \infty].
\end{equation}
        We introduce a shorthand notation $\sx, \cx$, as these particular functions will appear many times in our computation. In all cases, the radial $X$ integral can be performed analytically (in general $d$, if necessary).
\end{enumerate}
Let us now discuss the details of these operations.

\paragraph*{Details of step (1):} 
As part of this step, we will define dimensionless versions of all of the functions, denoted with tildes. For instance:
\begin{align}
    \PI{\mu \nu}(\hat{K}) \mapsto& \mE^{2} \widetilde{\Pi}^{\mu \nu}(\hat{K}), \\
    \DDN{n}(K) \mapsto& \mE^{-2} \widetilde{D}_{n}(K), \\
    G_X(K) \mapsto& \mE^{-2} \widetilde{G}_{X}(K), \qquad X \in \{ \text{T}, \text{L}\}
\end{align}
where $\widetilde{D}_{n}(K)$ and $\widetilde{G}_{X}(K)$ contain $\widetilde{\Pi}^{\mu \nu}(\hat{K})$ in place of $\PI{\mu \nu}(\hat{K})$. This rescaling also introduces a change in the integration measures; for example $\dNum{D} K \mapsto \mE^{D} \dNum{D} K$. Thus, the integrals change as
\begin{align}
    \intkpr \mapsto{}&  \mE^{2D} \intkpr \equiv \mE^{8} \intkprMod \nn
    ={}&  
    \mE^{8}
    \left (\frac{e^{\gamE}\Lh^2}{4\pi \mE^{2}}\right )^{4-D}
    \int \frac{\upd^D K}{(2\pi)^D} \int \frac{\upd^D P}{(2\pi)^D}  \int \frac{\upd^D R}{(2\pi)^D}(2\pi)^D \delta^{(D)}(K+P+R),
    \label{eq:rescaleIntkpr}
\end{align}
where we have defined a new integral (again, $\intkpMod$ can be defined by performing the trivial integral over $R$) by adding $\mE^{2}$ to the denominator of the factor out front. Thus, by factoring out $\mE^{8}$, and by simply ``placing tildes on everything else'', we can implement the desired rescaling of variables.  \emph{In what follows, we shall assume that we have already done this procedure everywhere, and hence drop the explicit tildes for convenience.}

\paragraph*{Details of step (2):} 
After performing the trivial integral over $R$, which sets $R \mapsto -(K + P)$, we wish to change variables to radial and angular parts. To this end, we introduce the following notations.  We first define the Euclidean spherical variables $(|K|,\Phi_K)$, where $\Phi_K \in [0,\pi]$ is the $D$-dimensional polar angle defined by
\begin{equation}
\label{eq:PhiKdef}
\tan \Phi_K = \frac{\vert \kt \vert}{K_0}, \quad\quad  (K_0, \vert \kt\vert) \mapsto (|K|\cos \Phi_K, |K|\sin \Phi_K).
\end{equation}
\label{eq:wDef}
The dot product between the Euclidean four-vectors $K$ and $P$ can be written as
\begin{equation}
K \cdot P = |K||P|\bigl (\cos \Phi_K \cos \Phi_P  + \sin \Phi_K \sin \Phi_P \cos \theta  \bigr ) \equiv |K||P| w,
\end{equation}
where the magnitudes $|K|$ and $|P|$ factorize from the angular part $w \equiv w(\Phi_K,\Phi_P,\theta)$ with $\cos \theta \equiv \hat{\vec{k}} \cdot \hat{\vec{p}}$. Note that this equation implies that $\hat{K} \cdot \hat{P} = w$, which will be used extensively in the computation below. 

Now, in these new coordinates, the integration measure $\intkp$ can be written
\begin{equation}
\label{eq:measureinDnewcoordinatesKP}
\begin{split}
    \intkp = 
    C(d) 
    \int_{\Omega}\,
    \int_{0}^{\infty} \ud |K| |K|^{d}\int_{0}^{\infty} \ud |P| |P|^{d} ,
\end{split}
\end{equation}
where the angular part of the integration measure is
\begin{equation}
\label{eq:def_omega}
\int_{\Omega} \equiv  
    \int_{0}^{\pi} \ud \Phi_K \sin^{d-1} \Phi_K 
    \int_{0}^{\pi} \ud \Phi_P \sin^{d-1} \Phi_P  
    \int_{0}^{\pi} \ud \theta \sin^{d-2} \theta,
\end{equation}
and we have introduced a compact notation for the dimensionless prefactor
\begin{equation}
\label{eq:cd}
C(d) \equiv  \left (\frac{e^{\gamE}\Lh^2}{4\pi \mE^{2}}\right )^{3-d}\frac{4\pi^{d-\frac{1}{2}}}{(2\pi)^{2d+2}}\frac{1}{\Gamma\left (\frac{d}{2} \right )\Gamma\left (\frac{d-1}{2}\right )}.
\end{equation}
The integrals over the azimuthal angles have been performed and included in $C(d)$, as all of our integrands are independent of those angles. For reference, we note here that $C(3) = 2 / (2 \pi)^{6}$, and that in $d = 3$, the integral $\int_\Omega = \pi^{2} / 2$.

\paragraph*{Details of step (3):}
Here, we stress that by moving to the $(X, \chi)$ coordinates, we only change the magnitudes $|K|, |P|$, and so $\hat{K}, \hat{P}$ remain unchanged. This change of variables gives rise to a nontrivial Jacobian, which leads to
\begin{equation}
    \intkp = 
    C(d) 
    \int_{\Omega}\,
    \int_{0}^{\pi / 2}  \ud \chi \, \sx^{d} \cx^{d} \int_{0}^{\infty} \ud X \, X^{2d + 1}.
\end{equation}
Finally, note that since we only integrate over the first quadrant in $(X, \chi)$ space, we have $\sx, \cx \geq 0$ everywhere in our integrals.

We are now in a position to evaluate our terms. We begin by identifying the independent integrals in $[\II{tot}]^{\text{UV}}$:
\begin{align}
    \II{A} \equiv{}& 
        \mE^{4} \intkpr 
        \frac{P^{2}}{R^{2}} [\hat{P} \cdot \DDN{1}(K) \cdot \hat{P}] 
        \tr{\DDN{1}(P)}, \\
    \II{B} \equiv{}& 
        \mE^{4} \intkpr 
        \tr{\DDN{1}(K)} 
        \tr{\DDN{1}(P)}, \\
    \II{C} \equiv{}& 
        \mE^{4} \intkpr 
        \frac{K^{2}}{P^{2} R^{2}} 
        \tr{\DDN{2}(K)}, \\
    \II{D} \equiv{}& 
        \mE^{4} \intkpr 
        \frac{d - 1}{R^{2}}
        \Bigl[
            \hat{P} \cdot \DDN{2}(K) \cdot \hat{P}
            \Bigl], \\
    \II{E} \equiv{}& \mE^{4} \intkpr 
        \frac{P^2}{R^{2}}
        \tr{\DDN{1}(K) \DDN{2}(P)}, \label{eq:IEdef} \\
    \II{F} \equiv{}& \mE^{4} \intkpr 
        \dGam{\mu \nu \rho}_{KPR}\dGam{\mu' \nu' \rho'}_{KPR}
        D^{\mu \mu'}(K)
        D^{\nu \nu'}(P)
        D^{\rho \rho'}(R),
    \label{eq:IF}
\end{align}
which combine as follows
\begin{equation}
    [\II{tot}]^{\text{UV}} \equiv 2 \II{A} - \frac{1}{4} \II{B} + 2 \II{C} +
    \II{D} + \II{E} + \frac{1}{12}\II{F}.
\label{eq:Itotal}
\end{equation}
Note that we have kept the $(d-1)$ inside $\II{D}$, since we want to be able to do the $\epsilon$ expansion for each of these integrals separately. Additionally, $\II{E}$ is not UV sensitive, as can be seen by counting the propagator subscripts as before. Therefore, we will now proceed as follows. First, as the forms of $\II{A}$--$\II{D}$ are so similar, and since they are all UV sensitive, we shall treat all of them together in \Sec\ref{sec:IAD} below, and define $\II{ABCD}$ to be the corresponding part of \eq\eqref{eq:Itotal}: 
\begin{equation}
    \II{ABCD} \equiv 2 \II{A} - \frac{1}{4} \II{B} + 2 \II{C} + \II{D},
\label{eq:ItotwithABCD}
\end{equation}
so that
\begin{equation}
    [\II{tot}]^{\text{UV}} = \II{ABCD} + \II{E} + \frac{1}{12}\II{F}.
\end{equation}
Second, as the structure in the remaining UV-sensitive term $\II{F}$ is so different from the others (in particular, the 3g HTL vertices cannot be removed completely by using the generalized Ward identities), we will evaluate it separately in \Sec\ref{sec:IF}. Finally, as all the remaining finite terms, including $\II{E}$, can be numerically evaluated in $d = 3$, we treat them all in \Sec\ref{sec:Ifinite}.

\subsection{\texorpdfstring{Integrals $\II{A}$--$\II{D}$}{Integrals I\_A--I\_D}}
\label{sec:IAD}

After performing the steps (1)--(3) described in the previous section, we are left with integrals over the angles in $\int_{\Omega}$, the separate angle $\chi$, and the radial coordinate $X$. As alluded to above, the $X$ integral can be performed analytically in general $d$. In $\II{A}$--$\II{D}$ only the following two master integrals in $X$ appear:
\begin{align}
    \label{eq:master1}
    \int_{0}^{\infty} \ud X &
    X^{2d - 3}
    \frac{\Pii{I}{K}}{X^{2} \sx^{2} + \Pii{I}{K}}
    \frac{\Pii{J}{P}}{X^{2} \cx^{2} + \Pii{J}{P}} =   \nonumber \\
    & 
    \Bigl[
        \frac{\pi}{2} \csc(d \pi)
        \Bigr]
    \sx^{1-d} \cx^{1-d} \Pii{I}{K} \Pii{J}{P}
    \biggl\{ 
    \frac{ \bigl[ \Pii{I}{K} \ctx \bigr]^{d - 2} - \bigl[ \Pii{J}{P} \tx \bigr]^{d - 2}}
    {\Pii{I}{K} \ctx - \Pii{J}{P} \tx}
    \biggr\}, \\[2ex]
    \label{eq:master2}
    \int_{0}^{\infty} \ud X &
    X^{2d - 5}
    \frac{\Pii{I}{K}^2}{X^{2} \sx^{2} + \Pii{I}{K}} =
    \Bigl[
        \frac{\pi}{2} \csc(d \pi)
    \Bigr]
    \Bigl[ 
        \sx^{4 - 2d} \Pii{I}{K}^{d - 1}
    \Bigr].
\end{align}
Here, $I, J \in \{\text{T}, \text{L}\}$ label the polarization components of the HTL self energy. We will now proceed through the integrals one by one, mostly showing details only for $\II{A}$.

\subsubsection{\texorpdfstring{Doing the integrals in general $d$}{Doing the integrals in general d}}

To unpack the notation in $\II{A}$, we use the definitions for the propagators $\DDN{n}$ and $D$ in \eqs\eqref{eq:DDN_explicit} and \eqref{eq:fullpropa}, respectively. This leads to the following form for $\II{A}$ in the $(X, \chi)$ coordinates
\begin{align}
    \II{A} ={}&  
    \mE^{4} C(d)  
    \int_{\Omega}
    \sum_{I,J}
    [\hat{P} \cdot \Proji{I}{K} \cdot \hat{P}]
    \tr{\Proji{J}{P}}
    \int_{0}^{\pi/2} \dInt \chi 
    \frac{\sx^{- 2}}{(1 + 2 \cx \sx w)} \nn
    & \qquad \qquad \qquad \times
    \sx^{d} \cx^{d} \int_{0}^{\infty} \dInt X 
    X^{2d - 3}
    \frac{\Pii{I}{K}}{X^{2} \sx^{2} + \Pii{I}{K}}
    \frac{\Pii{J}{P}}{X^{2} \cx^{2} + \Pii{J}{P}},
\end{align}
where $C(d)$ was defined in \eq\eqref{eq:cd}. Using the first master integral \eq\eqref{eq:master1}, this becomes
\begin{align}
    \II{A} ={}&  
        \mE^{4} C(d)  
        \Bigl[
            \frac{\pi}{2} \csc(d \pi)
        \Bigr]
        \int_{\Omega}
        \sum_{I,J}
        [\hat{P} \cdot \Proji{I}{K} \cdot \hat{P}]
        \tr{\Proji{J}{P}} \nn
        & \qquad \times\int_{0}^{\pi/2} \dInt \chi \,
        \frac{
            \bigl[
                \Pii{I}{K} \ctx 
            \bigr]
            \Pii{J}{P}
        }
        {1 + 2 \sx \cx w}
        \biggl\{ 
            \frac{ \bigl[ \Pii{I}{K} \ctx \bigr]^{d - 2} - \bigl[ \Pii{J}{P} \tx \bigr]^{d - 2}}
            {\Pii{I}{K} \ctx - \Pii{J}{P} \tx}
        \biggr\}.
\end{align}
The integral over $\chi$ here can be performed analytically, but it is unwieldy, and so we choose the following approach instead.  This integral contains a divergence in $d = 3$, from the region near $\chi = 0$. Near $\chi = 0$, the $\chi$ integrand is approximately
\begin{equation}
\begin{split}
        \frac{
            \bigl[
                \Pii{I}{K} \ctx 
            \bigr]
            \Pii{J}{P}
        }
        {1 + 2 \sx \cx w}
        &
        \biggl\{ 
            \frac{ \bigl[ \Pii{I}{K} \ctx \bigr]^{d - 2} - \bigl[ \Pii{J}{P} \tx \bigr]^{d - 2}}
            {\Pii{I}{K} \ctx - \Pii{J}{P} \tx}
        \biggr\} \\
        &\quad\quad\quad\quad\quad\quad
        \simeq
        \frac{
            \Pii{J}{P}
        }
        {1 + 2 \sx \cx w}
        \bigl[ \Pii{I}{K} \ctx \bigr]^{d - 2}.
\end{split}
\end{equation}
This means that we can isolate the $O(\epsilon^{-2})$ part of $\II{A}$ by evaluating
\begin{equation}
    \II{A}\Bigr|_{\chi \approx 0} =
        \mE^{4} C(d)  
        \Bigl[
            \frac{\pi}{2} \csc(d \pi)
        \Bigr]
        \int_{\Omega}
        [\hat{P} \cdot \Pi(\hat{K})^{d - 2} \cdot \hat{P}]
        \int_{0}^{\pi/2} \dInt \chi \,
        \frac{ \cot^{d - 2} \chi }
        {1 + 2 \sx \cx w},
\label{eq:IAchiE0}
\end{equation}
where in going to the second line, we used $\tr{\Pi} = 1$, which is true because we have factored out the mass scale $\mE$, and have dropped the tildes on all expressions. We have also introduced the notation
\begin{equation}
    [\Pi(\hat{K})^{d-2}]^{\mu\nu} = \sum_{I \in \{\text{T, L}\}} \Pii{I}{K}^{d-2} \Proji{I}{K}^{\mu \nu}.
\end{equation}
We will return to this $\chi \approx 0$ term in a bit. 

The remaining integral over $\chi$, which we define to be
\begin{equation}
    \Delta \II{A} \equiv \II{A} - \II{A}\Bigr|_{\chi \approx 0}
\end{equation}
in the following, will then be finite:
\begin{align}
    \Delta \II{A} ={}& 
        \mE^{4} C(d)  
        \Bigl[
            \frac{\pi}{2} \csc(d \pi)
        \Bigr]
        \int_{\Omega}
        \sum_{I,J}
        [\hat{P} \cdot \Proji{I}{K} \cdot \hat{P}]
        \tr{\Proji{J}{P}} \nn
        & \qquad \times\int_{0}^{\pi/2} \dInt \chi \,
        \frac{
            \bigl[
                \Pii{I}{K} \ctx 
            \bigr]
            \Pii{J}{P}
        }
        {1 + 2 \sx \cx w}
        \Biggl\{ 
            \frac{ \bigl[ \Pii{I}{K} \ctx \bigr]^{d - 2} - \bigl[ \Pii{J}{P} \tx \bigr]^{d - 2}}
            {\Pii{I}{K} \ctx - \Pii{J}{P} \tx} \nn
        & \qquad \phantom{\cdot\int_{0}^{\pi/2} \dInt \chi \,
        \frac{
            \Pii{J}{P}
            \bigl[
                \Pii{I}{K} \ctx 
            \bigr]
        }
        {1 + 2 \sx \cx w}
        \Biggl\{ \,\,\, }
            -
            \bigl[ \Pii{I}{K} \ctx \bigr]^{d - 3}
        \Biggr\}.
\end{align}
Note that this $\chi$ integral vanishes in $d = 3$, as the expression in braces equals $0$. Thus, we only need the $O(\epsilon)$ piece of the $\chi$ integral, and the $O(1 / \epsilon)$ piece of the divergent coefficient
\begin{align}
    \mE^{4} C(d)  
    \Bigl[
        \frac{\pi}{2} \csc(d \pi)
    \Bigr]
    \simeq{}& 
    \mE^{4} C(3)  
    \frac{1}{4 \epsilon} + O(\epsilon^{0}) \nn
    ={}&  \frac{\mE^{4}}{(2\pi)^{6}}  
    \frac{1}{2 \epsilon} + O(\epsilon^{0}).
\end{align}
The full expression up to $O(\epsilon^{0})$ is then the surprisingly simple
\begin{align}
    \Delta\II{A} \simeq{}& 
        \frac{\mE^{4}}{(2 \pi)^{6}}
        \int_{\Omega}
        \sum_{I,J}
        [\hat{P} \cdot \Proji{I}{K} \cdot \hat{P}]
        \tr{\Proji{J}{P}} \nn
        & \qquad \times\int_{0}^{\pi/2} \dInt \chi \,
        \frac{
            \Pii{I}{K} \Pii{J}{P}^{2}
        }
        {1 + 2 \sx \cx w}
        \Biggl\{ 
            \frac{ 
            \ln \bigl[ \Pii{J}{P} \tx \bigr]
            -
            \ln \bigl[ \Pii{I}{K} \ctx \bigr]}
            {\Pii{I}{K} \ctx - \Pii{J}{P} \tx} 
        \Biggl\},
\label{eq:IAconst}
\end{align}
where a cotangent and tangent cancelled in the $\Pii{I}{K} \Pii{J}{P}^{2}$ term. Here, one can analytically compute the $\chi$ integral in $d = 3$
\begin{align}
\label{eq:DeltaIA}
    \Delta\II{A} \simeq
        \frac{\mE^{4}}{(2 \pi)^{6}}
        \int_{\Omega}
        \sum_{I,J}
        &
        \frac{
            \bigl[\Pii{I}{K} + \Pii{J}{P}\bigr]
            \bigl[\hat{P} \cdot \Proji{I}{K} \cdot \hat{P} \bigr] 
            \tr{\Proji{J}{P}}
        } 
        {\bigl[\Pii{I}{K} + \Pii{J}{P} \bigr]^{2} + 4 \Pii{I}{K} \Pii{J}{P} w^{2}} \nn
        &
        \times \Biggl\{
            \arccos^{2}(w)  
            - \frac{1}{4} \ln^{2} \biggl[ \frac{\Pii{I}{K}}{\Pii{J}{P}} \biggr] 
            - \frac{\pi^{2}}{2} \Biggl[ 1 
            - 2 \frac{\sqrt{\Pii{I}{K} \Pii{J}{P}}}{\Pii{I}{K} + \Pii{J}{P}} w \Biggr] \nn
        & \quad \quad
            + \Biggl[ \frac{\Pii{I}{K} - \Pii{J}{P}}{\Pii{I}{K} + \Pii{J}{P}} \Biggr]
            \frac{w \arccos(w)}{\sqrt{1-w^{2}}}
            \ln \biggl[ \frac{\Pii{I}{K}}{\Pii{J}{P}} \biggr]
        \Biggr\},
\end{align}
which only leaves the integrals in $\int_{\Omega}$ to be computed numerically.

Let us now return to the $\chi \approx 0$ term in \eq\eqref{eq:IAchiE0}. We want to use tensor reduction to write the $\hat{P}$ integral inside of $\Omega$ in a simpler way. As written, the $\hat{P}$ integral in \eq\eqref{eq:IAchiE0} depends on $w = \hat{K} \cdot \hat{P}$, so we cannot use full $d$-dimensional symmetry, but only a restricted $(d - 1)$-dimensional symmetry, as detailed in \Ref\citep{Ee:2017}. However, we opt for another approach, which will yield a much simpler final expression. 

Let us consider the $\hat{P}$ integral first in \eq\eqref{eq:IAchiE0}, so that we focus on
\begin{align}
        \bigl[ \Pi(\hat{K})^{d - 2} \bigr]^{\mu \nu} 
        \int_{0}^{\pi/2} \dInt \chi
        \cot^{d - 2} \chi 
        \int_{\hat{P}}
        \frac{
            \hat{P}^{\mu} \hat{P}^{\nu}
        }
        {1 + 2 \sx \cx w}.
\end{align}
Now expand out the denominator in a geometric series (here, we omit the $\bigl[ \Pi(\hat{K})^{d - 2} \bigr]^{\mu \nu}$ for space), resulting in
\begin{align}
        \int_{0}^{\pi/2} \dInt \chi
        \cot^{d - 2} \chi 
        \int_{\hat{P}}
        \frac{
            \hat{P}^{\mu} \hat{P}^{\nu}
        }
        {1 + 2 \sx \cx w} ={}& 
        \int_{0}^{\pi/2} \dInt \chi
        \cot^{d - 2} \chi 
        \sum_{m = 0}^{\infty}
        \int_{\hat{P}}
        \hat{P}^{\mu} \hat{P}^{\nu}
        \bigl(
            - 2 \sx \cx w
        \bigr)^{m} \nn
        ={}& 
        \int_{0}^{\pi/2} \dInt \chi
        \cot^{d - 2} \chi 
        \sum_{m = 0}^{\infty}
        \int_{\hat{P}}
        \hat{P}^{\mu} \hat{P}^{\nu}
        \bigl(
            4 \sx^{2} \cx^{2} 
        \bigr)^{m}
        \bigl(
            \hat{K} \cdot \hat{P} 
        \bigr)^{2 m} \nn
        ={}& 
        \sum_{m = 0}^{\infty}
        4^{m}
        \int_{0}^{\pi/2} \dInt \chi
            \sx^{2m + (2 - d)} \cx^{2m + (d - 2)} 
        \int_{\hat{P}}
        \hat{P}^{\mu} \hat{P}^{\nu}
        \bigl(
            \hat{K} \cdot \hat{P} 
        \bigr)^{2 m}.
\end{align}
Here, in going to from the first to second line, we used the symmetry of the $\hat{P}$ integral to remove all the terms with an odd number of $\hat{P}$. Thus, we must evaluate a generic integral of the form
\begin{equation}
    \bigl[ \Pi(\hat{K})^{d-2} \bigr]^{\mu \nu} \int_{\hat{P}} \hat{P}^{\mu} \hat{P}^{\nu} 
    \bigl(
        \hat{K} \cdot \hat{P} 
    \bigr)^{2 m}.
\end{equation}
In \Ref\cite{Ee:2017}, they provide the result\footnote{see \eq (30) in \Ref\cite{Ee:2017}} (translated to our notation)
\begin{align}
\label{eq:angAve1}
    \int_{\hat{P}} (\hat{K} \cdot \hat{P})^{2m} ={}& 
    \frac{(2m - 1)!!}{D (D+2) \cdots (D + 2m - 2)} \int_{\hat{P}},
\end{align}
along with a recurrence relation between totally symmetric tensors\footnote{see \eq (21) in \Ref\cite{Ee:2017}}. Using the recurrence relation, and the Ward identity for $\Pi(\hat{K})$, we conclude 
\begin{align}
\label{eq:angAve2}
    \bigl[ \Pi(\hat{K})^{d-2} \bigr]^{\mu \nu} \int_{\hat{P}} \hat{P}^{\mu} \hat{P}^{\nu} 
    \bigl(
        \hat{K} \cdot \hat{P} 
    \bigr)^{2 m} =
    \tr{\Pi(\hat{K})^{d-2}} \frac{(2m - 1)!!}{D (D+2) \cdots (D + 2 m)}  \int_{\hat{P}}.
\end{align}

Additionally using
\begin{align}
\label{eq:eulerB}
    \int_{0}^{\pi / 2} \dInt \chi \,
    \sx^{a} \cx^{b} 
    =
    \frac{1}{2} \mathrm{B} \biggl( \frac{1 + a}{2} , \frac{1 + b}{2} \biggr),
\end{align}
where B is the Euler Beta function, we find in our case
\begin{align}
    \bigl[ \Pi(\hat{K})^{d-2} \bigr]^{\mu \nu} &
    \int_{0}^{\pi/2} \dInt \chi
    \cot^{d - 2} \chi 
    \int_{\hat{P}}
    \frac{
        \hat{P}^{\mu} \hat{P}^{\nu}
    }
    {1 + 2 \sx \cx w}  \nn
    ={}&
    \frac{
        \tr{\Pi(\hat{K})^{d - 2}}
    }
    {2}
    \sum_{m = 0}^{\infty}
    4^{m}
    \mathrm{B} \biggl( \frac{2m + (2 - d) + 1}{2} , \frac{2m + (d - 2) + 1}{2} \biggr) \nn
    & \qquad \qquad \qquad
    \times
    \frac{(2m - 1)!!}{D (D+2) \cdots (D + 2 m)}
    \int_{\hat{P}} \nn
    ={}& 
    - \tr{\Pi(\hat{K})^{d - 2}}
    \frac{\pi^{3 / 2}}{2^{d + 1}}
    \frac{
        \Gamma \bigl( \frac{d + 1}{2} \bigr)
    }
    {
        \Gamma \bigl( \frac{d}{2} \bigr)
    }
    \sec \Bigl( \frac{\pi d}{2} \Bigr)
    \int_{\hat{P}}.
\end{align}
Therefore, \eq\eqref{eq:IAchiE0} can be reduced to
\begin{align}
    \II{A}\Bigr|_{\chi \approx 0} ={}& 
        - \mE^{4} C(d)  
        \frac{\csc \bigl( \frac{\pi d}{2} \bigr)}{2^{d + 1}}
        \frac{
            \Gamma \bigl( \frac{1}{2} \bigr)
            \Gamma \bigl( \frac{d + 1}{2} \bigr)
        }
        {
            \Gamma \bigl( \frac{d}{2} \bigr)
        }
        \Bigl[
            \frac{\pi}{2} \sec \Bigl( \frac{\pi d}{2} \Bigr)
        \Bigr]^2
        \int_{\Omega}
        \tr{\Pi(\hat{K})^{d - 2}},
\label{eq:IAgenD}
\end{align}
so that we have reduced $\II{A}$ to the sum of the two integrals \eq\eqref{eq:IAgenD} and \eq\eqref{eq:IAconst}, the latter of which can be performed numerically in $d = 3$.

The remaining integrals $\II{B}$--$\II{D}$ can be performed similarly, and in fact, all of the remaining $\chi$ integrals that arise result in relatively simple analytical results in general $d$ . Using the two master integrals in \eqs\eqref{eq:master1}--\eqref{eq:master2}, the angular averages in \eqs\eqref{eq:angAve1}--\eqref{eq:angAve2} and the Euler-Beta-function expression in \eq\eqref{eq:eulerB}, we find the following results:
\begin{align}
    \II{B} ={}& 
        \mE^{4} C(d) 
        \Bigl[ 
            \frac{\pi}{2} \sec \Bigl( \frac{\pi  d}{2} \Bigl) 
        \Bigr]^{2}
        \int_{\Omega} 
        \tr{\Pi(\hat{K})^{\frac{d-1}{2}}} 
        \tr{\Pi(\hat{P})^{\frac{d-1}{2}}}, \\
    \II{C} ={}& 
        - 4 \mE^{4} C(d)  
        \frac{\csc \bigl( \frac{\pi d}{2} \bigr)}{2^{d + 1}}
        \frac{
            \Gamma \bigl( \frac{1}{2} \bigr)
            \Gamma \bigl( \frac{d + 1}{2} \bigr)
        }
        {
            \Gamma \bigl( \frac{d}{2} \bigr)
        }
        \Bigl[
            \frac{\pi}{2} \sec \Bigl( \frac{\pi d}{2} \Bigr)
        \Bigr]^2
        \int_{\Omega}
        \tr{\Pi(\hat{K})^{d - 1}},
\end{align}
and
\begin{align}
    \II{D} ={}& 
        (d-1) 
        \mE^{4} C(d)  
        \Bigl[
            \frac{\pi}{2} \csc(d \pi)
        \Bigr]^{2}
        \frac{\sin \bigl( \frac{\pi d}{2} \bigr)}{2^{d}}
        \frac{
            \Gamma \bigl( \frac{1}{2} \bigr)
            \Gamma \bigl( \frac{d + 1}{2} \bigr)
        }
        {
            \Gamma \bigl( 1 + \frac{d}{2} \bigr)
        }
        \int_{\Omega}
        \tr{ \Pi(\hat{K})^{d - 1} }.
\end{align}
Combining all of these results as in \eq\eqref{eq:ItotwithABCD}, we define
\begin{align}
    \label{eq:Iabcd}
    \bigl[ \II{ABCD}(d) \bigr]^{\text{UV}} \equiv{}& 
        - 
        \mE^{4} C(d)  
        \Bigl[
            \frac{\pi}{2} \sec \Bigl( \frac{\pi d}{2} \Bigr)
        \Bigr]^2
        \int_{\Omega}
        \Biggl(
            \frac{1}{4} 
            \tr{\Pi(\hat{K})^{\frac{d - 1}{2}}}
            \tr{\Pi(\hat{P})^{\frac{d - 1}{2}}} \nn
        & \;
        +
            \frac{
                \pi \csc \bigl( \frac{\pi d}{2} \bigr)
            }
            {
                4^{d}
            }
            \frac{
                \Gamma \bigl( d \bigr)
            }
            {
                \Gamma \bigl( \frac{d}{2} \bigr)^{2}
            }
            \left\{
                2 \tr{\Pi(\hat{K})^{d - 2}}
                + \left( \frac{1}{d} + 7 \right) \tr{\Pi(\hat{K})^{d - 1}}
            \right\} 
        \Biggr),
\end{align}
so that the total contribution from $\II{A}$--$\II{D}$, valid up to $O(\epsilon^{0})$, reads:
\begin{align}
    \II{ABCD} \simeq{}& 
        \bigl[ \II{ABCD}(d) \bigr]^{\text{UV}} + 2\Delta\II{A},
\end{align}
with $\Delta \II{A}$ as in \eq\eqref{eq:DeltaIA}. 

\subsubsection{\texorpdfstring{Extracting the coefficients of the $\epsilon$ expansion}{Extracting the coefficients of the epsilon expansion}}

We now turn to performing the $\epsilon$ expansions of $[\II{ABCD}(d)]^{\text{UV}}$ above to calculate the terms in the expansion of the pressure in \eq\eqref{eq:2loopSeriesSimple}. The only divergence is contained in the overall trigonometric coefficient, which has the expansion
\begin{equation}
    \Bigl[
        \frac{\pi}{2}
        \sec \Bigl( \frac{\pi  d}{2} \Bigr) 
    \Bigr]^{2}
    = \frac{1}{(2 \epsilon)^{2}} + \frac{\pi^{2}}{12} + O(\epsilon^{2}),
\label{eq:secExpand}
\end{equation}
and so we see that \eq\eqref{eq:Iabcd} above contributes to $p_{-2}$ in the expansion of the pressure. Moreover, from the $\epsilon$ expansion of the remainder of the integrand, this term likewise contributes to the remaining $p_{-1}$ and $p_{0}$ terms as well. Recall that the $\Pi$s appearing in this expression are $d$-dimensional self energies, and thus to calculate the subleading contributions, we must expand not only the explicit $d = 3 - 2 \epsilon$ appearing in the expression and the measure, but we also must expand the $d$-dimensional HTL self energies in a series 
\begin{equation}
\label{eq:piEpsExpansion}
    \Pi_{I}(\hat{K}) =  \Pi_{I,0}(\hat{K}) + \epsilon \Pi_{I,1}(\hat{K}) + \epsilon^{2} \Pi_{I,2}(\hat{K}) + O(\epsilon^{3}), \qquad I \in \{ \text{T}, \text{L} \}.
\end{equation}
The explicit details are shown in \app\ref{sec:HTLselfenergy}. Notice that the $\epsilon^{-2}$-divergence in \eq\eqref{eq:secExpand} means that we will need the expansion up to $\Pi_{I,2}$ in the calculation.

However, it turns out that we can avoid the explicit appearance of $\Pi_{I,2}$ in our expressions, using the following approach. We write down a simpler integral $\uvtext{I_{-2}(d)}$ that will (i) contain the full $O(\epsilon^{-2})$ behavior of $\uvtext{\II{ABCD}(d)}$, but (ii) is simple enough to be computed analytically in general $d$. This will mean that the difference $\uvtext{\II{ABCD}(d)} - \uvtext{I_{-2}(d)}$ only starts at $O(\epsilon^{-1})$, and so we will only need the terms up to $\Pi_{I,1}$ to calculate it. 

To construct this simpler integral, we examine \eq\eqref{eq:Iabcd}. Since \eq\eqref{eq:secExpand} contains the explicit $\epsilon^{-2}$, the coefficient $p_{-2}$ arises from setting $d = 3$ everywhere else. If we do this, we see that the first two terms of \eq\eqref{eq:Iabcd} will cancel (since $\tr{\Pi} = 1$), and only the term involving
\begin{equation}
    \tr{\Pi(\hat{K})^{d - 1}} \underset{d = 3}{\longrightarrow} \tr{\Pi(\hat{K})^{2}} 
\end{equation}
remains. Using this as motivation, we define
\begin{align}
\label{eq:IDL}
    \bigl[I_{-2}(d)\bigr]^{\text{UV}} \equiv{}& 
        - 
        \mE^{4} C(d)  
        \Bigl[
            \frac{\pi}{2} \sec \Bigl( \frac{\pi d}{2} \Bigr)
        \Bigr]^2
            \frac{
                \pi \csc \bigl( \frac{\pi d}{2} \bigr)
            }
            {
                4^{d}
            }
            \frac{
                \Gamma \bigl( d \bigr)
            }
            {
                \Gamma \bigl( \frac{d}{2} \bigr)^{2}
            }
            \left( \frac{1}{d} + 7 \right) 
        \int_{\Omega}
            \tr{\Pi(\hat{K})^{2}}.
\end{align}
As shown in \app\ref{sec:importantIntegrals}, this integral can be performed analytically, yielding
\begin{align}
\label{eq:IDLDone}
    \bigl[I_{-2}(d)\bigr]^{\text{UV}} ={}& 
        - 
        \mE^{4} C(d)  
        \Bigl[
            \frac{\pi}{2} \sec \Bigl( \frac{\pi d}{2} \Bigr)
        \Bigr]^2
            \left[
            \frac{
                \pi^{2} \csc \bigl( \frac{\pi d}{2} \bigr)
            }
            {
                (d - 1) 2^{d}
            }
            \right]
            \left( \frac{1}{d} + 7 \right) 
        \left[
        1 - \psi(d) + \psi \biggl( \frac{1+d}{2} \biggr)
        \right],
\end{align}
where $\psi$ is the digamma function
\begin{equation}
    \psi(x) \equiv \frac{\Gamma'(x)}{\Gamma(x)}.
\end{equation}

\begin{table}
\begin{ruledtabular}
\begin{tabular}{llll}
    & $\uvtext{I_{-2}(d)}$ & $\uvtext{\II{ABCD}(d)} - \uvtext{I_{-2}(d)}$ & $\Delta \II{A}$ \\
    \hline
    $p_{-2}$     & $\frac{11 \pi^{2}}{24}$    & $0$  & $0$\\
    $p_{-1}$     &   $\frac{\pi^{2}}{2} \bigl[\frac{1 + 11 \pi^{2}}{72}\bigr]$  & 
    $\int_{\Omega}
    \bigl(
        \frac{1}{4} 
        - \frac{11}{6} 
        \mathrm{Tr} \{
                \Pi_{0}(\hat{K})^{2} \ln \bigl[ \Pi_{0}(\hat{K}) \bigr]                 
            \}
    \bigr) \approx 3.74046$ & $0$ \\
    $p_{0}$   & $\frac{11 \pi^{2}}{16} [ \zeta(3)- \frac{1778 - 299 \pi^{2}}{1188}] $    & $\approx 10.84411$ & $\approx -3.71084$
\end{tabular}
\end{ruledtabular}
    \caption{\label{tab:ABCDcontributions} Contributions from $\II{ABCD}$ to the pressure after performing the necessary integrals. Here, $\zeta$ is the Riemann zeta function.}
\end{table}

With these new definitions, we can thus split $\II{ABCD}$ into three pieces, which contribute to \eq\eqref{eq:2loopSeriesSimple} in the following manner:
\begin{enumerate}
    \item $\uvtext{I_{-2}(d)}$ contributes to all coefficients $p_{-2}$, $p_{-1}$, and $p_{0}$,
    \item $\uvtext{\II{ABCD}(d)} - \uvtext{I_{-2}(d)}$ contributes to the coefficients $p_{-1}$, and $p_{0}$,
    \item $\Delta \II{A}$ contributes only to the coefficient $p_{0}$, which enters with a symmetry factor of $2$ from \eq\eqref{eq:ItotUV}.
\end{enumerate}
Performing the $\epsilon$ expansion in the three terms identified in this list, we arrive at a set of angular integrals to perform, only some of which can be performed analytically. A full list of these contributing integrals is given in \app\ref{sec:importantIntegrals}. In \tab\ref{tab:ABCDcontributions}, we summarize the computed contributions to the coefficients $p_{-2}$, $p_{-1}$, and $p_{0}$. As most of the one-dimensional integrals contributing to $p_{0}$ must be performed numerically, we mainly list numerical values for that row of the table. We note here that these numerical integrals can be easily calculated to high precision, and we show them all in \app\ref{sec:IABCD finite integrals}.
 The full contribution from $\II{ABCD}$ is thus the sum across the columns of this table. We have verified that we obtain the same total results if we do not introduce $\uvtext{I_{-2}(d)}$, but rather use the expansion of $\Pi_{I}$ in \eq\eqref{eq:piEpsExpansion} up to second order.

\subsection{\texorpdfstring{The $\II{F}$ term}{The I\_F term}}
\label{sec:IF}

We now turn to the integral $\II{F}$, given in \eq\eqref{eq:IF}, which we reproduce here:
\begin{equation}
    \II{F} \equiv \mE^{4} \intkpr 
        \dGam{\mu \nu \rho}_{KPR}\dGam{\mu' \nu' \rho'}_{KPR}
        D^{\mu \mu'}(K)
        D^{\nu \nu'}(P)
        D^{\rho \rho'}(R).
\end{equation}
The evaluation of this integral proceeds in much the same way as the previous ones. However, we must first perform the contractions, which are more complicated in this case: as remarked above, the two 3g HTL vertices cannot be fully removed by the generalized Ward identies, since they are only contracted with resummed propagators. 

Let us start by isolating the UV sensitivity in $\II{F}$. It is straightforward to verify from the power countings in \tab\ref{tab:vertexPowCount} and \eqs\eqref{eq:propExpandPowCountHard} and \eqref{eq:propExpandPowCountSoft} that $\II{F}$ has a UV logarithmic sensitivity only when all three momenta are hard. This leads to the following two conclusions for the UV sensitive part: Firstly, since the momenta $K$, $P$ (and $R$) cannot be scale-separated when they are all hard, there is really only a single independent integration momentum (the variable $X$), and thus we expect only a single $O(\epsilon^{-1})$ divergence. Secondly, when all the momenta are hard, the resummed propagators approach their bare versions (see \eq\eqref{eq:bareProp}), which leads to approximately direct contractions between the two 3g HTL vertices in this $O(\epsilon^{-1})$ term. This second point in particular motivates us to isolate the ``Kronecker-$\delta$-like part'' of the resummed propagators. We do this by unpacking the $\delta$ parts from the projection operators defined in \eqs\eqref{eq:projsDefs1}--\eqref{eq:projsDefs2} and rearranging the terms to arrive at
\begin{equation}
\label{eq:Dmod}
\begin{split}
    D^{\mu\mu'}(K) & = \sum_I \delta^{\mu\mu'}_I G_I(K) -  \hat{K}^{\mu}_\text{T}\hat{K}^{\mu'}_\text{T}G_{\text{TL}}(K) + \hat{K}^{\mu}\hat{K}^{\mu'}\biggl (\frac{1}{K^2} - G_\text{L}(K) \biggr ),
\end{split}
\end{equation}
where $I \in \{\text{T},\text{L}\}$ and $G_{\text{TL}}(K) \equiv G_\text{T}(K) - G_\text{L}(K)$. Here, we have also introduced the following notation:
\begin{equation}
\begin{split}   
    \hat{K}^{\mu}_\text{T}\hat{K}^{\mu'}_\text{T} & \equiv \delta^{\mu i}\delta^{\mu' j}\hat k^i \hat k^j,\\
    \delta^{\mu\mu'}_\text{T}+ \delta^{\mu\mu'}_\text{L} & \equiv \delta^{\mu\mu'}, 
\end{split}
\end{equation}
with $\delta^{\mu\mu'}_\text{T} \equiv \delta^{\mu i}\delta^{\mu' j}\delta^{ij}$ and $\delta^{\mu\mu'}_\text{L} \equiv \delta^{\mu 0}\delta^{\mu' 0}$. Hence, $\delta_\text{T}$ picks out spatial indices and $\delta_{\text{L}}$ picks out the temporal index. The UV-sensitivity will now arise from only taking the first term in \eq\eqref{eq:Dmod} from each propagator. The two remaining terms are seen to be more supressed in the UV, and so lead to finite contributions.

Using the symmetry properties of the integrand $\int_{KPR}$, we can now write $\II{F}$ in the form
\begin{equation}
\label{eq:IFsplit}  
    \II{F} = [\II{F}]^{\text{UV}} + \sum_{i=1}^{3} \Delta \II{F}^{(i)},
\end{equation}
where the UV part $\uvtext{\II{F}}$ is given by 
\begin{equation}
\label{eq:IFUV}
    \uvtext{\II{F}} = 
    \mE^4\int_{KPR} 
    \delta\Gamma^{\mu\nu\rho}_{KPR} 
    \delta\Gamma^{\mu'\nu'\rho'}_{KPR} 
    \sum_{I, J, W} 
    \biggl [
        \delta^{\mu\mu'}_I G_I(K) 
        \delta^{\nu\nu'}_J G_J(P) 
        \delta^{\rho\rho'}_W G_W(R) 
    \biggr ].
\end{equation}
The remaining finite parts $\Delta \II{F}^{(i)}$ are given by the following expressions 
\begin{align}
\Delta \II{F}^{(1)} = \mE^4\int_{KPR} 
    \delta\Gamma^{\mu\nu\rho}_{KPR}\delta\Gamma^{\mu'\nu'\rho'}_{KPR} \Biggl \{& -3\sum_{J,W} \biggl [\delta^{\nu\nu'}_J G_J(P) \delta^{\rho\rho'}_W G_W(R) \biggr ]\hat{K}^{\mu}_\text{T}\hat{K}^{\mu'}_\text{T} G_{\text{TL}}(K)\nn
    & + 3\biggl [\sum_W \delta^{\rho\rho'}_W G_W(R) \biggr ]
    \hat{K}^{\mu}_\text{T}\hat{K}^{\mu'}_\text{T}
    \hat{P}^{\nu}_\text{T}\hat{P}^{\nu'}_\text{T} 
    G_{\text{TL}}(K)G_{\text{TL}}(P)\nn
& - 
    \hat{K}^{\mu}_\text{T}\hat{K}^{\mu'}_\text{T} 
    \hat{P}^{\nu}_\text{T}\hat{P}^{\nu'}_\text{T}
    \hat{R}^{\rho}_\text{T}\hat{R}^{\rho'}_\text{T}
    G_{\text{TL}}(K)
    G_{\text{TL}}(P)
    G_{\text{TL}}(R)\Biggr \},
\label{eq:DIF1}
\end{align}\vspace{-1.5em}
\begin{align}
\Delta \II{F}^{(2)} = 6\mE^4\int_{KPR} & 
    \biggl [\sum_I \delta^{\mu\mu'}_I G_I(K) -  \hat{K}^{\mu}_\text{T}\hat{K}^{\mu'}_\text{T}
    G_{\text{TL}}(K) \biggr ] 
    \biggl [\sum_J \delta^{\nu\nu'}_J G_J(P) -  \hat{P}^{\nu}_\text{T}\hat{P}^{\nu'}_\text{T} G_{\text{TL}}(P)  \biggr ]\nn
& \times \Pi^{\mu\nu}(\hat K)
    \biggl [\Pi^{\mu'\nu'}(\hat K) - \Pi^{\mu'\nu'}(\hat P) \biggr ]
    \frac{\Pi_\text{L}(\hat R)}{R^4[R^2 + \Pi_\text{L}(\hat R)]},
\label{eq:DIF2}
\end{align}
and
\begin{align}
\Delta \II{F}^{(3)} = 3\mE^4\int_{KPR} &
    \biggl [\sum_I \delta^{\mu\mu'}_I G_I(K) -  \hat{K}^{\mu}_\text{T}\hat{K}^{\mu'}_\text{T} G_{\text{TL}}(K) \biggr ] \nn
& \times P^{\nu}P^{\nu'}\biggl [\Pi^{\mu\nu}(\hat K)\Pi^{\mu'\nu'}(\hat K)\biggr ]
    \frac{\Pi_\text{L}(\hat P)}{P^4[P^2 + \Pi_\text{L}(\hat P)]}
    \frac{\Pi_\text{L}(\hat R)}{R^4[R^2 + \Pi_\text{L}(\hat R)]}.
\label{eq:DIF3}
\end{align}
To obtain \eqs\nr{eq:DIF2} and \nr{eq:DIF3}, we first eliminated the 3g HTL vertices by using the generalized Ward identities and then plugged in the definition of $G_{\text{L}}$ from \eq\eqref{eq:gfuncdef}, where it occurs explicitly in \eq\eqref{eq:Dmod}. Note also that the term proportional to  $\delta\Gamma^{\mu\nu\rho}_{KPR}K^{\mu}P^{\nu}R^{\rho}$ has dropped out since it vanishes due to the Ward identity.

Before proceeding, we note the following fact, which we will use in evaluating these expressions. Namely, upon switching to the $(X, \chi)$ coordinates, the 3g HTL vertices scale as%
\begin{equation}
\label{eq:xscaling3gHTL}
    \delta \Gamma^{\mu \nu \rho}_{KPR} \equiv 
    \frac{1}{X} \delta \hat \Gamma^{\mu \nu \rho}_{KPR}(\chi,\Omega),
\end{equation}
where the $\delta \hat \Gamma^{\mu \nu \rho}_{KPR}(\chi,\Omega)$ defined here depends only on the angles.  This fact can be seen from the explicit integral representation of the 3g HTL vertex function in \app\ref{sec:HTLvertices}. 

We now proceed to the computation. As we did in the previous section, we shall only evalute the UV-sensitive term $\uvtext{\II{F}}$ in this section, and we defer the evaluation of the finite terms $\Delta \II{F}^{(i)}$ to \Sec\ref{sec:Ifinite} below.

\subsubsection{\texorpdfstring{The $X$ integral in $d$ dimensions}{The X integral in d dimensions}}

We proceed to evalute $\uvtext{\II{F}}$ using the steps introduced at the beginning of this section. After changing to the $(X, \chi)$ coordinates and using the scaling \eq\eqref{eq:xscaling3gHTL}, we see that we have only one master integral in $X$ to evaluate, namely
\begin{equation}
\label{eq:masterintXIFUV}
\begin{split}
 \int_{0}^{\infty} &  \frac{\ud X X^{2d-1}}{[X^2 + \hat \Pi_I(\hat K)][X^2 + \hat \Pi_J(\hat P)][X^2  + \hat \Pi_W(\hat R)]} \\
& = \Bigl[\pi\csc(\pi d)\Bigr] \frac{H_d(\chi,\Omega) }{2[\hat \Pi_I(\hat K) - \hat \Pi_J(\hat P)][\hat \Pi_I(\hat K) - \hat \Pi_W(\hat R)][\hat \Pi_J(\hat P)- \hat \Pi_W(\hat R)]}\\
& \equiv \Bigl[\pi\csc(\pi d)\Bigr] \, \mathcal{I}_d(I;J;W),
\end{split}
\end{equation}
where the function $H_d$ is given by the expression
\begin{equation}
\begin{split}
    H_d(\chi,\Omega)   ={}&  
    \hat \Pi_I(\hat K)^{d-1} \,\bigl[\hat \Pi_J(\hat P) - \hat \Pi_W(\hat R)\bigl] \\
    & + \hat \Pi_J(\hat P)^{d-1} \bigl[\hat \Pi_W(\hat R) - \hat \Pi_I(\hat K)\bigl]\\
    & + \hat \Pi_W(\hat R)^{d-1} \bigl[\hat \Pi_I(\hat K) - \hat \Pi_J(\hat P)\bigl].\\
\end{split}
\end{equation}
Here we have also introduced the notations
\begin{equation}
\hat \Pi_I(\hat K)  \equiv \frac{\Pi_I(\hat K)}{\sx^2}, \quad 
    \hat \Pi_I(\hat P) \equiv \frac{\Pi_I(\hat P)}{\cx^2}, \quad 
    \hat \Pi_I(\hat R) \equiv \frac{\Pi_I(\hat R)}{1 + 2\sx\cx w},
\end{equation}
and the $D$-dimensional polar angle $\Phi_R$ appearing in $\Pi(\hat R)$ is defined as 
\begin{equation}
\label{eq:phiR}
\begin{split}
\tan(\Phi_R) \equiv \frac{\vert \rt \vert}{R_0} & = - \sqrt{\frac{\vert \kt\vert^2 + \vert \pt\vert^2  + 2\kt \cdot \pt  }{(K_0 + P_0)^2}}\\
& = - \sqrt{\frac{\sx^2 \sin^2 \Phi_K + \cx^2 \sin^2 \Phi_P + 2\sx \cx \sin \Phi_K \sin \Phi_P \cos \theta}{\sx^2 \cos^2 \Phi_K + \cx^2 \cos^2 \Phi_P + 2\sx\cx\cos \Phi_K \cos\Phi_P}}.
\end{split}
\end{equation}
Note that the divergence is contained in the trigonometric coefficient
\begin{equation}
\label{eq:cscExpand}
    \pi \csc(\pi d) = \frac{1}{2 \epsilon} + \frac{\pi^{2}}{3} \epsilon + O(\epsilon^{2}),
\end{equation}

Using the above master integral, we find the following expression:
\begin{equation}
\label{eq:IFUVXchi}
\begin{split}
[I_F]^{\text{UV}}  =  \mE^4 C(d)& \Bigl[\pi\csc(\pi d)\Bigr]
    \int_{\Omega}\,
    \int_{0}^{\pi / 2}  \ud \chi \,\frac{ \sx^{d-2} \,\cx^{d-2} }{(1 + 2\sx\cx w)} \\
    & \times \biggl \{\left [\delta\hat \Gamma^{ijk}_{KPR}(\chi,\Omega)\right ]^2 \mathcal{I}_d(T;T;T)  + 3 \left [\delta \hat \Gamma^{i00}_{KPR}(\chi,\Omega) \right ]^2 \mathcal{I}_d(T;L;L)\\
 & +  3 \left [\delta \hat \Gamma^{ij0}_{KPR}(\chi,\Omega) \right ]^2 \mathcal{I}_d(T;T;L) +  \left [\delta \hat \Gamma^{000}_{KPR}(\chi,\Omega) \right ]^2 \mathcal{I}_d(L;L;L) \biggr \}.
    \end{split}
\end{equation}
Here, we have introduced a shorthand notation for contracted indices, e.g.,\ $[\delta\hat \Gamma^{ijk}_{KPR}(\chi,\Omega) ]^2 \equiv \delta\hat \Gamma^{ijk}_{KPR}(\chi,\Omega) \delta\hat \Gamma^{ijk}_{KPR}(\chi,\Omega)$.  We note that in general $d$, the functions $\mathcal{I}_d$ also depend on the angles $\chi$ and $\Omega$. 

\subsubsection{\texorpdfstring{Expanding in $\epsilon$ and doing the angular integrals}{Expanding in epsilon and doing the angular integrals}}

Following the procedure used to evaluate the other UV-sensitive contributions, we now turn to performing the $\epsilon$ expansion of $[I_F]^{\text{UV}}$ to calculate terms in the expansion of the pressure in \eq\eqref{eq:2loopSeriesSimple}. From the above expansion of the $d$-dimensional trigonometric coefficient, we can conclude that $[I_F]^{\text{UV}}$ contributes only to the $p_{-1}$ and $p_{0}$ terms of the pressure. Recall that, in order to obtain the correct $p_0$ coefficient, the HTL vertices appearing in the integrand are kept $d$-dimensional. Therefore, to calculate the subleading contributions, we must expand not only the explicit $d= 3 - 2\epsilon$ appearing in the measure and the expression, but we also must expand the $d$-dimensional 3g HTL vertices up to $O(\epsilon)$; schematically: 
\begin{equation}
    \delta \hat{\Gamma}^{\mu \nu \rho}_{KPR}(\chi, \Omega) = 
    \delta \hat{\Gamma}^{\mu \nu \rho}_{0,\,KPR}(\chi, \Omega) +
    \delta \hat{\Gamma}^{\mu \nu \rho}_{1,\,KPR}(\chi, \Omega) \epsilon 
    + O(\epsilon^{2}).
\end{equation}
The details of the terms in this expansion are given in \app\ref{sec:HTLvertices}, and the general approach is as follows:  The 3g HTL vertex is given by a $d$-dimensional integral representation. We introduce a modified Feynman parametrization, which allows us to do these $d$-dimensional integrals, at least order-by-order in $\epsilon$, leaving only the Feynman parameter left to be integrated over. Thus, we are able to write the functions in the above $\epsilon$ expansion as one-dimensional integral representations. As far as we are aware, such an explicit evaluation of the HTL vertices has never been performed before in the literature.

With this approach, the $p_{-1}$ and $p_{0}$ contributions from $\uvtext{\II{F}}$ [including the symmetry factor $1 / 12$ from \eq\eqref{eq:Itotal}] can be written as six-dimensional integrals (over $\chi$, $\Phi_K$, $\Phi_P$, $\theta$, and two Feynman parameters). We compute these integrals numerically using Monte Carlo integration routines provided by the CUBA library \cite{Hahn:2004fe}. Our results are summarized in \tab\ref{tab:Fcontributions}, where in the $p_{-1}$ term we see the anticipated direct contraction of the two 3g HTL vertices.

\begin{table}
\begin{ruledtabular}
\begin{tabular}{lll}
     & $p_{-1}$ & $p_{0}$ \\
    \hline
     $\uvtext{\II{F}}$    &   $\textstyle{
             \int_{\Omega}\,
    \int_{0}^{\pi / 2} \!\!\!\! \frac{\ud \chi \, \sx \,\cx}{12(1 + 2\sx\cx w)} 
    \left [\delta\hat \Gamma^{\mu \nu \rho}_{0,\,KPR}(\chi,\Omega)\right ]^2 \phantom{\Biggl (}\!\!\!\! = 0.4340(15)}$  & 
    0.2483(14)
      \\
\end{tabular}
\end{ruledtabular}
    \caption{\label{tab:Fcontributions} Contributions from $\uvtext{\II{F}}$ to the pressure after performing the necessary integrals.}
\end{table}

\subsection{The remaining finite terms}
\label{sec:Ifinite}

Having discussed every truly UV-divergent integral, we are left with the finite terms. These include the integrals $\II{E}$ and $\Delta \II{F}^{(i)}$ defined in \eqs\eqref{eq:IEdef} and \eqref{eq:DIF1}--\eqref{eq:DIF3} which are finite contributions to the potentially divergent UV term $[\II{tot}]^{\text{UV}}$, as well as the four integrals $[\IIUD{(0,0)}{3g}]_{111},\,[\IIUD{(0,H)}{3g}]_{110},\,[\IIUD{(0,H)}{3g}]_{111},$ and $[\IIUD{(H)}{4g}]_{11}$ defined in \eqs \eqref{eq:I3G00def}, \eqref{eq:I3G0Hdef}, and \eqref{eq:I4GHdef}, which were seen to be finite from the start. 

These terms all involve a coupling between the two loop-momenta to a degree that renders one unable to perform factorizations and other simplifications akin to those seen in \Sec\ref{sec:IAD}. However, the finite nature of these terms makes them simpler to evaluate numerically, as there is no need to extract the divergent parts when evaluating them, unlike in the previously considered contributions. For this reason, we automate the evaluation of these terms, performing any remaining non-Lorentz-invariant tensor contractions by adapting the implementation discussed in \Ref\cite{Shtabovenko:2016sxi} (although in Euclidean space). 

Various properties of the HTL vertex functions, in particular the generalized Ward identities in \eq\eqref{eq:WardVertices} and their tracelessness, are again used extensively to reduce everything down to terms containing either no vertex corrections or the components $\delta\Gamma^{000}_{KPR}$ or $\delta\Gamma^{0000}_{K,P,-K,-P}$, except in the case of $\Delta \II{F}^{(1)}$. For this one separate term, we cannot fully remove the spatial components of the 3g HTL vertex function, but we reduce it as much as possible by using 
\begin{equation}
\begin{split}
    \hat{K}_{\text{T}}^{\mu} \delta \Gamma^{\mu \nu \rho}_{KPR} ={}& 
    \frac{1}{|\vec{k}|} (K^{\mu} - \delta^{\mu 0}K_{0}) \delta \Gamma^{\mu \nu \rho}_{KPR} \\
    ={}& 
    \frac{1}{|\vec{k}|} (K^{\mu}\delta \Gamma^{\mu \nu \rho}_{KPR} - K_{0}\delta \Gamma^{0 \nu \rho}_{KPR}),
\end{split}
\end{equation}
and then applying the generalized Ward identity on the first term. This way, we can reduce the number of spatial indices appearing in our expressions, as the 3g HTL vertices with more time components are easier to compute numerically. In the end, with this $\Delta \II{F}^{(1)}$ term, we can reduce all expressions to terms involving the four independent expressions $\delta \Gamma^{000}_{KPR}$, $\delta \Gamma^{i00}_{KPR}$, $\delta \Gamma^{ij0}_{KPR}$, and $\delta \Gamma^{ijk}_{KPR}$, with their indices contracted either with a second occurrence of one of these terms, or with the self energy functions $\Pi^{00}(\hat{Y})$, $\Pi^{i0}(\hat{Y})$, and $\Pi^{ij}(\hat{Y})$, with $Y \in \{K,P,R\}$. More details about these manipulations can be found in \app\ref{subsec:3gHTLwithWard}.

\begin{table}
\begin{tabular}{@{\quad}l@{\quad\quad\quad}l@{\quad}}
\botrule
Contribution & $p_0$ \\
\colrule
$\II{E}$                        & $-2.9229(18)$ \\
$\Delta \II{F}^{(1)}$                & $-0.0810(7)$\\
$\Delta \II{F}^{(2)}$                & $+0.285(6)$\\
$\Delta \II{F}^{(3)}$                & $+0.0838(5)$\\
$\Bigl[\IIUD{(0,0)}{3g}\Bigr]_{111}$    & $-0.6772(12)$ \\
$\Bigl[\IIUD{(0,H)}{3g}\Bigr]_{110}$    & $-1.309(2)$\\
$\Bigl[\IIUD{(0,H)}{3g}\Bigr]_{111}$    & $+0.2027(6)$\\ 
$\Bigl[\IIUD{(H)}{4g}\Bigr]_{11}$       & $-0.669(2)$\\[1ex]
\botrule
\end{tabular}
    \caption{\label{tab:Ifinitecontributions} Contributions from the finite terms to the analytic part of the pressure $p_0$}
\end{table}

The 4g HTL vertex function appears in these finite terms for the first time; its properties are discussed in \app\ref{sec:HTLvertices}. It should also be noted that, while here we are able to work in an integer dimension, computations of the HTL integrals in general $d$ are easily automated using the method outlined in \app\ref{sec:htlauto}, and in practice this method was applied in our computation. 

With the integral scalarized, we simply follow the steps outlined in the beginning of \Sec\ref{sec:compute}, setting $d=3$ everywhere as these terms are finite. When changing to $(X, \chi)$ coordinates, we define new master integrals, which we do not show here but are easily computed. We also note here that similarly to the 3g HTL vertex, the 4g HTL vertex scales homogeneously as a function of $X$ 
\begin{equation}
\label{eq:xscaling4gHTL}
    \delta \Gamma^{\mu \nu \rho \sigma}_{K,P,-K,-P} \equiv 
    \frac{1}{X^{2}} \delta \hat \Gamma^{\mu \nu \rho \sigma}_{K,P,-K,-P}(\chi,\Omega),
\end{equation}
which can be seen from the explicit integral form (\eq\eqref{eq:4gHTL})

After this, the remaining nontrivial integrals (over $\chi$, $\Phi_K$, $\Phi_P$, $\theta$, and possible Feynman parameters in terms involving irreducible vertex corrections) are computed numerically using Monte Carlo methods \cite{Hahn:2004fe}. The numerical results for the finite contributions are displayed in \tab\ref{tab:Ifinitecontributions}. Note that in $[\IIUD{(H)}{4g}]_{11}$, the coefficient of the term  $\propto\delta\Gamma^{000}_{KPR}$ turns out to vanish, and as such the vertex corrections contain only a contribution $\propto\delta\Gamma^{0000}_{K,P,-K,-P}$. Finally, we remind the reader that there is a symmetry factor of $2$ for the $\II{3g}^{(0,H)}$ terms in this table (see \eq\eqref{eq:G4g3gsplit}), and a symmetry factor of $1 / 12$ for the $\Delta \II{F}^{(i)}$ terms.

\section{Results and conclusions}
\label{sec:pResult}

We are now in a position to present our final results. By adding the corresponding elements of \tabs\ref{tab:ABCDcontributions}, \ref{tab:Fcontributions}, and \ref{tab:Ifinitecontributions}, we find the following fully soft contributions to the pressure in \eq\eqref{eq:2loopSeriesSimple}:
{
    \allowdisplaybreaks
\begin{align}
    p_{-2} ={}& \frac{11}{6} \int_{\Omega} \tr{\Pi_0(\hat{K})^{2}} = \frac{11 \pi^{2}}{24},  \label{eq:doublelogvalue} \\
    p_{-1} ={}&  \int_{\Omega} \Biggl\{ \frac{19+11 \pi^{2}}{72} - \frac{11}{6} \tr{\Pi_0(\hat{K})^{2} \ln \bigl[ \Pi_{0}(\hat{K}) \bigr]} \nonumber \\*
    &\quad \quad \quad+ \int_{0}^{\pi / 2} \!\!\!\!\! \frac{\ud \chi \, \sin \chi \,\cos \chi}{12(1 + 2 \cdot w \sin \chi\cos \, \chi)} 
    \left [\delta\hat \Gamma^{\mu \nu \rho}_{0,\,KPR}(\chi,\Omega)\right ]^2 \Biggr\} \nonumber \\
    ={}&  11.6840(15), \label{eq:singlelogvalue} \\ 
    p_{0} ={}&  17.150(7).  \label{eq:constantvalue} 
\end{align}
}
$\hspace{-0.75em}$ Here, we note again for the reader that the angular integral $\int_\Omega = \pi^{2} / 2$ (see \eq\eqref{eq:def_omega} for a full definition of the measure), and we have used a shorthand $w=K\cdot P/(|K||P|)$. The HTL self energy $\Pi_0$ and the 3g vertex correction $\delta\hat{\Gamma}_0$ can be found in \apps \ref{sec:HTLselfenergy} and \ref{sec:3gHTL} respectively; note that we have also scaled out $\mE$ from them.

Though our final results can be written in a relatively compact form, the intermediate steps to arrive to this result involved many complex manipulations and techniques. To this end, to verify that our results are correct, we conducted many cross checks of the intermediate steps. Firstly, all of our numerical results were checked by multiple independent codes, each of which used a different implementation (e.g., full automatization versus partial simplifications by hand), which gives us strong confidence in our results. For the UV-sensitive terms in $\II{ABCD}$, we have some additional cross checks. As mentioned in the earlier section, we have verified that our results are unchanged if we do not introduce $\bigl[I_{-2}(d)\bigr]^{\text{UV}}$, but rather use the previous expressions containing $\Pi_{0}$ through $\Pi_{2}$. We have also performed a check of the highly nontrivial tensor reduction in \eqs\eqref{eq:angAve1}--\eqref{eq:angAve2}: we have verified that our results remain unchanged if we do not perform this averaging but instead do the $\epsilon$ expansion and then perform the more coupled angular integrals that arise.

In addition to these internal cross checks, we find that we reproduce the known coefficient of the leading logarithmic $O(\alpha_s^{3} \ln^{2} \alpha_s)$ contribution to the pressure obtained in \Ref\cite{Gorda:2018gpy}. That work obtained this coefficient by expanding the two-loop HTL pressure in the semisoft region and using a cutoff prescription. Here, on the other hand, we used the fully resummed expressions, without expansions, and used dimensional regularization to arrive at the result. Thus, the two techniques that yield agreement are quite independent. Note that if we use our result here for $p_{-2}$ in the expansion in \eq\eqref{eq:2loopSeriesExpanded}, we find that this result yields a coefficient for the $O(\alpha_s^{3} \ln^{2} \alpha_s)$ term that is a factor of two larger than the result in \Ref\cite{Gorda:2018gpy}\footnote{There is also a discrepancy in the overall sign. This is due to an error in \Ref\cite{Gorda:2018gpy}, where the free energy was calculated instead of the pressure for this term. We thank J.-L.~Kneur for bringing this issue to our attention.}. However, this is only the  $\ln^2 (\mE / \Lh)$ term arising in $\pDU{3}{s}$, which does not correspond to the full  $\ln^2 \alpha_s^{1/2}$ term in the sum over all regions. As we discuss in \app\ref{sec:four_regs_sum} using a simple example, it is natural to expect the  $\ln^2 \alpha_s^{1/2}$ coefficient to be precisely half of this $\ln^2 (\mE / \Lh)$ coefficient. Thus, we do indeed find agreement between our result and \Ref\cite{Gorda:2018gpy}.

As a further additional check with the literature, we have extended the formalism used in \Ref\cite{Gorda:2018gpy} to extract parts of the subleading logarithmic contribution, corresponding to pieces of $p_{-1}$. Within the cutoff regularization, there are two very distinct contributions to $p_{-1}$ from the two-loop HTL pressure: One contribution arises from the soft--semisoft region, obtained by expanding the two-loop HTL pressure for only a single loop momentum, and corresponds to the second term of \eq\eqref{eq:singlelogvalue}. The other contribution arises as a subleading correction to the semisoft--semisoft region (which in the $(X,\chi)$ coordinates has cutoffs on the radial $X$ integral). This correction is regulator dependent, as it is sensitive to the exact ratio of cutoffs that occur in the double logarithm, rather than just the parametric size of the ratio.  However, there is also a regulator-independent piece, corresponding in the language of the present paper to the last term of \eq\eqref{eq:singlelogvalue}, due to the unique scaling properties of the doubly contracted vertex correction. As with the $p_{-2}$ term, these regulator-independent pieces are found to be consistent between the two methods, giving us further confidence in our calculation.

We now turn to remarks upon our results. The first is that though we have rederived the same analytic value for the coefficient of the $O(\alpha_s^3 \ln^2 \alpha_s)$ term, the integral expression for that result was much more complicated than the remarkably simple integral expression for $p_{-2}$ above. In particular, in \Ref\cite{Gorda:2018gpy}, the integral expression did not depend on only one momentum $K$, but rather both $K$ and $P$. In fact, the integral given for $p_{-2}$ above is exactly the same angular integral that appears in the Freedman--McLerran $O(\alpha_s^{2} \ln \alpha_s)$ result. Looking back through our analysis, we find that the angular tensor reduction was what lead to this vast simplification, as such a decomposition was not used in \Ref\cite{Gorda:2018gpy}. We further note that, since this same integral appears in $p_{-2}$, the corresponding contribution to $p_{-1}$ [namely, the second term of \eq\eqref{eq:singlelogvalue}], is also an integral that appears in the $\apDU{2}{s}$ term.\footnote{This is the term involving the integral $\delta$ defined in \Ref\cite{Vuorinen:2003fs}.}

We also remark here upon the fact that in \Ref\cite{Gorda:2018gpy}, the authors found that the same final result would be obtained if they set $K$ on shell, with $\Pi_{\text{T},0} \mapsto \Pi_{\text{T}, 0}(i K_0 = |\vec{k}|, \vec{k}) = 1 / 2$ and $\Pi_{\text{L},0} \mapsto \Pi_{\text{L}, 0}(i K_0 = |\vec{k}|, \vec{k}) = 0$. This is also seen to work in the case of the $O(\alpha_s^{2} \ln \alpha_s)$ coefficient. (Here we have used our convention of factoring out $\mE$.) However, we now see from the analysis in the present work that the replacement working for the $O(\alpha_s^{3} \ln^2 \alpha_s)$ coefficient follows directly from it working for the lower-order result, since the integral expressions are exactly the same. It is tempting to speculate that the leading-logarithmic result at all orders may be related to the integral appearing in the  $\apDU{2}{s}$ term, and may allow for such a substitution.

As a further remark, we note that there have been many efforts to identify which diagrams give dominant contributions to an HTL calculation \cite{CaronHuot:2008uh,York:2012ib}. Here, we find, interestingly, that contributions containing irreducible HTL vertex corrections (i.e., those which cannot be removed using the generalized Ward identities) are clearly found to be smaller than those containing self energies.\footnote{Compare, e.g.\, the $p_{-1}$ result in \tab\ref{tab:Fcontributions} to the corresponding results in \tab\ref{tab:ABCDcontributions}.} We find that these corrections, though necessary for full, correct results, are only a few percent of the total. This observation is gauge invariant, though not unique, since it depends on the basis of irreducible vertex integrals chosen in the $p_{0}$ term.

Our result in \eqs\eqref{eq:doublelogvalue}--\eqref{eq:constantvalue} constitutes the fully soft contribution to the pressure: $\apDU{3}{s}$ in \eq\eqref{eq:schematics}. This physical result still has some scheme dependence arising from the ambiguity in splitting the semisoft modes to the hard or soft sectors, which can be seen from the residual dependence on the factorization scale $\Lh$ in the expressions. In our accompanying paper \cite{Gorda:2021znl}, we further discuss this point, as well as analyze the relative importance of this contribution to the pressure and the effect of this contribution on the convergence of the weak-coupling expansion.  Moreover, by undertaking the present calculation in dimensional regularization and by having the clear roadmap set forth in the Introduction, we believe that at least the further  $\apDU{3}{m}$ term can be obtained in a straightforward manner.  In particular, there is no difficulty in combining the contributions from different kinematic regions, since they are all regulated consistently in dimensional regularization. With the $\apDU{3}{m}$ term in hand, one could obtain the subleading $O(\alpha_s^{3} \ln \alpha_s)$ coefficient in the weak-coupling expansion, which may in turn allow one to use the principle of minimal sensitivity \cite{Stevenson:1982qw} to constrain the dependence of the pressure on the renormalization scale $\bar{\Lambda}$. This would potentially have important phenomenological implications for, e.g., the EOS of neutron-star matter \cite{Annala:2017llu,Annala:2019puf}. Additionally, the entire organizational overview presented in the Introduction may have important consequences in itself: it may be possible to resum these logarithmic contributions to the pressure in some systematic way. An investigation of these points is left to later work.

Finally, we remark that there are some possible generalizations of this work which may be greatly aided by the organization and machinery that we have developed here. For example, including nonzero quark masses \cite{Fraga:2004gz, Kurkela:2009gj}, or generalizations to nonzero temperature \cite{Kurkela:2016was} might be possible using our present techniques. Such endeavors are, however, also left for the future.

\acknowledgements{
RP, SS, and AV have been supported by the Academy of Finland grant no.~1322507, as well as by the European Research Council, grant no.~725369. This work was begun while TG was supported by the U.S.\ Department of Energy under Grant No.\ DE-SC0007984.
}

\newpage
\appendix

\section{Feynman rules}
\label{sec:feynman_rules}

The standard group-theory factors for the SU($N_c$) gauge group, some of which appear in the text, are given by
\begin{equation}
\label{eq:gtfactors}
\begin{split}
\da & \equiv \delta^{aa} = \nc^2 - 1,\\
C_{\rm A} \delta^{cd} & \equiv f^{abc}f^{abd} = \nc \delta^{cd},\\
C_{\rm F}\delta_{ij} & \equiv (t^a t^a)_{ij} = \frac{\nc^2 -1}{2\nc} \delta_{ij},
\end{split}
\end{equation}
where $\nc$ is the number of colors and $f^{abc}$ is the fully anti-symmetric structure constant. The generators of the fundamental representation $(t^a)_{ij}$ are normalized according to $\tr{t^a t^b} = \delta^{ab}/2$.

In the following we give the bare Feynman rules of QCD necessary for the computation of the diagrams under study. The vertices, with momentum flow towards the vertex, read:
\begin{equation}
\label{eq:3g_bare}
(\Gamma_0)^{\mu\nu\rho}_{abc}(P,Q,R)=
\vcenter{\hbox{\begin{tikzpicture}[scale=2*\picSc]
	\gluonLine{-1}{0}{0}{0}
	\gluonLine{0}{0}{{cos(45)}}{{sin(45)}}
	\gluonLine{{cos(-45)}}{{sin(-45)}}{0}{0}
	\node at (-1.15,0)(c){$P$};
	\node at ({1.15*cos(45)},{1.15*sin(45)})(c){$Q$};
	\node at ({1.15*cos(-45)},{1.15*sin(-45)})(c){$R$};
\end{tikzpicture}}} 
=  igf_{abc}\Big[ \delta^{\mu\nu}(P-Q)^{\rho} + \delta^{\nu\rho}(Q-R)^{\mu} + \delta^{\mu\rho}(R-P)^{\nu}\Big],
\end{equation}

\begin{equation}
\begin{split}
\label{eq:4g_bare}
(\Gamma_0)^{\mu\nu\rho\sigma}_{abcd}(P,Q,R,S)= 
\vcenter{\hbox{\begin{tikzpicture}[scale=2*\picSc]
	\gluonLine{0}{0}{{cos(135)}}{{sin(135)}}
	\gluonLine{{cos(-135)}}{{sin(-135)}}{0}{0}
	\gluonLine{0}{0}{{cos(45)}}{{sin(45)}}
	\gluonLine{{cos(-45)}}{{sin(-45)}}{0}{0}
	\node at ({1.15*cos(45)},{1.15*sin(45)})(c){$Q$};
	\node at ({1.15*cos(-45)},{1.15*sin(-45)})(c){$R$};
	\node at ({1.15*cos(135)},{1.15*sin(135)})(c){$P$};
	\node at ({1.15*cos(-135)},{1.15*sin(-135)})(c){$S$};
\end{tikzpicture}}} 
    = -g^2& \biggl [f_{eab}f_{ecd}\biggl (\delta^{\mu\rho}\delta^{\nu\sigma} - \delta^{\mu\sigma}\delta^{\nu\rho} \biggr )\\[-1cm]
& + f_{eac}f_{edb}\biggl (\delta^{\mu\sigma}\delta^{\nu\rho} - \delta^{\mu\nu}\delta^{\sigma\rho} \biggr )\\
& + f_{ead}f_{ebc}\biggl (\delta^{\mu\nu}\delta^{\sigma\rho} - \delta^{\mu\rho}\delta^{\sigma\nu} \biggr ) \biggr ],
\end{split}
\end{equation}

\begin{equation}
(V_0)^{\mu}_{a,ij}(P,Q,R)=
\vcenter{\hbox{\begin{tikzpicture}[scale=2*\picSc]
	\gluonLine{-1}{0}{0}{0}
	\quarkLine{0}{0}{{cos(45)}}{{sin(45)}}{0.5}
	\quarkLine{{cos(-45)}}{{sin(-45)}}{0}{0}{0.5}
	\node at (-1.28,0)(c){$P$};
	\node at ({1.35*cos(45)},{1.15*sin(45)})(c){$Q$};
	\node at ({1.35*cos(-45)},{1.15*sin(-45)})(c){$R$};
\end{tikzpicture}}} 
= g\gamma^{\mu}(t_a)_{ij},
\end{equation}

\begin{equation}
(W_0)^{\mu}_{abc}(P,Q,R)=
\vcenter{\hbox{\begin{tikzpicture}[scale=2*\picSc]
	\gluonLine{-1}{0}{0}{0}
	\ghostLine{0}{0}{{cos(45)}}{{sin(45)}}{0.5}
	\ghostLine{{cos(-45)}}{{sin(-45)}}{0}{0}{0.5}
	\node at (-1.28,0)(c){$P$};
	\node at ({1.35*cos(45)},{1.15*sin(45)})(c){$Q$};
	\node at ({1.35*cos(-45)},{1.15*sin(-45)})(c){$R$};
\end{tikzpicture}}} 
= ig f_{abc} Q^{\mu}.
\end{equation}

The corresponding propagators (with massless quarks), on the other hand, read

\begin{equation}
\label{eq:gluon_bare}
(\Delta_0)^{\mu\nu}_{ab}(P,Q)=
\vcenter{\hbox{\begin{tikzpicture}[scale=2*\picSc]
	\gluonLine{-1}{0}{1}{0}
	\node at (-1.25,0)(c){$P$};
	\node at (1.25,0)(c){$Q$};
\end{tikzpicture}}} 
= \delta_{ab}\delta^{(D)}(P+Q)\Bigg[\frac{\delta^{\mu\nu}}{P^2}-(1-\xi)\frac{P^{\mu}P^{\nu}}{P^4}\Bigg],
\end{equation}

\begin{equation}
(S_0)_{ij}(P,Q)=
\vcenter{\hbox{\begin{tikzpicture}[scale=2*\picSc]
	\quarkLine{-1}{0}{1}{0}{0.5}
	\node at (-1.25,0)(c){$P$};
	\node at (1.25,0)(c){$Q$};
\end{tikzpicture}}} 
= \delta_{ij}\delta^{(D)}(P+Q) \frac{-\cancel{P}}{P^2},
\end{equation}

\begin{equation}
(C_0)_{ab}(P,Q)=
\vcenter{\hbox{\begin{tikzpicture}[scale=2*\picSc]
	\ghostLine{-1}{0}{1}{0}{0.5}
	\node at (-1.25,0)(c){$P$};
	\node at (1.25,0)(c){$Q$};
\end{tikzpicture}}} 
= \delta_{ab}\delta^{(D)}(P+Q) \frac{1}{P^2}.
\end{equation}
We also define single-argument propagators to be two-argument propagators after enforcing the momentum-space delta function; for example $(\Delta_0)^{\mu\nu}_{ab}(P)=\delta^{\mu\nu}\delta_{ab}P^{-2}-(1-\xi)P^{\mu}P^{\nu}\delta_{ab}P^{-4}$. In \Sec\ref{sec:HTLappendix} below, we will write down some of the rules improved according to the standard HTL scheme.

\section{The HTL framework}
\label{sec:HTLappendix}

In this appendix, we outline in detail the Euclidean HTL framework used throughout this work.
To make our HTL appendices self-contained, we repeat here some definitions that are scattered throughout the text.  We leave the detailed evaluation of vertex functions to later appendices. We found the discussions in \Refs\cite{Bellac:2011kqa,Laine:2016hma,Andersen:2002ey} helpful when creating this appendix. 

\subsection{HTL propagator and self energy}
\label{sec:HTLselfenergy}

\paragraph{Propagator:} The HTL-resummed gluon propagator 
\begin{equation}
    \D{\mu \nu}_{ab}(P,K)  = \vcenter{ \hbox{\begin{tikzpicture}[scale=2*\picSc]
	\doubleGluonLine{-1}{0}{1}{0}
	\node at (-1.3,0)(c){$K$};
	\node at (1.3,0)(c){$P$};
\end{tikzpicture}}}\equiv \delta_{ab}\delta^{(D)}(P+K)\D{\mu \nu}(K)   
\end{equation} 
is defined in the covariant gauge as
\begin{equation}
\label{eq:APPfullpropa}
    \D{\mu\nu}(K) \equiv \ProjIX{\mu \nu}{T}{K} G_T(K) + \ProjIX{\mu \nu}{L}{K} G_L(K)  + \xi \frac{K^{\mu}K^{\nu}}{(K^2)^2},
\end{equation} 
where the parameter $\xi$ fixes the gauge and 
\begin{equation}
\label{eq:APPgfuncdef}
G_I(K) \equiv \frac{1}{K^2 + \Pi_{I}(K)}, \quad I \in \{ \text{T}, \text{L} \} .
\end{equation}
The computation performed in this paper is most efficient to carry out in the $\xi = 1$ gauge, which we use throughout the text. The $D$-dimensional transverse and longitudinal projection operators, $\prjoT^{\mu \nu}(\hat{K})$ and $\prjoL^{\mu \nu}(\hat{K})$, are defined as
\begin{equation}
    \label{eq:projsDefs1}
    \begin{split}
        \ProjIX{\mu\nu}{T}{K} & \equiv \delta^{\mu i}\delta^{\nu j}\biggl (\delta^{ij} - 
        \hat{k}^{i} \hat{k}^{j} \biggr ),\\
        \ProjIX{\mu \nu}{L}{K} & \equiv \ProjIX{\mu \nu}{$D$}{K}  - \ProjIX{\mu \nu}{T}{K}, 
    \end{split}
\end{equation}
with
\begin{equation}
\label{eq:projsDefs2}
    \ProjIX{\mu\nu}{$D$}{K}  \equiv \delta^{\mu\nu} - \hat{K}^{\mu} \hat{K}^{\nu},
\end{equation}
where $\hat{K} = K / |K|$, and $\hat{\vec{k}} = \vec{k} / |\vec{k}|$.  These projectors are $D$-dimensionally transverse to $K$
and obey the relations
\begin{equation}
\label{eq:APPprojrelations}
(\prjoT)^2 = \prjoT, \quad (\prjoL)^2 = \prjoL, \quad \prjoT^{\mu\alpha}(\hat{K})\prjoL^{\alpha\nu}(\hat{K})=0,
\end{equation}
and
\begin{equation}
\label{eq:APPprojtrrelations}
    \tr{\prjoT} = d - 1, \quad
    \tr{\prjoL} = 1, \quad
    \tr{\prjof} = d. 
\end{equation}

The $\Pi_{\text{T}}$ and $\Pi_\text{L}$ appearing in the denominators of the HTL propagators are components of the one-loop HTL self energy (or polarization tensor) $\Pi^{\mu\nu}_{ab} \equiv \delta_{ab}\Pi^{\mu\nu}$  of the gluon field. These functions satisfy the general relations
\begin{equation}
    \Pi^{\mu\nu}(K) = \ProjIX{\mu\nu}{T}{K} \Pi_\text{T}(K) + \ProjIX{\mu\nu}{L}{K} \Pi_\text{L}(K),
\end{equation}
and
\begin{equation}
\label{eq:APPpiTandpiL}
\begin{split}
\Pi^{\mu\mu}(K) & = (d-1)\Pi_\text{T}(K) + \Pi_\text{L}(K),\\
\Pi^{00}(K) & =  \frac{\vert \kt\vert^2}{K^2} \Pi_\text{L}(K).
\end{split}
\end{equation}
Additionally, the self energy satisfies the trivial Ward identity $K^{\mu} \Pi^{\mu \nu}(K) = 0$.

\paragraph{Machinery for manipulating the propagator:}
We let $\DN{0}^{\mu \nu}(K)$ denote the bare propagator with $\Pi_{I}(K) = 0$, $I \in \{\text{T}, \text{L}\}$. We can extend this notation to label the other terms in the expansion of the full propagator in powers of the self energy:
\begin{equation}
    \begin{split}
        \DN{n}^{\mu \nu}(K) \equiv&\, (-1)^{n} [\overbrace{\DZ(K)\cdot\Pi(K)\cdot\DZ(K)\cdots}^{n\,{\Pi}s}\DZ(K)]^{\mu \nu}, \qquad n\geq 0, \\
        ={}& (-1)^{n} \frac{[\Pi(K)^{n}]^{\mu \alpha}}{(K^{2})^{n}} \Dz{\alpha \nu}(K). 
    \end{split}
\end{equation}
Here, we use a dot to represent contraction of adjacent indices, and we use the notation  
\begin{equation}
    [\Pi(K)^{n}]^{\mu \nu} = \overbrace{\Pi^{\mu \alpha_1}(K) \Pi^{\alpha_1\alpha_2}(K) \cdots \Pi^{\alpha_{n-1}\nu}(K)}^{n \, \Pi\text{s}},
\end{equation}
and make the identification $[\Pi(K)^0]^{\mu \nu} = \delta^{\mu \nu}$ to match the leading term. Thus, 
\begin{equation}
    \D{\mu\nu}(K) \simeq\, \DN{0}^{\mu\nu}(K) + \DN{1}^{\mu\nu}(K) + \DN{2}^{\mu\nu}(K) + \cdots.
\end{equation}
We can now introduce the following notation for the resummed propagator with the $n$ leading terms removed:
\begin{equation}
    \DDN{n}^{\mu\nu}(K) \equiv\, \D{\mu\nu}(K) - \sum_{k = 0}^{n - 1} \DN{k}^{\mu \nu}(K), \qquad n \ge 1.
\end{equation}

Consequently, the $\DDN{n}^{\mu\nu}(K)$ are still resummed expressions, while the $\DN{n}^{\mu\nu}(K)$ are not. Note that both $\DN{n}^{\mu\nu}(K)$ and $\DDN{n}^{\mu\nu}(K)$ are $D$-dimensionally transverse for every $n$, and that the following relations hold for any $n\ge 1$:
\begin{align}
    \DDN{n}^{\mu \nu}(K) ={}&\, (-1)^{n} [\overbrace{\DZ(K)\cdot\Pi(K)\cdot\DZ(K)\cdots}^{n\,{\Pi}s} D(K)]^{\mu\nu},  \nn
        ={}& (-1)^{n} \frac{[\Pi(K)^{n}]^{\mu \alpha}}{(K^{2})^{n}} \D{\alpha \nu}(K), \label{eq:APPDDN_explicit}\\
    \D{\mu\nu}(K) ={}&\, \DN{0}^{\mu\nu}(K) + \DN{1}^{\mu\nu}(K) + \cdots + \DN{n-1}^{\mu\nu}(K) + \DDN{n}^{\mu\nu}(K) \label{eq:APPDDN_sum},\\
    \DDN{n-1}^{\mu\nu}(K) ={}&\, \DN{n-1}^{\mu\nu}(K) + \DDN{n}^{\mu\nu}(K), \label{eq:APPDDN_recursion} \\
    \DDN{n}^{\mu\nu}(K) \sim{}& \DN{n}^{\mu\nu}(K) \sim \mE^{2n} K^{-2(n+1)} \quad\text{in UV.} \label{eq:APPpowCount}
\end{align}
Notice also the full propagator at the end of both lines of \eq\eqref{eq:APPDDN_explicit}, and the fact that \eq\eqref{eq:APPDDN_sum} is not a partial sum of an infinite series, but is exact. The power-counting in \eq\eqref{eq:APPpowCount} allows one to use this notation to extract the form of the UV-sensitive terms in our calculations.

\paragraph{Self energy:}
In the \HTL\ approximation relevant for \CQM, the quark part of the one-loop gluon self energy is computed assuming that the momentum flowing along the quark lines is much larger than the external gluonic one. In this way, one obtains the result  
\begin{equation}
\label{eq:gluonselfHTL}
\Pi^{\mu\nu}(K) = m_\text{E}^2 \int_{\hat\vt}\biggl (\delta^{\mu 0}\delta^{\nu 0} - \frac{iK_0}{K \cdot V} V^{\mu}V^{\nu}\biggr ),
\end{equation}
where we have introduced the lightlike four-vector $V^{\mu} \equiv (-i, \hat \vt)$ with $\hat \vt$ a unit vector in $\mathbb{R}^d$. The integration measure in $d$ dimensions is defined as
\begin{equation}
\label{eq:intmesv}
\begin{split}
\int_{\hat\vt} & \equiv \frac{h(d)}{2}\int_{0}^{\pi} \ud \theta_v \sin^{d-2}(\theta_v) \\
& = \frac{h(d)}{2} \int_{-1}^{1}\ud z_v(1-z_v^2)^{\frac{d-3}{2}}, \quad\quad h(d) \equiv \frac{\Gamma\left (\frac{d}{2}\right )}{\Gamma\left (\frac{3}{2} \right )\Gamma\left (\frac{d-1}{2} \right )},
\end{split}
\end{equation}
where $z_v \equiv \hat{\kt} \cdot \hat\vt$; note that the measure here is normalized to integrate to unity. The $d$-dimensional in-medium effective mass scale $\mE$ is given by
\begin{equation}
\begin{split}
\mE^2 & = \sum_f g^2\mu_f^2\left (\frac{e^{\gamma_E}\Lh^2}{4\pi\mu_f^2} \right )^{\frac{(3-d)}{2}}\frac{4\Gamma(\frac{1}{2})}{(4\pi)^{\frac{d+1}{2}}\Gamma(\frac{d}{2})}\\
& = \sum_f \frac{g^2 \mu_f^2}{2\pi^2}  + O(\epsilon).
\end{split}
\end{equation}
This is the generalization of the effective mass scale to the case of multiple fermion flavors with different chemical potentials $\mu_f$ at zero temperature. Throughout our text, $\mE$ denotes its $d$-dimensional value, and is never expanded in $\epsilon$.

The scalar functions $\Pi_\text{T}$ and $\Pi_\text{L}$ can now be computed using the constraint equations in \eq\nr{eq:APPpiTandpiL} with the results
\begin{equation} \label{eq:dimregpi}
\begin{split}
\Pi^{\mu\mu}(K) & = \mE^2 \int_{\hat\vt} \delta^{00} = \mE^2,\\
\Pi^{00}(K) & = \mE^2\biggl [1 +  \int_{\hat\vt} \frac{iK_0}{-iK_0 + \vert \kt\vert z_v}\biggr ] = \mE^2\biggl [1 - {}_2F_1\left (\frac{1}{2},1,\frac{d}{2}; -\frac{\vert \kt\vert^2}{K_0^2} \right )\biggr ],
\end{split}
\end{equation}
where ${}_2F_1$ is the hypergeometric function, and where the final equality assumes $\vert \kt \vert/K_0 \in \mathbb{R}$ and $\Re(d) > 1$. If we now denote
\begin{equation}
\begin{split}
\label{eq:etaandkappa}
\eta_a(x) & \equiv \frac{\partial}{\partial z}~{}_2F_1\left (\frac{1}{2},1,z;x \right )\bigg\vert_{z=a},\\  
\kappa_a(x) & \equiv \frac{\partial^2}{\partial^2 z}~{}_2F_1\left (\frac{1}{2},1,z;x \right )\bigg\vert_{z=a},\\
\end{split}
\end{equation}
we find a very compact expression for the $\Pi^{00}$ integral
\begin{equation}
\Pi^{00}(K) =  \mE^2\biggl [1 + iK_0L(K) + \eta_{3/2}\left (-\frac{\vert \kt\vert^2}{K_0^2}\right )\varepsilon - \frac{1}{2}\kappa_{3/2}\left (-\frac{\vert \kt\vert^2}{K_0^2}\right )\varepsilon^2 + O(\varepsilon^3)\biggr ],
\end{equation}
with the notation 
\begin{equation}
\label{eq:Lfunction}
L(K) \equiv  -\frac{1}{2\vert \kt \vert}\ln \left (\frac{iK_0 + \vert \kt \vert}{iK_0 - \vert \kt \vert} \right ).
\end{equation}

Putting everything together, we find that the scalar functions $\Pi_\text{T}$ and $\Pi_\text{L}$, expanded up to $O(\varepsilon^2)$, can be expressed as
\begin{equation}
\label{eq:PiTandPepsexp}
 \Pi_{I}(K) =  \Pi_{I,0}(K) + \epsilon \Pi_{I,1}(K) + \epsilon^{2} \Pi_{I,2}(K) + O(\epsilon^{3}), \qquad I \in \{ \text{T}, \text{L} \}
\end{equation}
where the coefficients above are given by
\begin{equation}
\label{eq:PiTandPepsexpcoeff}
\begin{split}
\Pi_{\text{L},0}(K) & = \mE^2 \frac{K^2}{\vert \kt\vert^2}\biggl [1 + iK_0 L(K) \biggr ], \\
\Pi_{\text{L},1}(K) & = \mE^2 \frac{K^2}{\vert \kt\vert^2} \eta_{3/2}\left (-\frac{\vert \kt\vert^2}{K_0^2}\right ),\\
\Pi_{\text{L},2}(K) & = -\mE^2 \frac{K^2}{2\vert \kt\vert^2} \kappa_{3/2}\left (-\frac{\vert \kt\vert^2}{K_0^2}\right ),
\end{split}
\end{equation}
and 
\begin{equation}
\Pi_{\text{T},n}(K)  = \frac{1}{2}\left [\mE^2 - \sum_{i = 0}^{n} \Pi_{\text{L},i}(K)\right ].
\end{equation}
On occasion, we also denote $\Pi_n$, $n=0,1,2$, for the HTL self-energy \emph{tensor} truncated to the appropriate order. It turns out to be convenient to express these results in terms of the polar angle $\Phi_K$, which is defined in \eq\nr{eq:PhiKdef}. For example, in $D=4$ dimensions, we obtain:
\begin{equation}
\label{eq:PiTandLinpolar}
\begin{split}
\Pi_\text{T}(\Phi_K) & = \frac{\mE^2}{2} \cot(\Phi_K)\biggl [ \arctan[\tan(\Phi_K)] \csc^2(\Phi_K) - \cot(\Phi_K) \biggr ], \\
\Pi_\text{L}(\Phi_K) & = \mE^2 \csc^2(\Phi_K) \biggl [ 1 -  \arctan[\tan(\Phi_K)] \cot(\Phi_K) \biggr ].
\end{split}
\end{equation}
Note that $\arctan[\tan(\Phi_{K})] = \Phi_{K}$  only for $\Phi_{K} \in [0 , \pi / 2]$, Since we need expressions valid for the larger interval $\Phi_{K} \in [0 , \pi]$, we would need to replace $\arctan[\tan(\Phi_{K})]$ by ${\Phi_{K} - \pi \cdot \theta(\Phi_{K} - \pi / 2)}$, where $\theta$ denotes the Heaviside step function, if we wanted to further simplify \eq\eqref{eq:PiTandLinpolar}.

\subsection{HTL effective vertices}
\label{sec:HTLvertices}

As explained in the main text, treating the soft modes correctly within the HTL theory requires modifying not only the propagators, but also the $n$-point functions. This appendix contains the definitions of the three- (3g) and four-gluon  (4g) vertices appearing at zero temperature. 

\subsubsection{The three-gluon vertex}
\label{sec:3gvertex}

The effective 3g vertex is obtained by adding the HTL loop (which, at zero temperature, originates solely from the quark loop) to the bare vertex 
\begin{equation}
\Gamma^{\mu\nu\rho}_{abc}(P,Q,R) = \vcenter{\hbox{\begin{tikzpicture}[scale=2*\picSc]
	\doubleGluonLine{-1}{0}{0}{0}
	\doubleGluonLine{0}{0}{{cos(45)}}{{sin(45)}}
	\doubleGluonLine{{cos(-45)}}{{sin(-45)}}{0}{0}
	\node at (-1.15,0)(c){$P$};
	\node at ({1.15*cos(45)},{1.15*sin(45)})(c){$Q$};
	\node at ({1.15*cos(-45)},{1.15*sin(-45)})(c){$R$};
	\blobNode{0}{0}{3*\bigBlobSc}{$\,$}
\end{tikzpicture}}} 
= ig f_{abc}\Gamma^{\mu\nu\rho}(P,Q,R),
\end{equation}
with the decomposition 
\begin{equation}
\Gamma^{\mu\nu\rho}(P,Q,R) = \Gamma_{0}^{\mu\nu\rho}(P,Q,R) + \delta\Gamma^{\mu\nu\rho}(P,Q,R),
\end{equation}
where the bare 3g vertex $\Gamma_{0}^{\mu\nu\rho}$ can be read off from \eq\nr{eq:3g_bare}. The 3g HTL vertex function $\delta\Gamma^{\mu\nu\rho}$ is in turn given by the expression
\begin{equation}
\label{eq:3gHTL}
\delta\Gamma^{\mu\nu\rho}(P,Q,R) = \mE^2\int_{\hat\vt} V^{\mu}V^{\nu}V^{\rho}\biggl [\frac{iQ_0}{P\cdot V Q\cdot V} - \frac{iR_0}{P\cdot V R\cdot V} \biggr ].  
\end{equation}
The (tensor-valued) vertex function above is only defined when the sum of all of its arguments $P,Q$, and $R$ is zero, and it is totally symmetric in its (Lorentz) indices $(\mu, \nu, \rho)$ and traceless in any pair of indices, i.e. $\delta^{\mu\nu}\delta\Gamma^{\mu\nu\rho} =0$ since $V^2=0$. Furthermore, it is even (odd) under even (odd) permutations of ($P, Q$,$R$).

Contracting \eq\nr{eq:3gHTL} with one of the momenta, for example with $P^{\mu}$, yields
\begin{equation}
P^{\mu} \delta\Gamma^{\mu\nu\rho}(P,Q,R) = \mE^2 \int_{\hat\vt} V^{\nu}V^{\rho}\biggl [\frac{iQ_0}{Q\cdot V} - \frac{iR_0}{R\cdot V} \biggr ].
\end{equation}
Comparing this to \eq\nr{eq:gluonselfHTL}, we find that the 3g HTL vertex function obeys the generalized Ward identity given in \eq\nr{eq:WardVertices}.

\subsubsection{The four-gluon vertex} 
\label{sec:4gvertex}

The effective 4g vertex is given by the decomposition
\begin{equation}
\Gamma^{\mu\nu\rho\sigma}_{abcd}(P,Q,R,S)  =
\vcenter{\hbox{\begin{tikzpicture}[scale=2*\picSc]
	\doubleGluonLine{0}{0}{{cos(135)}}{{sin(135)}}
	\doubleGluonLine{{cos(-135)}}{{sin(-135)}}{0}{0}
	\doubleGluonLine{0}{0}{{cos(45)}}{{sin(45)}}
	\doubleGluonLine{{cos(-45)}}{{sin(-45)}}{0}{0}
    \node at ({1.15*cos(45)},{1.15*sin(45)})(c){$Q$};
	\node at ({1.15*cos(-45)},{1.15*sin(-45)})(c){$R$};
	\node at ({1.15*cos(135)},{1.15*sin(135)})(c){$P$};
	\node at ({1.15*cos(-135)},{1.15*sin(-135)})(c){$S$};
	\blobNode{0}{0}{3*\bigBlobSc}{$\,$}
\end{tikzpicture}}}= (\Gamma_{0})^{\mu\nu\rho\sigma}_{abcd}(P,Q,R,S) + \delta\Gamma^{\mu\nu\rho\sigma}_{abcd}(P,Q,R,S),
\end{equation}
where the bare 4g vertex $(\Gamma_{0})^{\mu\nu\rho\sigma}_{abcd}$ can be found in \eq\nr{eq:4g_bare}.  The general expression for the 4g HTL vertex function $\delta\Gamma^{\mu\nu\rho\sigma}_{abcd}$ is uniquely determined from the knowledge of its symmetries, the 3g HTL vertex, as well as the Ward identities in \eq\nr{eq:WardVertices}. Here, however, we limit our detailed discussion only to the special case that we need in \eq\nr{eq:G4g}. This is, we take $R = -P, S = -Q$ and we sum over two adjacent color indices. This gives for the 4g   vertex the expression

\begin{equation}
\Gamma_{abcb}^{\mu\nu\rho\sigma}(P,Q,-P,-Q) = -g^2f_{eba}f_{ebc}\Gamma^{\mu\nu\rho\sigma}(P,Q,-P,-Q) 
\end{equation}
with the decomposition 
\begin{equation}
\Gamma^{\mu\nu\rho\sigma}(P,Q,-P,-Q)  = \Gamma_{0}^{\mu\nu\rho\sigma}(P,Q,-P,-Q) + \delta\Gamma^{\mu\nu\rho\sigma}(P,Q,-P,-Q).
\end{equation}
In this special case, the 4g HTL vertex $\delta\Gamma^{\mu\nu\rho\sigma}$ is given by the expression
\begin{equation}
\label{eq:4gHTL}
\delta\Gamma^{\mu\nu\rho\sigma}(P,Q,-P,-Q) =  2\mE^2\int_{\hat\vt} \frac{V^{\mu}V^{\nu}V^{\rho}V^{\sigma}}{(P+Q)\cdot V (P-Q)\cdot V}\biggl [\frac{iQ_0}{Q\cdot V} - \frac{iP_0}{P\cdot V} \biggr ].
\end{equation}
Akin to the 3g vertex correction, the 4g vertex function is totally symmetric in its four (Lorentz) indices $(\mu, \nu, \rho, \sigma)$ and traceless in any pair of indices, i.e. $\delta^{\mu\nu}\delta\Gamma^{\mu\nu\rho\sigma} =0$ since $V^2=0$. Note that this vertex is also even under all permutations of the momenta ($P$, $Q$, $-P$, $-Q$). Lastly, we note that applying the Ward identities of \eq\nr{eq:WardVertices} twice to the four-point vertex correction yields the useful identity 
\begin{equation}
P^{\mu}P^{\nu} \delta\Gamma^{\mu\nu\rho\sigma}(P,Q,-P,-Q) = -2\Pi^{\rho\sigma}(Q) + \Pi^{\rho\sigma}(P+Q) + \Pi^{\rho\sigma}(P-Q).   
\end{equation}

\section{Evaluating HTL vertex functions}

While \eqs\eqref{eq:3gHTL} and \eqref{eq:4gHTL} suffice in principle for computing the vertex corrections, in practice further manipulations prove extremely helpful for numerical evaluations. We detail these manipulations in this appendix. Throughout this section, we scale out the explicit $\mE$ factors from all vertex expressions.

\subsection{Evaluation of the 3g HTL vertex function}
\label{sec:3gHTL}

We start the explicit evaluations of the HTL structures by considering the 3g vertex function $\delta\Gamma^{\mu\nu\rho}_{PQR}$. The generalized Ward identities [see \eq\nr{eq:WardVertices}] can often be used together with the tracelessness of the vertex correction to significantly simplify contributions containing the vertex correction. However, even for the two-loop HTL diagrams, the full structure of the HTL-corrected vertex is required due to the sunset diagram with \emph{two} vertex corrections, as seen in \eq\nr{eq:G3g}. As a specific example, the contraction of two vertices $(\delta\Gamma^{\mu\nu\rho})^2 \equiv \delta\Gamma^{\mu\nu\rho}\delta\Gamma^{\mu\nu\rho}$\footnote{We also use this compact square notation when some indices are fixed, or when only spatial indices are contracted.} includes every term allowed by the remnant $d$-dimensional rotational symmetry. As such, we must compute the following four independent vertex contributions $\delta\Gamma^{000}_{PQR}, \delta\Gamma^{i00}_{PQR}, \delta\Gamma^{ij0}_{PQR}$ and $\delta\Gamma^{ijk}_{PQR}$. 
  
It turns out to be convenient to rewrite the expression in \eq\nr{eq:3gHTL} in the more symmetric form
\begin{equation}
\label{eq:T3gHTLsym}
\delta\Gamma^{\mu\nu\rho}_{PQR} =  \int_{\hat\vt} V^{\mu}V^{\nu}V^{\rho}\biggl [\frac{iQ_0}{Q\cdot V(P+Q)\cdot V} - \frac{iP_0}{P\cdot V (P+Q)\cdot V} \biggr ].
\end{equation}
To evaluate the integral over the angles, one could try to combine the products in the two denominators into a single expression by using the ``standard'' Feynman parameterization
\begin{equation}
\label{eq:feynmanparm}
\begin{split}
\frac{1}{Q\cdot V (P+Q)\cdot V} & =  \int_{0}^{1}\ud u \frac{1}{[\left (uQ + (1-u)(P+Q)\right )\cdot V]^2}.
\end{split}
\end{equation}
However, in order to avoid the complications related to $uQ_0+(1-u)(P_0+Q_0)$  or $-uQ_0+(1-u)(P_0+Q_0)$ changing its sign at some value of $u$ within the unit interval, causing the denominator of the integrand of \eq\nr{eq:feynmanparm} to potentially vanish for some $P$ and $Q$, we need to generalize the way that Feynman parameters are introduced. Let us first introduce the ``symmetric'' form of the parametrization to reach the general form
\begin{equation}
\label{eq:genfeynmanparm}
\begin{split}
\frac{1}{Q\cdot V(P+Q)\cdot V}  = 
2\int_{-1}^{1}\ud u\frac{ \sigma_{Q,P+Q}}{\Big[(1+u)Q\cdot V+(1-u)\sigma_{Q,P+Q}(P+Q)\cdot V\Big]^2},\\
\end{split}
\end{equation}
where we have defined
\begin{equation}
     \sigma_{X,Y}\equiv\mathrm{sgn}[\Im(X\cdot V)\Im(Y\cdot V)] = \mathrm{sgn}(X_0Y_0).
\end{equation}
Evidently, the definition requires nonzero imaginary parts of $Q\cdot V$ and $(P+Q)\cdot V$.\footnote{However, it can be generalized to momenta with vanishing zero-components; see \app\ref{sec:generalisedFP}.} With this assumption, the modified parametrization can be shown to be equivalent to the standard form not only whenever the denominator of the latter is strictly nonvanishing, but also to yield $1/(AB)$ when the standard form does display divergent behaviour. We show this explicitly in \app\ref{sec:generalisedFP}. Similarly, for the second term in \eq\nr{eq:T3gHTLsym}, we find  
\begin{equation}
\begin{split}
\frac{1}{P\cdot V(P+Q)\cdot V}  = 
2\int_{-1}^{1}\ud u\frac{ \sigma_{P,P+Q}}{\Big[(1+u)P\cdot V+(1-u)\sigma_{P,P+Q} (P+Q)\cdot V\Big]^2}.\\
\end{split}
\end{equation}

The 3g HTL vertex function now takes the form 
\begin{equation}
\label{eq:T3gHTLsymfp}
\delta\Gamma^{\mu\nu\rho}_{PQR} =  2\int_{-1}^{1}\ud u \int_{\hat\vt} V^{\mu}V^{\nu}V^{\rho}\biggl [\frac{iQ_0 \sigma_{Q,P+Q}}{(T\cdot V)^2} - \frac{iP_0 \sigma_{P,P+Q}}{(S\cdot V)^2} \biggr ],
\end{equation}
where the four-vectors $T$ and $S$ are defined as:
\begin{equation}
\begin{split}
T & \equiv (1+u)Q + (1-u) \sigma_{Q,P+Q}(P + Q), \\
S & \equiv (1+u)P + (1-u) \sigma_{P,P+Q}(P + Q).
\end{split}
\end{equation}
For further discussion, we will need the angular integral computed in \eq\eqref{eq:generalHTLintegral}; we list here in $d=3-2\varepsilon$ dimensions some special cases as master integrals that will be used repeatedly in the following section
\begin{equation}
\label{eq:elementary}
\begin{split}
\int_{\hat\vt} & = 1,\\
\int_{\hat\vt}  (S\cdot V) & = -iS_0,\\
\int_{\hat\vt} \frac{1}{(S\cdot V)^2} & = - \frac{1}{S^2} + \frac{a(S)}{S^2}2\varepsilon+ O(\varepsilon^2),\\
\int_{\hat\vt} \frac{1}{(S\cdot V)}  & = L(S)  + b(S) 2\varepsilon  + O(\varepsilon^2), 
\end{split}
\end{equation}
where the master integrals are expanded to order $\epsilon$ and we have introduced the compact notation
\begin{equation}
\begin{split}
a(S) & \equiv 1 + S_0iL(S),\\
b(S) & \equiv \biggl [\ln(2) - 1 + \ln\left (\vert \st\vert/S\right )\biggr ]L(S)
+ \frac{1}{4\vert \st\vert}\biggl [\text{Li}_2\left(\frac{iS_0-\vert \st\vert}{iS_0+\vert \st\vert}\right) - \text{Li}_2\left(\frac{iS_0+\vert \st\vert}{iS_0-\vert \st\vert}\right) \biggr ].
\end{split}
\end{equation}
Here, the function $L(S)$ is defined in \eq\eqref{eq:Lfunction} and $\mathrm{Li}_2$ is the dilogarithm function.

In the following subsections, we describe how to further evaluate the four independent vertex contributions $\delta\Gamma^{000}_{PQR}, \delta\Gamma^{i00}_{PQR}, \delta\Gamma^{ij0}_{PQR}$ and $\delta\Gamma^{ijk}_{PQR}$. We also compute the $O(\varepsilon)$ corrections to these vertex functions, which are needed in order to obtain the full $O(\varepsilon^0)$ contribution to the UV term presented in \Sec\ref{sec:IF}.

\subsubsection{\texorpdfstring{The $\delta \Gamma^{000}_{PQR}$ function}{The dGamma\^{}\{000\}\_(P,Q,R) function}}
\label{subsec:dG000}

Let us first concentrate on the $\delta\Gamma^{000}_{PQR}$ vertex function. By using the general expression in \eq\nr{eq:T3gHTLsymfp}, we easily obtain
\begin{equation}
\label{eq:T000fp}
\begin{split}
\delta \Gamma^{000}_{PQR} & = -2\int_{-1}^{1}\ud u  \int_{\hat \vt} \biggl [\frac{Q_0\sigma_{(Q,P+Q)}}{(T\cdot V)^2} - \frac{P_0\sigma_{(P,P+Q)}}{(S\cdot V)^2} \biggr ]\\
& = +2\int_{-1}^{1}\ud u  \biggl [\biggl (\frac{Q_0\sigma_{(Q,P+Q)}}{T^2} - \frac{P_0\sigma_{(P,P+Q)}}{S^2}\biggr )\\
& \quad\quad\quad\quad - \biggl (\frac{Q_0\sigma_{(Q,P+Q)}}{T^2}a(T) - \frac{P_0\sigma_{(P,P+Q)}}{S^2}a(S)\biggr )2\varepsilon\biggr ],
\end{split}
\end{equation}
where in the second line the $\int_{\hat\vt}$ integral is performed by using a master integral listed in \eq\nr{eq:elementary}. Correspondingly, the $\delta \Gamma^{000}_{PQR}$ function squared can now be easily computed by using the expression above with two Feynman parameters $u_1$ and $u_2$,
\begin{equation}
\label{eq:dG000sqrfp}
\begin{split}
(\delta\Gamma^{000}_{PQR})^2 = 4 \int_{-1}^{1} \int_{-1}^{1} \ud  u_1 \ud  u_2 &  \Biggl \{ \biggl [\biggl (\frac{Q_0\sigma_{(Q,P+Q)}}{T_1^2} - \frac{P_0\sigma_{(P,P+Q)}}{S_1^2}\biggr )\\
& \quad\quad - \biggl (\frac{Q_0\sigma_{(Q,P+Q)}}{T_1^2}a(T_1) - \frac{P_0\sigma_{(P,P+Q)}}{S_1^2}a(S_1)\biggr )2\varepsilon\biggr ]\Biggr \}\\
& \hspace{0.15cm} \times \Biggl \{ \biggl [\biggl (\frac{Q_0\sigma_{(Q,P+Q)}}{T_2^2} - \frac{P_0\sigma_{(P,P+Q)}}{S_2^2}\biggr ) \\
&\quad\quad  - \biggl (\frac{Q_0\sigma_{(Q,P+Q)}}{T_2^2}a(T_2) - \frac{P_0\sigma_{(P,P+Q)}}{S_2^2}a(S_2)\biggr )2\varepsilon\biggr ]\Biggr \},
\end{split}
\end{equation}
where the variables $T_i$ and $S_i$ for $i=1,2$ are defined as follows: 
\begin{equation}
\label{eq:kinematicsTiandSi}
\begin{split}
T_i  & \equiv (1+u_i)Q + (1-u_i)\sigma_{(Q,P+Q)}(P + Q),\\
S_i  & \equiv (1+u_i)P + (1-u_i)\sigma_{(P,P+Q)}(P + Q).
\end{split}
\end{equation}

\subsubsection{\texorpdfstring{The $\delta \Gamma^{i00}_{PQR}, \delta \Gamma^{ij0}_{PQR}$ and $\delta \Gamma^{ijk}_{PQR}$ functions}{The dGamma\^{}\{i00\}(P,Q,R), dGamma\^{}\{ij0\}(P,Q,R) and dGamma\^{}\{ijk\}\_(P,Q,R) functions}}
\label{subsec:dGi00anddGij0anddGijk}

We then proceed to describe the evaluation of the vertex functions $\delta \Gamma^{i00}_{PQR}, \delta \Gamma^{ij0}_{PQR}$ and $\delta \Gamma^{ijk}_{PQR}$. By using the general expression in \eq\nr{eq:T3gHTLsymfp}, we obtain:
\begin{equation}
\label{eq:Ttensors2}
\begin{split}
\delta \Gamma^{i00}_{PQR} & = -2i\int_{-1}^{1}\ud u \int_{\hat \vt} \hat v^{i}  \biggl [\frac{Q_0\sigma_{(Q,P+Q)}}{(T\cdot V)^2} - \frac{P_0\sigma_{(P,P+Q)}}{(S\cdot V)^2} \biggr ],\\
\delta \Gamma^{ij0}_{PQR} & = 2\int_{-1}^{1}\ud u \int_{\hat \vt} \hat v^{i} \hat v^{j}  \biggl [\frac{Q_0\sigma_{(Q,P+Q)}}{(T\cdot V)^2} - \frac{P_0\sigma_{(P,P+Q)}}{(S\cdot V)^2} \biggr ],\\
\delta \Gamma^{ijk}_{PQR} & = 2i\int_{-1}^{1}\ud u  \int_{\hat \vt} \hat v^{i} \hat v^{j} \hat v^{k}  \biggl [\frac{Q_0\sigma_{(Q,P+Q)}}{(T\cdot V)^2} - \frac{P_0\sigma_{(P,P+Q)}}{(S\cdot V)^2} \biggr ].\\
\end{split}
\end{equation}
The angular integrals appearing in \eq\nr{eq:Ttensors2} can be dealt with by using the $d$-dimensional rotational symmetry. For instance, the rank-one integral can be written as
\begin{equation}
i\int_{\hat \vt} \frac{\hat v^i}{(S \cdot V)^2} = s^i f_0(S_0,\vert \st \vert).
\end{equation}
Contracting both sides with the vector $s^i$, and noting that 
\begin{equation}
\st \cdot \hat \vt = \st \cdot \hat \vt + S_0V_0 - S_0V_0 = S\cdot V + iS_0,
\end{equation}
we find for the reduction coefficient $f_0$ the following form
\begin{equation}
\label{eq:f0withangularint}
\begin{split}
f_0(S_0,\vert \st \vert) & \equiv f_0(S)  = \frac{i}{\vert \st\vert ^2}\int_{\hat \vt} \frac{\st \cdot \hat \vt}{(S \cdot V)^2} = \frac{i}{\vert \st\vert ^2}\int_{\hat \vt} \frac{1}{S \cdot V} - \frac{S_0}{\vert \st\vert ^2}\int_{\hat \vt} \frac{1}{(S \cdot V)^2}.
\end{split}
\end{equation}
After performing the remaining angular integrals in \eq\nr{eq:f0withangularint} by using the master integrals listed in \eq\nr{eq:elementary}, we find the result 
\begin{equation}
\label{eq:coefficientf0}
f_0(S) = \frac{1}{\vert \st\vert^2}\biggl [\frac{S_0}{S^2} + iL(S)\biggr ] - \frac{1}{\vert \st\vert^2}\biggl [\frac{S_0}{S^2}a(S) - ib(S)\biggr ]2\varepsilon + O(\varepsilon^2).
\end{equation}
All in all, the vertex function $\delta \Gamma^{i00}_{PQR}$ then takes the form
\begin{equation}
\label{eq:dGi00}
\delta \Gamma^{i00}_{PQR} = -2\int_{-1}^{1}\ud u \biggl [Q_0\sigma_{(Q,P+Q)} t^i f_0(T) - P_0\sigma_{(P,P+Q)} s^i f_0(S) \biggr ].
\end{equation}

To evaluate the vertex functions $\delta \Gamma^{ij0}_{PQR}$ and $\delta \Gamma^{ijk}_{PQR}$ further, we use the following tensor-integral reduction
\begin{equation}
\label{eq:tensorreduction}
\begin{split}
\int_{\hat \vt} \frac{\hat v^i \hat v^j}{(S \cdot V)^2} & = \delta^{ij} f_{00}(S_0,\vert \st \vert) + s^i s^j  f_{12}(S_0,\vert \st \vert),\\
i\int_{\hat \vt} \frac{\hat v^i \hat v^j \hat v^k}{(S \cdot V)^2} & = \{\delta s\}^{ijk} f_{000}(S_0,\vert \st \vert) + s^i s^j s^k f_{123}(S_0,\vert \st \vert),\\
\end{split}
\end{equation}
where the notation $\{\delta s\}^{ijk} \equiv \delta^{ij}s^k + \delta^{ik}s^j + \delta^{jk}s^i$ has been introduced. Contracting \eq\nr{eq:tensorreduction} with the Kronecker delta and vector $s^i$, it is straightforward to show that the reduction coefficients above can be written as
\begin{equation}
\label{eq:f0andf00f12}
\begin{split}
f_{00}(S) & = -\frac{1}{\vert \st\vert^2}\biggl [1 + S_0iL(S)\biggr ] -\frac{S_0}{\vert \st\vert^2}ib(S)2\varepsilon + O(\varepsilon^2) ,\\
f_{12}(S) & = +\frac{1}{\vert \st \vert^4}\biggl [2 + \frac{S_0^2}{S^2} + 3S_0iL(S) \biggr ] -\frac{1}{\vert \st \vert^4}\biggl [\frac{S_0^2}{S^2}a(S) - 3S_0ib(S) \biggr ]2\varepsilon + O(\varepsilon^2),
\end{split}
\end{equation}
and 
\begin{equation}
\label{eq:f000f123}
\begin{split}
f_{000}(S) & = \frac{1}{2\vert \st \vert^4}\biggl [3S_0 + (S^2+2S_0^2)iL(S)\biggr ]\\
& \quad\quad\quad + \frac{1}{2\vert \st \vert^4}\biggl [\frac{3S_0}{2} - S_0 a(S) + (S^2+2S_0^2)\left (\frac{i}{2}L(S) + ib(S) \right )\biggr ]2\varepsilon + O(\varepsilon^2),\\
f_{123}(S) & = -\frac{1}{2S^2\vert \st \vert^6}\biggl [2S_0^3 + 13S_0S^2 + 3(S^4 + 4S_0^2S^2)iL(S) \biggr ]\\
& \quad\quad\quad\quad -\frac{1}{2S^2\vert \st \vert^6}\biggl [\frac{15S_0S^2}{2} - S_0(6S^2 + 2S_0^2)a(S)\\
& \quad\quad\quad\quad\quad\quad\quad\quad\quad\quad\quad + 3(S^4 + 4S_0^2S^2)\left (\frac{i}{2}L(S) + ib(S) \right ) \biggr ]2\varepsilon + O(\varepsilon^2).
\end{split}
\end{equation}
Finally, inserting these results into \eq\nr{eq:Ttensors2}, we obtain for the vertex function $\delta \Gamma^{ij0}_{PQR}$ the following expression
\begin{equation}
\label{eq:dGij0}
\begin{split}
\delta \Gamma^{ij0}_{PQR}  & = 2\int_{-1}^{1}\ud u  \biggl [Q_0\sigma_{(Q,P+Q)}\biggl (\delta^{ij}f_{00}(T) + t^it^j f_{12}(T) \biggr )\\
& \quad\quad\quad\quad - P_0\sigma_{(P,P+Q)}\biggl (\delta^{ij}f_{00}(S) + s^is^j f_{12}(S) \biggr )\biggr ].\\
\end{split}
\end{equation}
Similarly, for the vertex function $\delta \Gamma^{ijk}_{PQR}$, we obtain
\begin{equation}
\label{eq:dGijk}
\begin{split}
\delta\Gamma^{ijk}_{PQR}   & = 2\int_{-1}^{1} \ud u \biggl [Q_0\sigma_{(Q,P+Q)}\biggl (\{\delta t\}^{ijk}f_{000}(T)  + t^it^jt^k f_{123}(T) \biggr )\\
& \quad\quad\quad\quad\quad   - P_0\sigma_{(P,P+Q)}\biggl (\{\delta s\}^{ijk}f_{000}(S) + s^is^js^k f_{123}(S) \biggr )\biggr ].
\end{split}
\end{equation}

Having all these results at hand, we can now turn to computing the functions $(\delta \Gamma^{i00}_{PQR})^2$, $(\delta \Gamma^{ij0}_{PQR})^2$ and $(\delta \Gamma^{ijk}_{PQR})^2$. First, the function $(\delta \Gamma^{i00}_{PQR})^2$ can be written as
\begin{equation}
\label{eq:dGi00sqr}
(\delta \Gamma^{i00}_{PQR})^2 = 4\int_{-1}^{1}\int_{-1}^{1}\ud u_1 \ud u_2 \biggl [Q_0^2 X_1 - P_0Q_0\sigma_{(P,P+Q)}\sigma_{(Q,P+Q)}(X_2 + X_3) + P_0^2X_4 \biggr ], 
\end{equation}
where the coefficients $X_i$ are defined as
\begin{equation}
\begin{split}
X_1 & \equiv (\ttt_1 \cdot \ttt_2) f_0(T_1)f_0(T_2), \\
X_2 & \equiv (\ttt_1 \cdot \st_2) f_0(T_1)f_0(S_2), \\
X_3 & \equiv (\ttt_2 \cdot \st_1) f_0(S_1)f_0(T_2), \\
X_4 & \equiv (\st_1 \cdot \st_2) f_0(S_1)f_0(S_2). \\
\end{split}
\end{equation}
Similarly, the functions $(\delta \Gamma^{ij0}_{PQR})^2$ and $(\delta \Gamma^{ijk}_{PQR})^2$ can be written as 
\begin{equation}
\label{eq:Tij0sqrandTijksqrv2}
\begin{split}
(\delta \Gamma^{ij0}_{PQR})^2 = & 4\int_{-1}^{1}\int_{-1}^{1}\ud u_1 \ud u_2 \biggl [Q_0^2 Y_1 - P_0Q_0\sigma_{(P,P+Q)}\sigma_{(Q,P+Q)} \left (Y_2 + Y_3\right ) + P_0^2 Y_4\biggr ],\\
(\delta \Gamma^{ijk}_{PQR})^2 = & 4\int_{-1}^{1}\int_{-1}^{1}\ud u_1 \ud u_2 \biggl [Q_0^2 Z_1 - P_0Q_0\sigma_{(P,P+Q)}\sigma_{(Q,P+Q)} \left (Z_2 + Z_3\right ) + P_0^2 Z_4\biggr ],
\end{split}
\end{equation}
where the coefficients $Y_i$ and $Z_i$ are defined as
\begin{equation}
\begin{split}  
Y_1 & \equiv (3-2\varepsilon)f_{00}(T_1)f_{00}(T_2) + \vert \ttt_2\vert^2 f_{00}(T_1)f_{12}(T_2)\\
& \quad\quad\quad\quad + \vert \ttt_1\vert^2f_{00}(T_2)f_{12}(T_1) + (\ttt_1\cdot \ttt_2)^2 f_{12}(T_1)f_{12}(T_2),\\
Y_2 & \equiv (3-2\varepsilon)f_{00}(S_1)f_{00}(T_2) + \vert \ttt_2\vert^2 f_{00}(S_1)f_{12}(T_2)\\
& \quad\quad\quad\quad + \vert \st_1\vert^2 f_{00}(T_2)f_{12}(S_1) + (\ttt_2 \cdot \st_1)^2 f_{12}(S_1)f_{12}(T_2),\\
Y_3 & \equiv (3-2\varepsilon)f_{00}(T_1)f_{00}(S_2) + \vert \st_2\vert^2 f_{00}(T_1)f_{12}(S_2)\\
& \quad\quad\quad\quad + \vert \ttt_1\vert^2 f_{00}(S_2)f_{12}(T_1) + (\ttt_1 \cdot \st_2)^2 f_{12}(T_1)f_{12}(S_2),\\
Y_4 & \equiv (3-2\varepsilon)f_{00}(S_1)f_{00}(S_2) + \vert \st_2\vert^2 f_{00}(S_1)f_{12}(S_2)\\
& \quad\quad\quad\quad + \vert \st_1\vert^2f_{00}(S_2)f_{12}(S_1) + (\st_1\cdot \st_2)^2 f_{12}(S_1)f_{12}(S_2), \\
\end{split}
\end{equation}
and 
\begin{equation}
\begin{split}  
Z_1 & \equiv 3(5-2\varepsilon)(\ttt_1 \cdot \ttt_2) f_{000}(T_1)f_{000}(T_2) + 3\vert \ttt_2\vert^2 (\ttt_1 \cdot \ttt_2) f_{000}(T_1)f_{123}(T_2)\\
& \quad\quad\quad\quad\quad + 3\vert \ttt_1\vert^2 (\ttt_1 \cdot \ttt_2) f_{000}(T_2)f_{123}(T_1) + (\ttt_1\cdot \ttt_2)^3 f_{123}(T_1)f_{123}(T_2),\\
Z_2 & \equiv 3(5-2\varepsilon)(\ttt_1 \cdot \st_2)  f_{000}(T_1)f_{000}(S_2) +  3\vert \st_2\vert^2 (\ttt_1 \cdot \st_2) f_{000}(T_1)f_{123}(S_2)\\
& \quad\quad\quad\quad\quad +  3\vert \ttt_1\vert^2 (\ttt_1 \cdot \st_2) f_{000}(S_2)f_{123}(T_1) + (\ttt_1 \cdot \st_2)^3 f_{123}(T_1)f_{123}(S_2),\\
Z_3 & \equiv  3(5-2\varepsilon)(\ttt_2 \cdot \st_1)  f_{000}(S_1)f_{000}(T_2) + 3\vert \st_1\vert^2 (\ttt_2 \cdot \st_1) f_{000}(T_2)f_{123}(S_1)\\
& \quad\quad\quad\quad\quad + 3\vert \ttt_2\vert^2 (\ttt_2 \cdot \st_1)  f_{000}(S_1)f_{123}(T_2) + (\ttt_2 \cdot \st_1)^3 f_{123}(S_1)f_{123}(T_2),\\
Z_4 & \equiv  3(5-2\varepsilon)(\st_1 \cdot \st_2)f_{000}(S_1)f_{000}(S_2) + 3\vert \st_2\vert^2 (\st_1 \cdot \st_2) f_{000}(S_1)f_{123}(S_2)\\
& \quad\quad\quad\quad\quad + 3\vert \st_1\vert^2 (\st_1 \cdot \st_2) f_{000}(S_2)f_{123}(S_1) + (\st_1\cdot \st_2)^3 f_{123}(S_1)f_{123}(S_2). \\
\end{split}
\end{equation}

\subsubsection{3g HTL vertices contracted with external momenta}
\label{subsec:3gHTLwithWard}

In this section, we show how to evaluate the 3g HTL vertices contracted with external momenta. These techniques are used extensively in \Sec\ref{sec:Ifinite}. Let us first consider the case where the 3g HTL vertex $\delta \Gamma^{\mu\nu\rho}_{PQR}$ is contracted with  $P_{\text{T}}^{\mu} = \delta^{\mu i} p^i$. For this term, we cannot fully remove the spatial components of the 3g HTL vertex, but we can reduce it as much as possible using the identity
\begin{equation}
\label{eq:3gHTLwithP}
\begin{split}
    P_{\text{T}}^{\mu} \delta \Gamma^{\mu \nu \rho}_{PQR} ={}& 
    \left (P^{\mu} - \delta^{\mu 0}P_{0} \right ) \delta \Gamma^{\mu \nu \rho}_{PQR} \\
    ={}& 
    P^{\mu}\delta \Gamma^{\mu \nu \rho}_{PQR} - P_{0}\delta \Gamma^{0 \nu \rho}_{PQR}.
\end{split}
\end{equation}
To reduce the number of spatial indices appearing in our expressions above, we use the generalized Ward identity in \eq\nr{eq:WardVertices} on the first term. This gives 
\begin{equation}
P_{\text{T}}^{\mu} \delta \Gamma^{\mu \nu \rho}_{PQR} =  \Pi^{\nu\rho}(R) - \Pi^{\nu\rho}(Q) -  P_{0}\delta \Gamma^{0 \nu \rho}_{PQR}.
\end{equation}

This method can be easily applied to the more complicated cases  $P_{\text{T}}^{\mu}Q_{\text{T}}^{\nu} \delta \Gamma^{\mu \nu \rho}_{PQR}$ and $P_{\text{T}}^{\mu}Q_{\text{T}}^{\nu}R_{\text{T}}^{\rho}  \delta \Gamma^{\mu \nu \rho}_{PQR}$. By using \eq\nr{eq:3gHTLwithP} and Ward identities, the following relations can be derived:
\begin{equation}
\begin{split}
P_{\text{T}}^{\mu} Q_{\text{T}}^{\nu}  \delta \Gamma^{\mu \nu \rho}_{PQR} & = Q_0 \Pi^{0\rho}(Q) + q^i \Pi^{i\rho}(R) - P_0 \left [\Pi^{0\rho}(P) - \Pi^{0\rho}(R) \right ] + P_0Q_0  \delta \Gamma^{00 \rho}_{PQR},
\end{split}
\end{equation}
and
\begin{equation}
\begin{split}
P_{\text{T}}^{\mu} Q_{\text{T}}^{\nu} R_{\text{T}}^{\rho}  \delta \Gamma^{\mu \nu \rho}_{PQR} & = Q_0 r^k\Pi^{0k}(Q) - P_0r^k \Pi^{0k}(P) - R_0 q^i \Pi^{0i}(R) - P_0R_0 \Pi^{00}(R)\\
& + P_0 Q_0 \left [\Pi^{00}(Q) - \Pi^{00}(P) \right ] - P_0Q_0R_0  \delta \Gamma^{000}_{PQR}.
\end{split}
\end{equation}
Here, the different components $\Pi^{00}(Y), \Pi^{0i}(Y)$ and $\Pi^{ij}(Y)$ of the self energy $\Pi^{\mu\nu}(Y)$ are given by 
\begin{equation}
\label{eq:SEcomponents}
\begin{split}
\Pi^{00}(Y) & = \frac{\vert \yt\vert^2}{Y^2}\Pi_{\text{L}}(\hat Y),\\ 
\Pi^{0i}(Y) & = -\frac{Y_0 \vert \yt \vert }{Y^2} \hat y^i \Pi_{\text{L}}(\hat Y),\\   
\Pi^{ij}(Y) & = \delta^{ij} \Pi_{\text{T}}(\hat Y) -\hat y^i \hat y^j \biggl [\Pi_{\text{T}}(\hat Y) - \frac{Y_0^2}{Y^2}\Pi_{\text{L}}(\hat Y) \biggr ],
\end{split}
\end{equation}
where $Y \in \{P,Q,R\}$. Note that the 3g HTL vertices with more time components are easier to compute numerically.

\subsection{Evaluation of the 4g HTL vertex function}
\label{sec:4gHTL}

Following the discussion on the 3g HTL vertex correction, we will next consider the 4g vertex correction. \emph{A priori}, it is considerably more complicated, and in order to handle the vertex in its full generality, a sensible option would be to turn to automation (see \app\ref{sec:htlauto}). However, for the \NLO{3} pressure, there is only a single resummed graph involving the 4g vertex correction, and it includes only a single vertex. It is easy to see that applying the symmetries and the Ward identities of the vertex correction along the same lines as in the previous section can only lead to a \emph{single} irreducible term containing the 4g vertex correction $\delta\Gamma^{0000}_{P,Q,-P,-Q}$\footnote{Recall that we have scaled away the mass $\mE$.}: 

\begin{equation}
\delta\Gamma^{0000}_{P,Q,-P,-Q}=\left(-i\right)^{4}2\int_{\hat\vt}\frac{1}{\left(P+Q\right)\cdot V\left(P-Q\right)\cdot V}\left[\frac{iQ_{0}}{Q\cdot V}-\frac{iP_{0}}{P\cdot V}\right].
\end{equation}

Following the 3g computation, we will apply a Feynman parametrization to make the numerics more tractable. However, as before, we must generalize the parametrization, following the discussion of \app\ref{sec:generalisedFP} [see also \eq\eqref{eq:genfeynmanparm}]. We combine the two common factors in the denominators via 
\begin{equation}
\frac{1}{\left(P+Q\right)\cdot V\left(P-Q\right)\cdot V}=2\int_{-1}^{1}\ud u_{1}\frac{\sigma_{P+Q,P-Q}}{\left[\left(1+u_{1}\right)\left(P+Q\right)\cdot V+\left(1-u_{1}\right)\sigma_{P+Q,P-Q}\left(P-Q\right)\cdot V\right]^{2}},
\end{equation}
and to include the third factor, we denote
\begin{equation}
U\equiv\left(1+u_{1}\right)\left(P+Q\right)+\left(1-u_{1}\right)\sigma_{P+Q,P-Q}\left(P-Q\right),
\end{equation}
to obtain
\begin{equation}
\begin{split}
\frac{1}{Q\cdot V\left(U\cdot V\right)^{2}}&=-\frac{\partial}{\partial (U\cdot V)}\frac{1}{U\cdot VQ\cdot V} \\
&=-\frac{\partial}{\partial (U\cdot V)}\int_{-1}^{1}\ud u_{2}\frac{2\sigma_{U,Q}}{\left[\left(1+u_{2}\right)U\cdot V+\left(1-u_{2}\right)\sigma_{U,Q}Q\cdot V\right]^{2}}\\
&=4\int_{-1}^{1}\ud u_{2}\frac{\left(1+u_{2}\right)\sigma_{U,Q}}{\left[\left(1+u_{2}\right)U\cdot V+\left(1-u_{2}\right)\sigma_{Q,U}Q\cdot V\right]^{3}}.
\end{split}
\end{equation}
Hence, the full generalized Feynman parametrization for three-term denominator reads
\begin{equation}
\frac{1}{\left(P+Q\right)\cdot V\left(P-Q\right)\cdot V Q\cdot V}=8\int_{-1}^{1}\ud u_{1}\int_{-1}^{1}\ud u_{2}\frac{\left(1+u_{2}\right)\sigma_{P+Q,P-Q}\sigma_{U,Q}}{\left[\left(1+u_{2}\right)U\cdot V+\left(1-u_{2}\right)\sigma_{Q,U}Q\cdot V\right]^{3}}
\end{equation}
and analogously for $P\cdot V$. Given this, $\delta\Gamma^{0000}_{P,Q,-P,-Q}$ admits a representation
\begin{equation}
\begin{split}
\delta\Gamma^{0000}_{P,Q,-P,-Q}&=2^{4}\sigma_{P+Q,P-Q}\int_{-1}^{1}\ud u_{1}\int_{-1}^{1}\ud u_{2}\left(1+u_{2}\right) \\
&\times\Bigg\{iQ_{0}\sigma_{U,Q}\int_{\hat\vt}\frac{1}{\left[((1+u_2)U+(1-u_2)\sigma_{Q,U}Q)\cdot V\right]^{3}} \\
&\quad-iP_{0}\sigma_{U,P}\int_{\hat\vt}\frac{1}{\left[((1+u_2)U+(1-u_2)\sigma_{P,U}P)\cdot V\right]^{3}} \Bigg\},
\end{split}
\end{equation}
where the $\hat\vt$-integral can be obtained from \eq\eqref{eq:generalHTLintegral}, and reads 
\begin{equation}
\int_{\hat\vt}\frac{1}{\left(P\cdot V\right)^{3}}=-\frac{iP_{0}}{P^{4}}+O\left(\varepsilon\right).
\end{equation}
Substituting the leading-order term, we get
\begin{equation}
\begin{split}
\delta&\Gamma^{0000}_{P,Q,-P,-Q}=16\sigma_{P+Q,P-Q}\int_{-1}^{1}\ud u_{1}\int_{-1}^{1}\ud u_{2}\left(1+u_{2}\right)\\
\times&\left\{ \frac{\left(\left(1+u_{2}\right)U_{0}\sigma_{Q,U}+\left(1-u_{2}\right)Q_{0}\right)Q_{0}}{\left[\left(1+u_{2}\right)U+\left(1-u_{2}\right)\sigma_{Q,U}Q\right]^{4}}-\frac{\left(\left(1+u_{2}\right)U_{0}\sigma_{P,U}+\left(1-u_{2}\right)P_{0}\right)P_{0}}{\left[\left(1+u_{2}\right)U+\left(1-u_{2}\right)\sigma_{P,U}P\right]^{4}}\right\}+O\left(\varepsilon\right).
\end{split}    
\end{equation}
For us, setting $\varepsilon=0$ suffices, as the 4g vertex correction only appears in the finite term \eq\eqref{eq:I4GHdef}.

\subsection{Evaluating higher-rank HTL integrals} \label{sec:htlauto}

In \apps \ref{sec:3gHTL} and \ref{sec:4gHTL} we have discussed special cases of calculations involving HTL vertices. However, for example for the purposes of automation and possible future, more complicated, computations, it is useful to be able to discuss the integrals that arise on a more general level. The prototypical tensor integrals arising in HTL calculations are of the form $\int_{\hat\vt} V^{\mu_1}\ldots V^{\mu_r} (K\cdot V)^{-n}$ where $r$ is the tensor rank. Recalling that $V^0=-i$ is a constant, they are equivalent to
\begin{equation}
    H^{i_1 \ldots i_r}_n (K) = \int_{\hat\vt} \frac{\hat\vt^{i_1}\ldots \hat\vt^{i_r}}{(K \cdot V)^n}.
\end{equation}
Here, we outline a more general method, useful for larger values of $r$ and convenient when working in arbitrary dimensions. 

To begin with, the tensor $ H^{i_1 \ldots i_r}_n (K)$ is decomposed in a basis $\lbrace \mathcal{H}^{i_1\ldots i_r}_{r,b}(K)\rbrace_{b\in\mathcal{B}}$ consisting of rank $r$ tensors respecting the symmetries of the system, with $\mathcal{B}$ some finite index set enumerating the basis elements. With $ H^{i_1 \ldots i_r}_n (K)$ purely spatial, it retains the full $\mathrm{SO}(d)$ symmetry, and is furthermore fully symmetric in all indices. The basis for fixed $r$ can then be easily constructed using the external spatial vector $k^i$ as well as the spatial metric $\delta^{ij}$. For an example of an explicit construction, see \Ref\citep{Ee:2017}. In order to solve the coefficients $\lbrace H_{r,b,n}(K)\rbrace_{b\in\mathcal{B}}$ relative to this basis, we simply solve the following equation for each $b'\in\mathcal{B}$: 
\begin{equation}
     H^{i_{1}\ldots i_{r}}_n\left(K\right)\left(\mathcal{H}_{r,b'}\right)_{i_{1}\ldots i_{r}}\left(K\right)=\sum_{b\in\mathcal{B}}H_{r,b,n}\left(K\right)\mathcal{H}_{r,b}^{i_{1}\ldots i_{r}}\left(K\right)\left(\mathcal{H}_{r,b'}\right)_{i_{1}\ldots i_{r}}\left(K\right).
\end{equation}
Now the tensor is given explicitly as a combination of the basis elements and scalar integrals of the form 

\begin{equation}
\label{eq:generalHTLintegral}
\begin{split}
    h_{n}^{l}\left(K\right) \equiv \int_{\hat\vt}\frac{\left(\kt\cdot\hat\vt\right)^{l}}{\left(K\cdot V\right)^{n}}&=\frac{\Gamma\left(\frac{d}{2}\right)}{2\sqrt{\pi}}\frac{|\kt|^{l}}{\left(-ik_{0}\right)^{n}}\Biggl\{\left[1+\left(-1\right)^{l}\right]\frac{\Gamma\left(\frac{1}{2}+\frac{l}{2}\right)}{\Gamma\left(\frac{d}{2}+\frac{l}{2}\right)} \\
& \times {}_{3}F_{2}\left(\frac{1}{2}+\frac{n}{2},\frac{n}{2},\frac{1}{2}+\frac{l}{2};\frac{1}{2},\frac{d}{2}+\frac{l}{2};-\tan^{2}\Phi_{K}\right) \\
&\hspace{2.8em}-in\tan\Phi_{K}\left[1-\left(-1\right)^{l}\right]\frac{\Gamma\left(1+\frac{l}{2}\right)}{\Gamma\left(\frac{1}{2}+\frac{d}{2}+\frac{l}{2}\right)} \\
& \times {}_{3}F_{2}\left(\frac{1}{2}+\frac{n}{2},1+\frac{n}{2},1+\frac{l}{2};\frac{3}{2},\frac{d}{2}+\frac{1}{2}+\frac{l}{2};-\tan^{2}\Phi_{K}\right)\Biggr\}.
\end{split}
\end{equation}
Note that this is essentially a generalization of the integrals appearing in the self energy in \eq\eqref{eq:dimregpi}.

\subsection{Proof of the generalized Feynman parametrization}
\label{sec:generalisedFP}

To finish the discussion of the evaluation of the HTL vertices, we prove the generalized Feynman parametrization used in \apps\ref{sec:3gHTL} and \ref{sec:4gHTL}. The standard Feynman parametrization reads
\begin{equation}
\frac{1}{AB}=\int_{0}^{1}\ud t \frac{1}{\left[tB+\left(1-t\right)A\right]^{2}},
\end{equation}
where $A,B \in\mathbb{C}$. The representation is valid in many commonly encountered situations, in particular when $A,B \in \mathbb{R}_{+}$.
However, when the denominator of the right-hand-side vanishes, the right-hand side is no longer strictly convergent. This occurs when there exists a $t\in (0,1)$ such that $tB+\left(1-t\right)A$ vanishes,
that is, when the origin
is contained in the shortest line segment connecting $A$ and $B$ in the complex plane. We will denote this shortest line segment by $\gamma\left(A, B\right)$. In the present paper, we have encountered the need to combine a factorized denominator to obtain a Feynman-like parametrization for integrals such as \eq\eqref{eq:T3gHTLsym}. They involve \emph{arbitrary} points $A,B$ for which a nonvanishing denominator is not guaranteed. 

Here we show that a suitable generalized Feynman parametrization is 
\begin{equation}
F\left(A,B\right)\equiv\int_{-1}^{1}\ud u\frac{2\sigma}{\left[\left(1+u\right)A+\sigma\left(1-u\right)B\right]^{2}},
\end{equation}
by means of a detailed proof that $F\left(A,B\right)=1/\left(AB\right)$
for any $A,B\in\mathbb{C}\backslash\mathbb{R}$. The missing case of real $A,B$ will be briefly covered near the end of the section. In the following, we denote $\sigma\equiv\mathrm{sgn}(\Im A \, \Im B)$ in analogy with the $\sigma$ defined in \app\ref{sec:3gHTL}.

First, we assume $\sigma=+1$. Then necessarily $0\notin\gamma\left(A, B\right)$, so that we merely check that the identity holds in this standard case. Now, replace $u\mapsto1-2t$, so that $\left(-1,1\right)\mapsto (0,1)$ with the orientation reversed, leading to
\begin{equation}
\begin{split}
F\left(A,B\right)&=\left(-1\right)^2 2\int_{0}^{1}\ud t\frac{2}{\left[2\left(1-t\right)A+2tB\right]^{2}}\\
&=\int_{0}^{1}\ud t\frac{1}{\left[tB+\left(1-t\right)A\right]^{2}}=\frac{1}{AB},
\end{split}
\end{equation}
where we see \emph{a posteriori} that the change of variables is permitted
by assumption of $0\notin\gamma\left(A, B\right)$.
Note that this includes the special case $B=A$. 

Next, consider $\sigma=-1$, this time without restrictions on whether
or not origin is contained in $\gamma\left(A, B\right)$,
but by first assuming $A+B\neq0$. This case requires slightly more
care. For $\sigma=-1$ to be true, $A,B$ must lie on opposite half-planes. As we exclude reals, barring the aforementioned $A=-B$ we will have then covered all cases where
$0\in\gamma\left(A, B\right)$. As a consequence of
the assumptions, the denominator must always have a nonzero imaginary
part for all $u\in\left(-1,1\right)$, keeping it from vanishing on the
interval. To see this, recall first that by assumption $\Im A\neq0$
(and $\Im B\neq0$). Should the denominator vanish for some $u_*\in\left(-1,1\right)$,
we would be lead to the equality 
\begin{equation}
\left(1+u_*\right)\Im A-\left(1-u_*\right)\Im B=0\iff\frac{1+u_*}{1-u_*}=\frac{\Im B}{\Im A}.
\end{equation}
The condition $\sigma=-1$ sets $\Im B/\Im A<0$, and we immediately
see that there does not exist a $u_*\in\left(-1,1\right)$ such that $\left(1+u_*\right)/\left(1-u_*\right)\leq0$,
a contradiction. Therefore the integrand is finite on  $\left(-1,1\right)$,
and any divergence could only appear at the endpoints $\left\{ -1,1\right\}$. However, a direct calculation shows that the antiderivative is regular at them  (in the manipulation of what follows, recall that
we assume $A+B\neq0$):
\begin{align}
F\left(A,B\right)&=\int_{-1}^{1}\ud u\frac{-2}{\left[\left(1+u\right)A-\left(1-u\right)B\right]^{2}}=-\frac{-2}{\left(A+B\right)}\left[\frac{1}{\left(1+u\right)A+\left(u-1\right)B}\right]_{u=-1}^{u=1}\nn
&=\frac{2}{\left(A+B\right)}\left[\frac{1}{2A}+\frac{1}{2B}\right]=\frac{A+B}{A+B}\frac{1}{AB}=\frac{1}{AB}.
\end{align}

Lastly, we cover the missing case $B=-A$: Here, we see right away that $0\in\gamma\left(A, B\right)\implies\sigma=-1$ holds always and the integration is trivial, even though the above steps would no longer be valid: 
\begin{equation}
\begin{split}
F\left(A,B\right)&=\int_{-1}^{1}\ud u\frac{2\sigma}{\left[\left(1+u\right)A+\sigma\left(1-u\right)B\right]^{2}} \\ &=-\frac{1}{A^{2}}\int_{-1}^{1}\ud u\frac{2}{\left[\left(1+u\right)+\left(1-u\right)\right]^{2}}=\frac{1}{A\left(-A\right)}.
\end{split}
\end{equation}

In summary, we have shown that $F\left(A,B\right)=1/\left(AB\right)$ for any $A,B\in\mathbb{C}\backslash\mathbb{R}$. The result can even be shown to extend to all nonzero complex numbers by defining $\sigma=\mathrm{sgn}[\Theta(A)\Theta(B)]$ where $\Theta(A)=\Im A$ for $\Im A\neq 0$ and $\Theta(A)=A$ otherwise. We will not cover this in detail, as it is unnecessary for us and the proof mostly consists of reapplying the above steps together with the implications $\sigma=-1$ leads to in these special cases. Furthermore, an extension to multiple denominators is straightforward by differentiation and an iterative application of the two-point formula, with an explicit example covered in \app\ref{sec:4gHTL}. In such cases, the geometric interpretation of the origin lying in the line segment connecting the two factors of the denominator is naturally generalised to the origin being within the convex hull of the points of the factorised denominator \cite{Folland:2008zz}. This condition serves as a check to see whether or not a generalized parametrization is necessary.

\section{Summary of the contributing integrals}
\label{sec:importantIntegrals}

Many of the contributions appearing in the evaluation of the UV-sensitive terms in \Sec\ref{sec:IAD} involve nontrivial integrals, which we will discuss here. Throughout this appendix, we shall use the following notation for the angular average in $d$ spatial dimensions of some function $f(K,P)$
\begin{equation}
    \Ave{d}{f(K,P)} \equiv 
    \frac{\int_{\Omega} f(K,P)}
    {\int_{\Omega}},
\end{equation}
where $\Omega$ is as in \eq\eqref{eq:def_omega}. We define $\Ave{3}{f(K,P)}$ similarly in the obvious way. Observe that if $f$ only depends on some proper subset of the variables integrated over, this still corresponds to the angular average of $f$ over the angles upon which it depends. 

We also note that in this appendix we use the convention where we have rescaled all momenta by $\mE$ and then removed the tildes as in the later parts of our main text. With the integrals that we will evaluate here, this is equivalent to simply setting $\mE = 1$.

\subsection{\texorpdfstring{The integral in $\uvtext{I_{-2}(d)}$}{The integral in I\_{-2}(d)]\^{}UV}}

In this section, we demonstrate that
\begin{equation}
    \Ave{d}{\tr{\Pi(\hat{K})^{2}}} =
    1 - \psi(d) + \psi \biggl( \frac{1+d}{2} \biggr),
\label{eq:genDavePiSq}
\end{equation}
which was used in \eq\eqref{eq:IDLDone} above. To perform this averaging, we use the definition of the self energy given in \eq\eqref{eq:gluonselfHTL}, leading to
\begin{equation}
\begin{split}
    \tr{\Pi(\hat{K})^{2}} ={}&  
    \Pi^{\mu\nu}(\hat{K}) \Pi^{\nu\mu}(\hat{K}) \\
    ={}& 
    \int_{\hat\ut, \hat\vt}
    \biggl (\delta^{\mu 0}\delta^{\nu 0} - \frac{i \hat{K} \cdot \hat{N}}{\hat{K} \cdot U} U^{\mu}U^{\nu}\biggr ) 
    \biggl (\delta^{\mu 0}\delta^{\nu 0} - \frac{i\hat{K} \cdot \hat{N}}{\hat{K} \cdot V} V^{\mu}V^{\nu}\biggr ) \\
    ={}& 
    \int_{\hat\ut, \hat\vt}
    \biggl [ 1 + 2 \frac{i \hat{K} \cdot \hat{N}}{\hat{K} \cdot U} - \frac{(\hat{K} \cdot \hat{N})^2 (U \cdot V)^2}{(\hat{K} \cdot U)(\hat{K} \cdot V)} \biggr ],
\end{split}
\end{equation}
where we use the lightlike four-vectors $V$ defined in \app\ref{sec:HTLselfenergy} as well as the analogously defined $U^{\mu} \equiv (-i, \hat \ut)$ with $\hat\ut\in\mathbb{R}^d$ a unit vector. Additionally, we have defined the unit vector $\hat{N}$ in the temporal direction. In going from the second to the third line above, we have changed variables in one term, $\hat{\ut} \leftrightarrow \hat{\vt}$, to combine two terms.

Now observe the identity
\begin{equation}
    1 + 2 \frac{i \hat{K} \cdot \hat{N}}{\hat{K} \cdot U} = 
    \frac{\hat{K} \cdot U}{\hat{K} \cdot U} + 2 \frac{i \hat{K} \cdot \hat{N}}{\hat{K} \cdot U} =
    \frac{\hat{K} \cdot U^*}{\hat{K} \cdot U},
\end{equation}
where we used the definition of $U$, and the fact that $i$ only appears in the temporal part. But then inside an angular average over $\hat{K}$, by multiplying the numerator and denominator by $\hat{K} \cdot U^{*}$, we find
\begin{equation}
    \Ave{d}{1 + 2 \frac{i \hat{K} \cdot \hat{N}}{\hat{K} \cdot U}} = 
    \Ave{d}{\frac{\hat{K} \cdot U^*}{\hat{K} \cdot U}} = 
    U^{*\mu}U^{*\nu}\Ave{d}{\frac{\hat{K}^{\mu}\hat{K}^{\nu}}{|\hat{K} \cdot U|^{2}}}.
\end{equation}
We note that the function we are averaging over depends only on the components of $\hat{K}$ within $\Span(\hat{N}, \hat{\ut})$ (and we note that $\{ \hat{N}, \hat{\ut} \}$ form a perpendicular basis for this subspace). If we split $\hat{K} = K_{||}+ K_{\perp}$, with $K_{||} \in \Span(\hat{N}, \hat{\ut})$ and $K_{\perp} \cdot \hat{N} = K_{\perp} \cdot \hat{\ut} = 0$, then
\begin{equation}
    \frac{\hat{K}^{\mu}\hat{K}^{\nu}}{|\hat{K} \cdot U|^{2}} 
    = \frac{K_{||}^{\mu}K_{||}^{\nu}}{|K_{||} \cdot U|^{2}}
    = \frac{K_{||}^{\mu}K_{||}^{\nu}}{(K_{||} \cdot \hat{N})^{2} + (K_{||} \cdot \hat{\ut})^{2}} 
    = \hat{K_{||}}^{\mu}\hat{K_{||}}^{\nu},
\end{equation}
where we have recognized the denominator simply as $|K_{||}|^{2}$. We therefore see that the angular average involved above is the average of a unit vector over all of its directions, which simply leads to a constant multiple of the identity within its span $\delta_{||}^{\mu\nu}$. However, $U$ only depends on vectors within that span as well, and so we see
\begin{equation}
    \Ave{d}{1 + 2 \frac{i \hat{K} \cdot \hat{N}}{\hat{K} \cdot U}} = \frac{U^{*\mu}U^{*\nu} \delta_{||}^{\mu \nu}}{\text{const.}} = 0.
\end{equation}
Note that the entire analysis of this term could be conducted in the subspace $\Span(\hat{N}, \hat{\ut})$.

\begin{figure}[t]
    \includegraphics[width=0.5\textwidth]{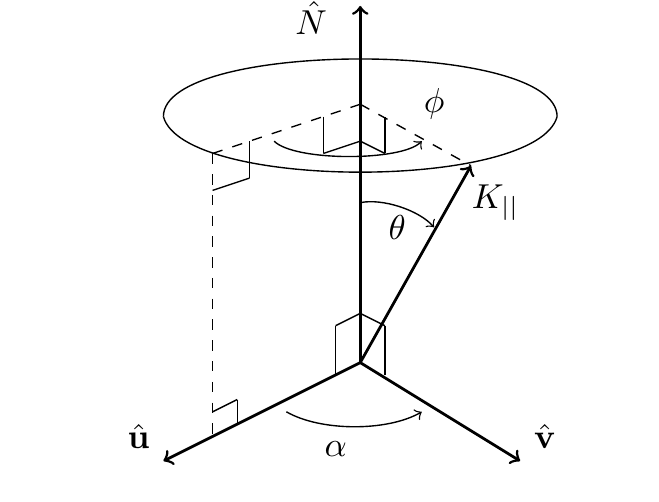}
    \caption{Our choice of coordinates for the angular average in \eq\eqref{eq:piSq_ave_simpl}. The angle between $\hat{\ut}$ and $\hat{\vt}$ here is a polar angle in the $d$-dimensional spatial subspace.}
\label{fig:ang_ave_coords}
\end{figure}

Using the above results, our original average simplifies to 
\begin{equation}
    \Ave{d}{\tr{\Pi(\hat{K})^{2}}} 
    = 
    - \int_{\hat\ut, \hat\vt}
    \Ave{d}{\frac{(\hat{K} \cdot \hat{N})^2 (U \cdot V)^2}{(\hat{K} \cdot U)(\hat{K} \cdot V)}}.
\end{equation}
We shall now analyze this within the three-dimensional subspace $\Span(\hat{N}, \hat{\ut}, \hat{\vt})$. Let us first split $\hat{K} = K_{||} + K_{\perp}$ as above, but this time with $K_{||} \in \Span(\hat{N}, \hat{\ut}, \hat{\vt})$ and $K_{\perp} \cdot \hat{N} = K_{\perp} \cdot \hat{\ut} = K_{\perp} \cdot \hat{\vt} = 0$. Because $\hat{\ut}$ and $\hat{\vt}$ are perpendicular to $\hat{N}$, we can set up a three-dimensional coordinate system as depicted in \fig\ref{fig:ang_ave_coords} to perform the integral. Note that because of the geometry within this subspace, only the angle between $\hat{\ut}$ and $\hat{\vt}$ is in the $d$-dimensional spatial subspace. The calculation proceeds as
%
%
\begin{align}
\label{eq:piSq_ave_simpl}
    \Ave{d}{\tr{\Pi(\hat{K})^{2}}} 
    ={}& 
    - \int_{\hat\ut, \hat\vt}
    \Ave{d}{\frac{(K_{||} \cdot \hat{N})^2 (U \cdot V)^2}{(K_{||} \cdot U)(K_{||} \cdot V)}} \nn
    \begin{split}
    ={}& 
    \frac{\Gamma(d / 2)}{\sqrt{\pi} \, \Gamma[(d-1) / 2] }
    \int_0^{\pi} \ud \alpha \sin^{d-2} \alpha 
    (1 - \cos \alpha)^2
    \int_0^{\pi} \frac{\ud \theta \sin \theta}{2} 
     \\
        &\quad\times\int_{0}^{2\pi} 
        \frac{\ud \phi / (2 \pi)}{[1 + i \tan \theta \cos \phi][ 1 + i \tan \theta \cos( \phi - \alpha)]}
    \end{split} \nn
    ={}& 
    \frac{\Gamma(d / 2)}{\sqrt{\pi} \, \Gamma[(d-1) / 2] }
    \int_0^{\pi} \ud \alpha \sin^{d-2} \alpha 
    (1 - \cos \alpha)^2
    \int_0^{\pi} 
    \frac{\ud \theta \sin \theta |\cos \theta|}{2 + (1 + \cos \alpha) \tan ^2(\theta )} \nn
    ={}& 
    \frac{\Gamma(d / 2)}{\sqrt{\pi} \, \Gamma[(d-1) / 2] }
    \int_0^{\pi} \ud \alpha \sin^{d-2} \alpha 
    \left[
    1 - \cos \alpha +2 (1 + \cos \alpha ) \ln \left( \cos \frac{\alpha }{2}\right)
    \right] \nn
    ={}& 
    1 - \psi(d) + \psi \biggl( \frac{1+d}{2} \biggr),
\end{align}
where, we used the fact that $\int_{\hat{\ut}} = 1$. This is the desired result.

\subsection{\texorpdfstring{The integrals in $\uvtext{\II{ABCD}} - \uvtext{I_{-2}}$}{The integrals in [I\_\{ABCD\}]\^{}UV - [I\_\{-2\}]\^{}UV}}
\label{sec:IABCD finite integrals}

In this subsection, we list all the one-dimensional integrals appearing in the $p_0$ contribution from $\uvtext{\II{ABCD}} - \uvtext{I_{-2}}$. We start by deriving an analytic angular average occurring in that term, which involves the function $\Pi_{1}$.  Using the angular average in \eq\eqref{eq:genDavePiSq}, expanding for small $\epsilon$, and setting the $O(\epsilon)$ terms equal on each side of the equality yields
\begin{align}
    \Bigl\langle
        &
        2 \tr{
            \Pi_0(\hat{K}) \Pi_1(\hat{K})
        }
        -
        \Pi_{\text{T},0}(\hat{K})^{2} \nn
        &-
        \tr{
            \Pi_{0}(\hat{K})^{2}
        }
        \left( 
            \ln 2 
            + \ln \bigl[\sin(\theta) \sin (\Phi_{P}) \sin (\Phi_{K})  \bigr] 
        \right)
    \Bigr\rangle_3
    =
    \frac{\pi^{2} - 9}{6}.
\end{align}
All of the integrals involving only $\Pi_{X,0}$, $X \in \{ \text{T}, \text{L}\}$ can be performed analytically. In particular, if we average over the angles (in $d = 3$), the following identities hold:
\begin{align}
    \AveT{\tr{\Pi_{0}(\hat{K})^{2}}} ={}&  \frac{1}{2},  \label{eq:APPangAve1}\\
    \AveT{\ln \bigl[ \sin(\Phi_{P}) \bigr] } ={}&  \frac{1}{2} - \ln 2, \\
    \AveT{\ln \bigl[ \sin(\theta) \bigr] } ={}&  - 1 + \ln 2, \\
    \AveT{\ln \bigl[ \sin(\Phi_{K}) \bigr] \tr{\Pi_{0}(\hat{K})^{2}}} ={}&  -\frac{1}{12} + \frac{\pi^{2}}{12} - \frac{7}{6} \ln 2, \\
    \AveT{ \Pi_{\text{T},0}(\hat{K})^{2}} ={}&  - \frac{5}{12} + \frac{2}{3} \ln 2.
\label{eq:angAveN}
\end{align}
Substituting these, we can deduce the following analytic equality:
\begin{equation}
    \AveT{
        \Pi_0(\hat{K}) \Pi_1(\hat{K})
    }
    =
    \frac{\pi^{2} - 9}{6}.
\label{eq:Pi1Int}
\end{equation}

In addition to this analytic integral, the above contribution to $p_{0}$ has many additional sources that can only be computed numerically. After many manipulations and angular averages, one can deduce that the $p_0$ contribution from $\uvtext{\II{ABCD}}-\uvtext{I_{-2}}$ is the sum of the terms
\begin{equation}
\label{eq:p0_IABCD_minus_Im2}
    \begin{split}
        p_{0} ={}  \frac{\pi^{2}}{2} \biggl[&
            \frac{11}{12} I_{\alpha} 
            + \frac{11}{6} I_{\beta} 
            + \left( -\frac{133}{36} + \frac{11}{6} \ln 2 \right) I_{\gamma} \\
            &+ \frac{11}{6} I_{\delta} 
            - \frac{11}{6} I_{\epsilon} 
            + \frac{1}{8} I_{\zeta} 
            - \frac{1}{4} I_{\eta} 
            - \frac{1}{8} (I_{\eta})^{2}
            + c_{\theta} \biggr],
    \end{split}
\end{equation}
where these contributions are mainly one-dimensional integrals, defined as
{
    \allowdisplaybreaks
\begin{align}
    I_{\alpha} ={}& 
    \AveT{\tr{\Pi_{0}(\hat{K})^2 \ln^{2}\bigl[\Pi_{0}(\hat{K})] }} \approx 0.2694302213,  \\
    I_{\beta} ={}& 
    \AveT{\Pi_{\text{T},0}(\hat{K})^2 \ln\bigl[\Pi_{\text{T},0}(\hat{K})] }
    \approx -0.06345103322, \\
    I_{\gamma} ={}& \AveT{ \tr{ \Pi_{0}^{2}(\hat{K}) \ln \bigl[ \Pi_{0}(\hat{K}) \bigr] } }
    \approx -0.2770771704, \\
    I_{\delta} ={}& \AveT{ \ln \bigl[ \sin( \Phi_{K} ) \bigr]  \tr{ \Pi_{0}^{2}(\hat{K}) \ln \bigl[ \Pi_{0}(\hat{K}) \bigr] } } 
    \approx 0.06818971068, \\
    I_{\epsilon} ={}& \AveT{ \tr{ \Pi_{0}(\hat{K}) \Pi_{1}(\hat{K}) \ln \bigl[ \Pi_{0}(\hat{K}) \bigr] } } 
    \approx -0.1249403621, \\
    I_{\zeta} ={}& \AveT{ \tr{ \Pi_{0}(\hat{K}) \ln^{2} \bigl[ \Pi_{0}(\hat{K}) \bigr] } } 
    \approx 1.122871337, \\
    I_{\eta} ={}& \AveT{ \tr{ \Pi_{0}(\hat{K}) \ln \bigl[ \Pi_{0}(\hat{K}) \bigr] } } 
    \approx -0.8577878295, \\
    c_{\theta} ={}& \frac{15}{8} - \frac{\pi^{2}}{9} 
    \approx 0.7783772888.
\end{align}
}%
Substituting these numerical values into \eq\eqref{eq:p0_IABCD_minus_Im2}, one reproduces the result of $10.84411$ found in \tab\ref{tab:ABCDcontributions}.  In deriving these results, we have used the angular average involving $\Pi_1$ in \eq\eqref{eq:Pi1Int}, some of the averages in \eqs\eqref{eq:APPangAve1}--\eqref{eq:angAveN}, as well as the following two averages:
\begin{align}
    \AveT{\ln^2 \bigl[ \sin(\Phi_{P}) \bigr] } ={}& -\frac{1}{2} + \frac{\pi^2}{12} - \ln 2 + \ln^2 2, \\
    \AveT{\ln^{2} \bigl[ \sin(\theta) \bigr] } ={}&  2 - \frac{\pi^2}{12} - 2 \ln 2 + \ln^2 2.
\end{align}

\section{Cancellations of the factorization scale and associated divergences among the different regions}
\label{sec:four_regs_sum}

In this appendix, we work through a simple factorized, two-dimensional integral expression to demonstrate the following two points about the the sum over the soft, hard, and mixed regions contributing to the physical double logarithms of the coupling at \NLO{3}: First, we show how all of the $O(\epsilon^{-2})$ and $O(\epsilon^{-1})$ terms cancel when one sums over these three regions, explicitly illustrating the cancellations of the divergences in the column labelled ``two-loop HTL'' in the right panel of \fig\ref{fig:visual_organization}, using a single $\epsilon$ regulator in all four regions. Second, we show how the coefficient of the $O(\epsilon^0)$, $\ln^2( \mE / \Lh )$ term in the soft region is exactly double the coefficient of the final $\ln^2 \alpha^{1/2}$ term obtained by summing over the three regions and cancelling spurious double-logarithmic terms.

Let us begin by showing briefly why spurious double logarithms arise in the soft region. To this end, we consider the simple factorized example integral
\begin{equation}
    I^{2} \equiv \left[ (\Lh)^{3-d} \int_0^\infty \dInt P\,P^{d - 2} \frac{\mE^2}{P^{2} + \mE^2} \right]^{2},
\end{equation}
with $d = 3 - 2\epsilon$, as before. We think of this as the soft part of a full calculation, which we detail further below. If we examine what one of these integrals yields upon integration, namely
\begin{align}
    I ={}&  - \frac{\pi}{2} \mE^{2} \left( \frac{\mE}{\Lh} \right)^{d - 3} \sec \Bigl( \frac{d \pi}{2} \Bigr) \nn
    \simeq{}&  \frac{\mE^{2}}{2 \varepsilon} - \mE^2 \ln \Bigl( \frac{\mE}{\Lh} \Bigr) + \epsilon \left[ \mE^2 \ln^2 \Bigl( \frac{\mE}{\Lh} \Bigr) \right] + O(\epsilon^{2}),
\end{align}
we observe the following: When this term is squared, there is both the square of the original $O(\epsilon^0)$ (single) logarithm, and the cross term between the $O(\epsilon^{-1})$ divergence and the $O(\epsilon^1)$ double logarithm. Only the former of these logarithms contributes in the end to the $\ln^2 \alpha_s^{1/2}$ term, and the second one is spurious.  These spurious logarithms must cancel when one sums over all three regions, since the $O(\epsilon^{-2})$ and $O(\epsilon^{-1})$ will cancel out in this sum. We now show this explicitly in our simple example.

To begin, let us note that if we were to resum the gluonic lines with the \emph{full} one-loop quark kinematics in the gluonic polarization tensors (instead of just resumming with HTL-resummed kinematics), there would be no divergences at all. This follows from the fact that the quark contribution to the one-loop gluonic self energies behaves like 
\begin{equation}
    \Pi_{\substack{\text{1-loop,} \\ \text{quark}}}(K) \simeq 
    \frac{\#}{K^{2}} + O(K^{-4})
\end{equation}
in the UV (see \eqs(A7) and (A9) in \Ref\cite{Kurkela:2009gj}), and thus the matter integrals would be completely UV convergent, and no divergences could arise at all. 

Let us thus consider such a UV-convergent full expression, and examine the contributions from each kinematic region. Let us choose
\begin{align}
    \II{finite}^{2} \equiv{}&  \left[ \int_0^\infty \dInt P\,P \frac{\mE^2}{P^{2} + \mE^2} \exp(- P / \mu) \right]^{2}  \nn
    ={}& 
	\mE^4 \left\{ \frac{1}{2} \left[ \pi -2 \text{Si}\Bigl(\frac{\mE}{\mu}\Bigr) \right] \sin \Bigl(\frac{\mE}{\mu}\Bigr)-\text{Ci}\Bigl(\frac{\mE}{\mu}\Bigr) \cos \Bigl(\frac{\mE}{\mu}\Bigr) \right\}^2 \nn
	={}& \mE^4 \ln^2\Bigl(\frac{\mE}{\mu}\Bigr)+2 \gamma_{\text{E}}  \mE^4 \ln \Bigl(\frac{\mE}{\mu}\Bigr)+\gamma_{\text{E}} ^2 \mE^4 + O(\mE^{5}),
\label{eq: reg}
\end{align}
which is a modification of our original $I^{2}$. Let us split this up into hard and soft regions, regulating them all in dimensional regularization. If we do this to the $\II{finite}$ itself first, we have the following:

\paragraph{Soft region:}

This is the same as the original example $I$ above, as we can ignore the $\exp(- P / \mu)$ term, so we again find
\begin{equation}
    \II{finite, s} \simeq{} \frac{\mE^{2}}{2 \varepsilon} - \mE^2 \ln \Bigl( \frac{\mE}{\Lh} \Bigr) + \epsilon \left[ \mE^2 \ln^2 \Bigl( \frac{\mE}{\Lh} \Bigr) \right] + O(\epsilon^{2}).
\end{equation}
In order to make this simpler to multiply with the hard part, let us rewrite this as 
\begin{align}
    \II{finite, s} \simeq{}&  \frac{\mE^{2}}{2 \varepsilon} - \mE^2 \left[ \ln \Bigl( \frac{\mE}{\mu} \Bigr) + \ln \Bigl( \frac{\mu}{\Lh} \Bigr) \right] + \epsilon \cdot \mE^2 \left[ \ln \Bigl( \frac{\mE}{\mu} \Bigr) + \ln \Bigl( \frac{\mu}{\Lh} \Bigr) \right]^{2} + O(\epsilon^{2}).
\end{align}

\paragraph{Hard region:}
 
Here, we can ignore the $\mE^{2}$ in the denominator:
\begin{align}
    \II{finite, h} ={}&  (\Lh)^{3 - d} \int_0^{\infty} \dInt P\,  P^{d-4} \exp(-P / \mu) =  \mE^{2} \left( \frac{\mu}{\Lh} \right)^{d - 3} \Gamma(d - 3) \nn
    \simeq{}& - \frac{\mE^{2}}{2 \epsilon} - \mE^{2} \left[ \gamma_{\text{E}} - \ln \Bigl( \frac{\mu}{\Lh} \Bigr) \right] - \epsilon \cdot \mE^{2} \left\{ \frac{\pi^{2}}{6}  + \left[ \gamma_{\text{E}} - \ln \Bigl( \frac{\mu}{\Lh} \Bigr) \right]^{2} \right\} + O(\epsilon^{2})
\end{align}

We thus see that the full result for the regulated 1d integrand is
\begin{align}
    \II{finite} = \II{finite, s} + \II{finite, h} = - \mE^{2} \left[ \gamma_{\text{E}} + \ln \Bigl( \frac{\mE}{\mu} \Bigr) \right],
\label{eq: I1d}
\end{align}
which, when squared,  agrees with \eq\eqref{eq: reg} above at $O(\mE^{4})$. Let us now inspect each of the kinematic regions, which we for simplicity denote by ss, sh, hs, and hh according to which combinations of the soft and hard contributions are combined: 
\begin{equation}
    \II{ss} + \II{sh} + \II{hs} + \II{hh} = \II{ss} + 2\II{sh} + \II{hh}.
\end{equation}
We find:
\begin{align}
    \II{ss} ={}&  \II{finite, s} \cdot \II{finite, s} \nn
    \simeq{}&  
    \frac{\mE^{4}}{4 \varepsilon^{2}} - \frac{\mE^{4}}{\epsilon} \left[ \ln \Bigl( \frac{\mE}{\mu} \Bigr) + \ln \Bigl( \frac{\mu}{\Lh} \Bigr) \right]
    + 2 \mE^4 \left[ \ln \Bigl( \frac{\mE}{\mu} \Bigr) + \ln \Bigl( \frac{\mu}{\Lh} \Bigr) \right]^2 + O(\epsilon), \\[2ex]
    2 \II{sh} ={}&  2 \II{finite, s} \cdot \II{finite, h}  \nn
    \simeq{}&  
    - \frac{\mE^{4}}{2 \varepsilon^{2}}  + \frac{\mE^4}{\epsilon} \left[ -\gamma_{\text{E}}  + \ln \Bigl( \frac{\mE}{\mu} \Bigr) + 2 \ln \Bigl( \frac{\mu}{\Lh} \Bigr) \right] \nn
    &\quad- \mE^{4} \left\{ \frac{\pi^{2}}{6} + \left[ -\gamma_{\text{E}}  + \ln \Bigl( \frac{\mE}{\mu} \Bigr) + 2 \ln \Bigl( \frac{\mu}{\Lh} \Bigr) \right]^2 \right\} + O(\epsilon),
\end{align}
and
\begin{align}
    \II{hh} ={}&  \II{finite, h} \cdot \II{finite, h} \nn
    \simeq{}&  
    \frac{\mE^{4}}{4 \varepsilon^{2}} + \frac{\mE^{4}}{\epsilon} \left[ \gamma_{\text{E}} - \ln \Bigl( \frac{\mu}{\Lh} \Bigr) \right]
    + \mE^4 \left\{ \frac{\pi^{2}}{6} + 2 \left[ \gamma_{\text{E}} - \ln \Bigl( \frac{\mu}{\Lh} \Bigr) \right]^2 \right\} + O(\epsilon) .
\end{align}
One may easily verify that the sum of these terms as written here reproduces \eq\eqref{eq: reg}, as it must. In particular, all of the divergent terms and the terms involving the fictitious mass scale $\Lh$ cancel. From this exercise, we also see that the double-logarithmic (DL) terms that contain the ratio $\mE / \mu$ of physical scales, are distributed as
\begin{align}
    \Bigl( \II{ss}, 2\II{sh}, \II{hh} \Bigr) \Bigr|_{\text{DL}} = 
    \left( 2 \mE^4 \ln^{2} \Bigl( \frac{\mE}{\mu} \Bigr), - \mE^4 \ln^{2} \Bigl( \frac{\mE}{\mu} \Bigr), 0 \right),
\end{align}
which indeed sum to the correct result. Importantly, we see that the coefficient of the ss contribution matches the coefficient of the $\ln^2(\mE / \Lh)$ there and is indeed twice the value that is obtained from summing over all four regions, as claimed.

\bibliography{references}

\begin{thebibliography}{48}%
\makeatletter
\providecommand \@ifxundefined [1]{%
 \@ifx{#1\undefined}
}%
\providecommand \@ifnum [1]{%
 \ifnum #1\expandafter \@firstoftwo
 \else \expandafter \@secondoftwo
 \fi
}%
\providecommand \@ifx [1]{%
 \ifx #1\expandafter \@firstoftwo
 \else \expandafter \@secondoftwo
 \fi
}%
\providecommand \natexlab [1]{#1}%
\providecommand \enquote  [1]{``#1''}%
\providecommand \bibnamefont  [1]{#1}%
\providecommand \bibfnamefont [1]{#1}%
\providecommand \citenamefont [1]{#1}%
\providecommand \href@noop [0]{\@secondoftwo}%
\providecommand \href [0]{\begingroup \@sanitize@url \@href}%
\providecommand \@href[1]{\@@startlink{#1}\@@href}%
\providecommand \@@href[1]{\endgroup#1\@@endlink}%
\providecommand \@sanitize@url [0]{\catcode `\\12\catcode `\$12\catcode
  `\&12\catcode `\#12\catcode `\^12\catcode `\_12\catcode `\%12\relax}%
\providecommand \@@startlink[1]{}%
\providecommand \@@endlink[0]{}%
\providecommand \url  [0]{\begingroup\@sanitize@url \@url }%
\providecommand \@url [1]{\endgroup\@href {#1}{\urlprefix }}%
\providecommand \urlprefix  [0]{URL }%
\providecommand \Eprint [0]{\href }%
\providecommand \doibase [0]{http://dx.doi.org/}%
\providecommand \selectlanguage [0]{\@gobble}%
\providecommand \bibinfo  [0]{\@secondoftwo}%
\providecommand \bibfield  [0]{\@secondoftwo}%
\providecommand \translation [1]{[#1]}%
\providecommand \BibitemOpen [0]{}%
\providecommand \bibitemStop [0]{}%
\providecommand \bibitemNoStop [0]{.\EOS\space}%
\providecommand \EOS [0]{\spacefactor3000\relax}%
\providecommand \BibitemShut  [1]{\csname bibitem#1\endcsname}%
\let\auto@bib@innerbib\@empty
\bibitem [{\citenamefont {Gorda}\ \emph {et~al.}(2021)\citenamefont {Gorda},
  \citenamefont {Kurkela}, \citenamefont {Paatelainen}, \citenamefont
  {S\"appi},\ and\ \citenamefont {Vuorinen}}]{Gorda:2021znl}%
  \BibitemOpen
  \bibfield  {author} {\bibinfo {author} {\bibfnamefont {T.}~\bibnamefont
  {Gorda}}, \bibinfo {author} {\bibfnamefont {A.}~\bibnamefont {Kurkela}},
  \bibinfo {author} {\bibfnamefont {R.}~\bibnamefont {Paatelainen}}, \bibinfo
  {author} {\bibfnamefont {S.}~\bibnamefont {S\"appi}}, \ and\ \bibinfo
  {author} {\bibfnamefont {A.}~\bibnamefont {Vuorinen}},\ }\href {\doibase
  10.1103/PhysRevLett.127.162003} {\bibfield  {journal} {\bibinfo  {journal}
  {Phys. Rev. Lett.}\ }\textbf {\bibinfo {volume} {127}},\ \bibinfo {pages}
  {162003} (\bibinfo {year} {2021})},\ \Eprint
  {http://arxiv.org/abs/2103.05658} {arXiv:2103.05658 [hep-ph]} \BibitemShut
  {NoStop}%
\bibitem [{\citenamefont {Ghiglieri}\ \emph {et~al.}(2020)\citenamefont
  {Ghiglieri}, \citenamefont {Kurkela}, \citenamefont {Strickland},\ and\
  \citenamefont {Vuorinen}}]{Ghiglieri:2020dpq}%
  \BibitemOpen
  \bibfield  {author} {\bibinfo {author} {\bibfnamefont {J.}~\bibnamefont
  {Ghiglieri}}, \bibinfo {author} {\bibfnamefont {A.}~\bibnamefont {Kurkela}},
  \bibinfo {author} {\bibfnamefont {M.}~\bibnamefont {Strickland}}, \ and\
  \bibinfo {author} {\bibfnamefont {A.}~\bibnamefont {Vuorinen}},\ }\href
  {\doibase 10.1016/j.physrep.2020.07.004} {\bibfield  {journal} {\bibinfo
  {journal} {Phys. Rept.}\ }\textbf {\bibinfo {volume} {880}},\ \bibinfo
  {pages} {1} (\bibinfo {year} {2020})},\ \Eprint
  {http://arxiv.org/abs/2002.10188} {arXiv:2002.10188 [hep-ph]} \BibitemShut
  {NoStop}%
\bibitem [{\citenamefont {Kajantie}\ \emph
  {et~al.}(2003{\natexlab{a}})\citenamefont {Kajantie}, \citenamefont {Laine},
  \citenamefont {Rummukainen},\ and\ \citenamefont
  {Schroder}}]{Kajantie:2002wa}%
  \BibitemOpen
  \bibfield  {author} {\bibinfo {author} {\bibfnamefont {K.}~\bibnamefont
  {Kajantie}}, \bibinfo {author} {\bibfnamefont {M.}~\bibnamefont {Laine}},
  \bibinfo {author} {\bibfnamefont {K.}~\bibnamefont {Rummukainen}}, \ and\
  \bibinfo {author} {\bibfnamefont {Y.}~\bibnamefont {Schroder}},\ }\href
  {\doibase 10.1103/PhysRevD.67.105008} {\bibfield  {journal} {\bibinfo
  {journal} {Phys. Rev. D}\ }\textbf {\bibinfo {volume} {67}},\ \bibinfo
  {pages} {105008} (\bibinfo {year} {2003}{\natexlab{a}})},\ \Eprint
  {http://arxiv.org/abs/hep-ph/0211321} {arXiv:hep-ph/0211321} \BibitemShut
  {NoStop}%
\bibitem [{\citenamefont {Kajantie}\ \emph
  {et~al.}(2003{\natexlab{b}})\citenamefont {Kajantie}, \citenamefont {Laine},
  \citenamefont {Rummukainen},\ and\ \citenamefont
  {Schroder}}]{Kajantie:2003ax}%
  \BibitemOpen
  \bibfield  {author} {\bibinfo {author} {\bibfnamefont {K.}~\bibnamefont
  {Kajantie}}, \bibinfo {author} {\bibfnamefont {M.}~\bibnamefont {Laine}},
  \bibinfo {author} {\bibfnamefont {K.}~\bibnamefont {Rummukainen}}, \ and\
  \bibinfo {author} {\bibfnamefont {Y.}~\bibnamefont {Schroder}},\ }\href
  {\doibase 10.1088/1126-6708/2003/04/036} {\bibfield  {journal} {\bibinfo
  {journal} {JHEP}\ }\textbf {\bibinfo {volume} {04}},\ \bibinfo {pages} {036}
  (\bibinfo {year} {2003}{\natexlab{b}})},\ \Eprint
  {http://arxiv.org/abs/hep-ph/0304048} {arXiv:hep-ph/0304048} \BibitemShut
  {NoStop}%
\bibitem [{\citenamefont {Di~Renzo}\ \emph {et~al.}(2006)\citenamefont
  {Di~Renzo}, \citenamefont {Laine}, \citenamefont {Miccio}, \citenamefont
  {Schroder},\ and\ \citenamefont {Torrero}}]{DiRenzo:2006nh}%
  \BibitemOpen
  \bibfield  {author} {\bibinfo {author} {\bibfnamefont {F.}~\bibnamefont
  {Di~Renzo}}, \bibinfo {author} {\bibfnamefont {M.}~\bibnamefont {Laine}},
  \bibinfo {author} {\bibfnamefont {V.}~\bibnamefont {Miccio}}, \bibinfo
  {author} {\bibfnamefont {Y.}~\bibnamefont {Schroder}}, \ and\ \bibinfo
  {author} {\bibfnamefont {C.}~\bibnamefont {Torrero}},\ }\href {\doibase
  10.1088/1126-6708/2006/07/026} {\bibfield  {journal} {\bibinfo  {journal}
  {JHEP}\ }\textbf {\bibinfo {volume} {07}},\ \bibinfo {pages} {026} (\bibinfo
  {year} {2006})},\ \Eprint {http://arxiv.org/abs/hep-ph/0605042}
  {arXiv:hep-ph/0605042} \BibitemShut {NoStop}%
\bibitem [{\citenamefont {Arnold}\ and\ \citenamefont
  {Zhai}(1994)}]{Arnold:1994ps}%
  \BibitemOpen
  \bibfield  {author} {\bibinfo {author} {\bibfnamefont {P.~B.}\ \bibnamefont
  {Arnold}}\ and\ \bibinfo {author} {\bibfnamefont {C.-X.}\ \bibnamefont
  {Zhai}},\ }\href {\doibase 10.1103/PhysRevD.50.7603} {\bibfield  {journal}
  {\bibinfo  {journal} {Phys. Rev. D}\ }\textbf {\bibinfo {volume} {50}},\
  \bibinfo {pages} {7603} (\bibinfo {year} {1994})},\ \Eprint
  {http://arxiv.org/abs/hep-ph/9408276} {arXiv:hep-ph/9408276} \BibitemShut
  {NoStop}%
\bibitem [{\citenamefont {Zhai}\ and\ \citenamefont
  {Kastening}(1995)}]{Zhai:1995ac}%
  \BibitemOpen
  \bibfield  {author} {\bibinfo {author} {\bibfnamefont {C.-X.}\ \bibnamefont
  {Zhai}}\ and\ \bibinfo {author} {\bibfnamefont {B.~M.}\ \bibnamefont
  {Kastening}},\ }\href {\doibase 10.1103/PhysRevD.52.7232} {\bibfield
  {journal} {\bibinfo  {journal} {Phys. Rev. D}\ }\textbf {\bibinfo {volume}
  {52}},\ \bibinfo {pages} {7232} (\bibinfo {year} {1995})},\ \Eprint
  {http://arxiv.org/abs/hep-ph/9507380} {arXiv:hep-ph/9507380} \BibitemShut
  {NoStop}%
\bibitem [{\citenamefont {Hietanen}\ \emph {et~al.}(2005)\citenamefont
  {Hietanen}, \citenamefont {Kajantie}, \citenamefont {Laine}, \citenamefont
  {Rummukainen},\ and\ \citenamefont {Schroder}}]{Hietanen:2004ew}%
  \BibitemOpen
  \bibfield  {author} {\bibinfo {author} {\bibfnamefont {A.}~\bibnamefont
  {Hietanen}}, \bibinfo {author} {\bibfnamefont {K.}~\bibnamefont {Kajantie}},
  \bibinfo {author} {\bibfnamefont {M.}~\bibnamefont {Laine}}, \bibinfo
  {author} {\bibfnamefont {K.}~\bibnamefont {Rummukainen}}, \ and\ \bibinfo
  {author} {\bibfnamefont {Y.}~\bibnamefont {Schroder}},\ }\href {\doibase
  10.1088/1126-6708/2005/01/013} {\bibfield  {journal} {\bibinfo  {journal}
  {JHEP}\ }\textbf {\bibinfo {volume} {01}},\ \bibinfo {pages} {013} (\bibinfo
  {year} {2005})},\ \Eprint {http://arxiv.org/abs/hep-lat/0412008}
  {arXiv:hep-lat/0412008} \BibitemShut {NoStop}%
\bibitem [{\citenamefont {Hietanen}\ and\ \citenamefont
  {Kurkela}(2006)}]{Hietanen:2006rc}%
  \BibitemOpen
  \bibfield  {author} {\bibinfo {author} {\bibfnamefont {A.}~\bibnamefont
  {Hietanen}}\ and\ \bibinfo {author} {\bibfnamefont {A.}~\bibnamefont
  {Kurkela}},\ }\href {\doibase 10.1088/1126-6708/2006/11/060} {\bibfield
  {journal} {\bibinfo  {journal} {JHEP}\ }\textbf {\bibinfo {volume} {11}},\
  \bibinfo {pages} {060} (\bibinfo {year} {2006})},\ \Eprint
  {http://arxiv.org/abs/hep-lat/0609015} {arXiv:hep-lat/0609015} \BibitemShut
  {NoStop}%
\bibitem [{\citenamefont {Kajantie}\ \emph {et~al.}(1997)\citenamefont
  {Kajantie}, \citenamefont {Laine}, \citenamefont {Rummukainen},\ and\
  \citenamefont {Shaposhnikov}}]{Kajantie:1997tt}%
  \BibitemOpen
  \bibfield  {author} {\bibinfo {author} {\bibfnamefont {K.}~\bibnamefont
  {Kajantie}}, \bibinfo {author} {\bibfnamefont {M.}~\bibnamefont {Laine}},
  \bibinfo {author} {\bibfnamefont {K.}~\bibnamefont {Rummukainen}}, \ and\
  \bibinfo {author} {\bibfnamefont {M.~E.}\ \bibnamefont {Shaposhnikov}},\
  }\href {\doibase 10.1016/S0550-3213(97)00425-2} {\bibfield  {journal}
  {\bibinfo  {journal} {Nucl. Phys. B}\ }\textbf {\bibinfo {volume} {503}},\
  \bibinfo {pages} {357} (\bibinfo {year} {1997})},\ \Eprint
  {http://arxiv.org/abs/hep-ph/9704416} {arXiv:hep-ph/9704416} \BibitemShut
  {NoStop}%
\bibitem [{\citenamefont {Braaten}\ and\ \citenamefont
  {Nieto}(1995)}]{Braaten:1995cm}%
  \BibitemOpen
  \bibfield  {author} {\bibinfo {author} {\bibfnamefont {E.}~\bibnamefont
  {Braaten}}\ and\ \bibinfo {author} {\bibfnamefont {A.}~\bibnamefont
  {Nieto}},\ }\href {\doibase 10.1103/PhysRevD.51.6990} {\bibfield  {journal}
  {\bibinfo  {journal} {Phys. Rev. D}\ }\textbf {\bibinfo {volume} {51}},\
  \bibinfo {pages} {6990} (\bibinfo {year} {1995})},\ \Eprint
  {http://arxiv.org/abs/hep-ph/9501375} {arXiv:hep-ph/9501375} \BibitemShut
  {NoStop}%
\bibitem [{\citenamefont {Braaten}\ and\ \citenamefont
  {Nieto}(1996)}]{Braaten:1995jr}%
  \BibitemOpen
  \bibfield  {author} {\bibinfo {author} {\bibfnamefont {E.}~\bibnamefont
  {Braaten}}\ and\ \bibinfo {author} {\bibfnamefont {A.}~\bibnamefont
  {Nieto}},\ }\href {\doibase 10.1103/PhysRevD.53.3421} {\bibfield  {journal}
  {\bibinfo  {journal} {Phys. Rev. D}\ }\textbf {\bibinfo {volume} {53}},\
  \bibinfo {pages} {3421} (\bibinfo {year} {1996})},\ \Eprint
  {http://arxiv.org/abs/hep-ph/9510408} {arXiv:hep-ph/9510408} \BibitemShut
  {NoStop}%
\bibitem [{\citenamefont {Manuel}(1996)}]{Manuel:1995td}%
  \BibitemOpen
  \bibfield  {author} {\bibinfo {author} {\bibfnamefont {C.}~\bibnamefont
  {Manuel}},\ }\href {\doibase 10.1103/PhysRevD.53.5866} {\bibfield  {journal}
  {\bibinfo  {journal} {Phys. Rev. D}\ }\textbf {\bibinfo {volume} {53}},\
  \bibinfo {pages} {5866} (\bibinfo {year} {1996})},\ \Eprint
  {http://arxiv.org/abs/hep-ph/9512365} {arXiv:hep-ph/9512365} \BibitemShut
  {NoStop}%
\bibitem [{\citenamefont {Gerhold}\ \emph {et~al.}(2004)\citenamefont
  {Gerhold}, \citenamefont {Ipp},\ and\ \citenamefont
  {Rebhan}}]{Gerhold:2004tb}%
  \BibitemOpen
  \bibfield  {author} {\bibinfo {author} {\bibfnamefont {A.}~\bibnamefont
  {Gerhold}}, \bibinfo {author} {\bibfnamefont {A.}~\bibnamefont {Ipp}}, \ and\
  \bibinfo {author} {\bibfnamefont {A.}~\bibnamefont {Rebhan}},\ }\href
  {\doibase 10.1103/PhysRevD.70.105015} {\bibfield  {journal} {\bibinfo
  {journal} {Phys. Rev. D}\ }\textbf {\bibinfo {volume} {70}},\ \bibinfo
  {pages} {105015} (\bibinfo {year} {2004})},\ \Eprint
  {http://arxiv.org/abs/hep-ph/0406087} {arXiv:hep-ph/0406087} \BibitemShut
  {NoStop}%
\bibitem [{\citenamefont {Annala}\ \emph {et~al.}(2018)\citenamefont {Annala},
  \citenamefont {Gorda}, \citenamefont {Kurkela},\ and\ \citenamefont
  {Vuorinen}}]{Annala:2017llu}%
  \BibitemOpen
  \bibfield  {author} {\bibinfo {author} {\bibfnamefont {E.}~\bibnamefont
  {Annala}}, \bibinfo {author} {\bibfnamefont {T.}~\bibnamefont {Gorda}},
  \bibinfo {author} {\bibfnamefont {A.}~\bibnamefont {Kurkela}}, \ and\
  \bibinfo {author} {\bibfnamefont {A.}~\bibnamefont {Vuorinen}},\ }\href
  {\doibase 10.1103/PhysRevLett.120.172703} {\bibfield  {journal} {\bibinfo
  {journal} {Phys. Rev. Lett.}\ }\textbf {\bibinfo {volume} {120}},\ \bibinfo
  {pages} {172703} (\bibinfo {year} {2018})},\ \Eprint
  {http://arxiv.org/abs/1711.02644} {arXiv:1711.02644 [astro-ph.HE]}
  \BibitemShut {NoStop}%
\bibitem [{\citenamefont {Annala}\ \emph {et~al.}(2020)\citenamefont {Annala},
  \citenamefont {Gorda}, \citenamefont {Kurkela}, \citenamefont {N\"attil\"a},\
  and\ \citenamefont {Vuorinen}}]{Annala:2019puf}%
  \BibitemOpen
  \bibfield  {author} {\bibinfo {author} {\bibfnamefont {E.}~\bibnamefont
  {Annala}}, \bibinfo {author} {\bibfnamefont {T.}~\bibnamefont {Gorda}},
  \bibinfo {author} {\bibfnamefont {A.}~\bibnamefont {Kurkela}}, \bibinfo
  {author} {\bibfnamefont {J.}~\bibnamefont {N\"attil\"a}}, \ and\ \bibinfo
  {author} {\bibfnamefont {A.}~\bibnamefont {Vuorinen}},\ }\href {\doibase
  10.1038/s41567-020-0914-9} {\bibfield  {journal} {\bibinfo  {journal} {Nature
  Phys.}\ } (\bibinfo {year} {2020}),\ 10.1038/s41567-020-0914-9},\ \Eprint
  {http://arxiv.org/abs/1903.09121} {arXiv:1903.09121 [astro-ph.HE]}
  \BibitemShut {NoStop}%
\bibitem [{\citenamefont {Freedman}\ and\ \citenamefont
  {McLerran}(1977{\natexlab{a}})}]{Freedman:1976dm}%
  \BibitemOpen
  \bibfield  {author} {\bibinfo {author} {\bibfnamefont {B.~A.}\ \bibnamefont
  {Freedman}}\ and\ \bibinfo {author} {\bibfnamefont {L.~D.}\ \bibnamefont
  {McLerran}},\ }\href {\doibase 10.1103/PhysRevD.16.1147} {\bibfield
  {journal} {\bibinfo  {journal} {Phys. Rev. D}\ }\textbf {\bibinfo {volume}
  {16}},\ \bibinfo {pages} {1147} (\bibinfo {year}
  {1977}{\natexlab{a}})}\BibitemShut {NoStop}%
\bibitem [{\citenamefont {Freedman}\ and\ \citenamefont
  {McLerran}(1977{\natexlab{b}})}]{Freedman:1976ub}%
  \BibitemOpen
  \bibfield  {author} {\bibinfo {author} {\bibfnamefont {B.~A.}\ \bibnamefont
  {Freedman}}\ and\ \bibinfo {author} {\bibfnamefont {L.~D.}\ \bibnamefont
  {McLerran}},\ }\href {\doibase 10.1103/PhysRevD.16.1169} {\bibfield
  {journal} {\bibinfo  {journal} {Phys. Rev. D}\ }\textbf {\bibinfo {volume}
  {16}},\ \bibinfo {pages} {1169} (\bibinfo {year}
  {1977}{\natexlab{b}})}\BibitemShut {NoStop}%
\bibitem [{\citenamefont {Kurkela}\ and\ \citenamefont
  {Vuorinen}(2016)}]{Kurkela:2016was}%
  \BibitemOpen
  \bibfield  {author} {\bibinfo {author} {\bibfnamefont {A.}~\bibnamefont
  {Kurkela}}\ and\ \bibinfo {author} {\bibfnamefont {A.}~\bibnamefont
  {Vuorinen}},\ }\href {\doibase 10.1103/PhysRevLett.117.042501} {\bibfield
  {journal} {\bibinfo  {journal} {Phys. Rev. Lett.}\ }\textbf {\bibinfo
  {volume} {117}},\ \bibinfo {pages} {042501} (\bibinfo {year} {2016})},\
  \Eprint {http://arxiv.org/abs/1603.00750} {arXiv:1603.00750 [hep-ph]}
  \BibitemShut {NoStop}%
\bibitem [{\citenamefont {Gorda}\ \emph {et~al.}(2018)\citenamefont {Gorda},
  \citenamefont {Kurkela}, \citenamefont {Romatschke}, \citenamefont
  {S\"appi},\ and\ \citenamefont {Vuorinen}}]{Gorda:2018gpy}%
  \BibitemOpen
  \bibfield  {author} {\bibinfo {author} {\bibfnamefont {T.}~\bibnamefont
  {Gorda}}, \bibinfo {author} {\bibfnamefont {A.}~\bibnamefont {Kurkela}},
  \bibinfo {author} {\bibfnamefont {P.}~\bibnamefont {Romatschke}}, \bibinfo
  {author} {\bibfnamefont {M.}~\bibnamefont {S\"appi}}, \ and\ \bibinfo
  {author} {\bibfnamefont {A.}~\bibnamefont {Vuorinen}},\ }\href {\doibase
  10.1103/PhysRevLett.121.202701} {\bibfield  {journal} {\bibinfo  {journal}
  {Phys. Rev. Lett.}\ }\textbf {\bibinfo {volume} {121}},\ \bibinfo {pages}
  {202701} (\bibinfo {year} {2018})},\ \Eprint
  {http://arxiv.org/abs/1807.04120} {arXiv:1807.04120 [hep-ph]} \BibitemShut
  {NoStop}%
\bibitem [{\citenamefont {Alford}\ \emph {et~al.}(2008)\citenamefont {Alford},
  \citenamefont {Schmitt}, \citenamefont {Rajagopal},\ and\ \citenamefont
  {Sch\"afer}}]{Alford:2007xm}%
  \BibitemOpen
  \bibfield  {author} {\bibinfo {author} {\bibfnamefont {M.~G.}\ \bibnamefont
  {Alford}}, \bibinfo {author} {\bibfnamefont {A.}~\bibnamefont {Schmitt}},
  \bibinfo {author} {\bibfnamefont {K.}~\bibnamefont {Rajagopal}}, \ and\
  \bibinfo {author} {\bibfnamefont {T.}~\bibnamefont {Sch\"afer}},\ }\href
  {\doibase 10.1103/RevModPhys.80.1455} {\bibfield  {journal} {\bibinfo
  {journal} {Rev. Mod. Phys.}\ }\textbf {\bibinfo {volume} {80}},\ \bibinfo
  {pages} {1455} (\bibinfo {year} {2008})},\ \Eprint
  {http://arxiv.org/abs/0709.4635} {arXiv:0709.4635 [hep-ph]} \BibitemShut
  {NoStop}%
\bibitem [{\citenamefont {Blaizot}\ and\ \citenamefont
  {Iancu}(2002)}]{Blaizot:2001nr}%
  \BibitemOpen
  \bibfield  {author} {\bibinfo {author} {\bibfnamefont {J.-P.}\ \bibnamefont
  {Blaizot}}\ and\ \bibinfo {author} {\bibfnamefont {E.}~\bibnamefont
  {Iancu}},\ }\href {\doibase 10.1016/S0370-1573(01)00061-8} {\bibfield
  {journal} {\bibinfo  {journal} {Phys. Rept.}\ }\textbf {\bibinfo {volume}
  {359}},\ \bibinfo {pages} {355} (\bibinfo {year} {2002})},\ \Eprint
  {http://arxiv.org/abs/hep-ph/0101103} {arXiv:hep-ph/0101103} \BibitemShut
  {NoStop}%
\bibitem [{\citenamefont {Andersen}\ \emph {et~al.}(2000)\citenamefont
  {Andersen}, \citenamefont {Braaten},\ and\ \citenamefont
  {Strickland}}]{Andersen:1999sf}%
  \BibitemOpen
  \bibfield  {author} {\bibinfo {author} {\bibfnamefont {J.~O.}\ \bibnamefont
  {Andersen}}, \bibinfo {author} {\bibfnamefont {E.}~\bibnamefont {Braaten}}, \
  and\ \bibinfo {author} {\bibfnamefont {M.}~\bibnamefont {Strickland}},\
  }\href {\doibase 10.1103/PhysRevD.61.014017} {\bibfield  {journal} {\bibinfo
  {journal} {Phys. Rev. D}\ }\textbf {\bibinfo {volume} {61}},\ \bibinfo
  {pages} {014017} (\bibinfo {year} {2000})},\ \Eprint
  {http://arxiv.org/abs/hep-ph/9905337} {arXiv:hep-ph/9905337} \BibitemShut
  {NoStop}%
\bibitem [{\citenamefont {Braaten}\ and\ \citenamefont
  {Pisarski}(1990)}]{Braaten:1989mz}%
  \BibitemOpen
  \bibfield  {author} {\bibinfo {author} {\bibfnamefont {E.}~\bibnamefont
  {Braaten}}\ and\ \bibinfo {author} {\bibfnamefont {R.~D.}\ \bibnamefont
  {Pisarski}},\ }\href {\doibase 10.1016/0550-3213(90)90508-B} {\bibfield
  {journal} {\bibinfo  {journal} {Nucl. Phys. B}\ }\textbf {\bibinfo {volume}
  {337}},\ \bibinfo {pages} {569} (\bibinfo {year} {1990})}\BibitemShut
  {NoStop}%
\bibitem [{\citenamefont {Laine}\ and\ \citenamefont
  {Vuorinen}(2016)}]{Laine:2016hma}%
  \BibitemOpen
  \bibfield  {author} {\bibinfo {author} {\bibfnamefont {M.}~\bibnamefont
  {Laine}}\ and\ \bibinfo {author} {\bibfnamefont {A.}~\bibnamefont
  {Vuorinen}},\ }\href {\doibase 10.1007/978-3-319-31933-9} {\emph {\bibinfo
  {title} {{Basics of Thermal Field Theory}}}},\ Vol.\ \bibinfo {volume} {925}\
  (\bibinfo  {publisher} {Springer},\ \bibinfo {year} {2016})\ \Eprint
  {http://arxiv.org/abs/1701.01554} {arXiv:1701.01554 [hep-ph]} \BibitemShut
  {NoStop}%
\bibitem [{\citenamefont {Ipp}\ \emph {et~al.}(2006)\citenamefont {Ipp},
  \citenamefont {Kajantie}, \citenamefont {Rebhan},\ and\ \citenamefont
  {Vuorinen}}]{Ipp:2006ij}%
  \BibitemOpen
  \bibfield  {author} {\bibinfo {author} {\bibfnamefont {A.}~\bibnamefont
  {Ipp}}, \bibinfo {author} {\bibfnamefont {K.}~\bibnamefont {Kajantie}},
  \bibinfo {author} {\bibfnamefont {A.}~\bibnamefont {Rebhan}}, \ and\ \bibinfo
  {author} {\bibfnamefont {A.}~\bibnamefont {Vuorinen}},\ }\href {\doibase
  10.1103/PhysRevD.74.045016} {\bibfield  {journal} {\bibinfo  {journal} {Phys.
  Rev. D}\ }\textbf {\bibinfo {volume} {74}},\ \bibinfo {pages} {045016}
  (\bibinfo {year} {2006})},\ \Eprint {http://arxiv.org/abs/hep-ph/0604060}
  {arXiv:hep-ph/0604060} \BibitemShut {NoStop}%
\bibitem [{\citenamefont {Kajantie}\ \emph {et~al.}(2002)\citenamefont
  {Kajantie}, \citenamefont {Laine},\ and\ \citenamefont
  {Schroder}}]{Kajantie:2001hv}%
  \BibitemOpen
  \bibfield  {author} {\bibinfo {author} {\bibfnamefont {K.}~\bibnamefont
  {Kajantie}}, \bibinfo {author} {\bibfnamefont {M.}~\bibnamefont {Laine}}, \
  and\ \bibinfo {author} {\bibfnamefont {Y.}~\bibnamefont {Schroder}},\ }\href
  {\doibase 10.1103/PhysRevD.65.045008} {\bibfield  {journal} {\bibinfo
  {journal} {Phys. Rev. D}\ }\textbf {\bibinfo {volume} {65}},\ \bibinfo
  {pages} {045008} (\bibinfo {year} {2002})},\ \Eprint
  {http://arxiv.org/abs/hep-ph/0109100} {arXiv:hep-ph/0109100} \BibitemShut
  {NoStop}%
\bibitem [{\citenamefont {Manuel}\ \emph {et~al.}(2016)\citenamefont {Manuel},
  \citenamefont {Soto},\ and\ \citenamefont {Stetina}}]{Manuel:2016wqs}%
  \BibitemOpen
  \bibfield  {author} {\bibinfo {author} {\bibfnamefont {C.}~\bibnamefont
  {Manuel}}, \bibinfo {author} {\bibfnamefont {J.}~\bibnamefont {Soto}}, \ and\
  \bibinfo {author} {\bibfnamefont {S.}~\bibnamefont {Stetina}},\ }\href
  {\doibase 10.1103/PhysRevD.94.025017} {\bibfield  {journal} {\bibinfo
  {journal} {Phys. Rev. D}\ }\textbf {\bibinfo {volume} {94}},\ \bibinfo
  {pages} {025017} (\bibinfo {year} {2016})},\ \bibinfo {note} {[Erratum:
  Phys.Rev.D 96, 129901 (2017)]},\ \Eprint {http://arxiv.org/abs/1603.05514}
  {arXiv:1603.05514 [hep-ph]} \BibitemShut {NoStop}%
\bibitem [{\citenamefont {Carignano}\ \emph {et~al.}(2018)\citenamefont
  {Carignano}, \citenamefont {Manuel},\ and\ \citenamefont
  {Soto}}]{Carignano:2017ovz}%
  \BibitemOpen
  \bibfield  {author} {\bibinfo {author} {\bibfnamefont {S.}~\bibnamefont
  {Carignano}}, \bibinfo {author} {\bibfnamefont {C.}~\bibnamefont {Manuel}}, \
  and\ \bibinfo {author} {\bibfnamefont {J.}~\bibnamefont {Soto}},\ }\href
  {\doibase 10.1016/j.physletb.2018.03.012} {\bibfield  {journal} {\bibinfo
  {journal} {Phys. Lett. B}\ }\textbf {\bibinfo {volume} {780}},\ \bibinfo
  {pages} {308} (\bibinfo {year} {2018})},\ \Eprint
  {http://arxiv.org/abs/1712.07949} {arXiv:1712.07949 [hep-ph]} \BibitemShut
  {NoStop}%
\bibitem [{\citenamefont {Carignano}\ \emph {et~al.}(2020)\citenamefont
  {Carignano}, \citenamefont {Carrington},\ and\ \citenamefont
  {Soto}}]{Carignano:2019ofj}%
  \BibitemOpen
  \bibfield  {author} {\bibinfo {author} {\bibfnamefont {S.}~\bibnamefont
  {Carignano}}, \bibinfo {author} {\bibfnamefont {M.~E.}\ \bibnamefont
  {Carrington}}, \ and\ \bibinfo {author} {\bibfnamefont {J.}~\bibnamefont
  {Soto}},\ }\href {\doibase 10.1016/j.physletb.2019.135193} {\bibfield
  {journal} {\bibinfo  {journal} {Phys. Lett. B}\ }\textbf {\bibinfo {volume}
  {801}},\ \bibinfo {pages} {135193} (\bibinfo {year} {2020})},\ \Eprint
  {http://arxiv.org/abs/1909.10545} {arXiv:1909.10545 [hep-ph]} \BibitemShut
  {NoStop}%
\bibitem [{\citenamefont {Gynther}\ \emph {et~al.}(2009)\citenamefont
  {Gynther}, \citenamefont {Kurkela},\ and\ \citenamefont
  {Vuorinen}}]{Gynther:2009qf}%
  \BibitemOpen
  \bibfield  {author} {\bibinfo {author} {\bibfnamefont {A.}~\bibnamefont
  {Gynther}}, \bibinfo {author} {\bibfnamefont {A.}~\bibnamefont {Kurkela}}, \
  and\ \bibinfo {author} {\bibfnamefont {A.}~\bibnamefont {Vuorinen}},\ }\href
  {\doibase 10.1103/PhysRevD.80.096002} {\bibfield  {journal} {\bibinfo
  {journal} {Phys. Rev. D}\ }\textbf {\bibinfo {volume} {80}},\ \bibinfo
  {pages} {096002} (\bibinfo {year} {2009})},\ \Eprint
  {http://arxiv.org/abs/0909.3521} {arXiv:0909.3521 [hep-ph]} \BibitemShut
  {NoStop}%
\bibitem [{\citenamefont {Blaizot}\ \emph {et~al.}(2001)\citenamefont
  {Blaizot}, \citenamefont {Iancu},\ and\ \citenamefont
  {Rebhan}}]{Blaizot:2000fc}%
  \BibitemOpen
  \bibfield  {author} {\bibinfo {author} {\bibfnamefont {J.~P.}\ \bibnamefont
  {Blaizot}}, \bibinfo {author} {\bibfnamefont {E.}~\bibnamefont {Iancu}}, \
  and\ \bibinfo {author} {\bibfnamefont {A.}~\bibnamefont {Rebhan}},\ }\href
  {\doibase 10.1103/PhysRevD.63.065003} {\bibfield  {journal} {\bibinfo
  {journal} {Phys. Rev.}\ }\textbf {\bibinfo {volume} {D63}},\ \bibinfo {pages}
  {065003} (\bibinfo {year} {2001})},\ \Eprint
  {http://arxiv.org/abs/hep-ph/0005003} {arXiv:hep-ph/0005003 [hep-ph]}
  \BibitemShut {NoStop}%
\bibitem [{\citenamefont {Andersen}\ \emph
  {et~al.}(2011{\natexlab{a}})\citenamefont {Andersen}, \citenamefont
  {Leganger}, \citenamefont {Strickland},\ and\ \citenamefont
  {Su}}]{Andersen:2010wu}%
  \BibitemOpen
  \bibfield  {author} {\bibinfo {author} {\bibfnamefont {J.~O.}\ \bibnamefont
  {Andersen}}, \bibinfo {author} {\bibfnamefont {L.~E.}\ \bibnamefont
  {Leganger}}, \bibinfo {author} {\bibfnamefont {M.}~\bibnamefont
  {Strickland}}, \ and\ \bibinfo {author} {\bibfnamefont {N.}~\bibnamefont
  {Su}},\ }\href {\doibase 10.1016/j.physletb.2010.12.070} {\bibfield
  {journal} {\bibinfo  {journal} {Phys. Lett. B}\ }\textbf {\bibinfo {volume}
  {696}},\ \bibinfo {pages} {468} (\bibinfo {year} {2011}{\natexlab{a}})},\
  \Eprint {http://arxiv.org/abs/1009.4644} {arXiv:1009.4644 [hep-ph]}
  \BibitemShut {NoStop}%
\bibitem [{\citenamefont {Andersen}\ \emph
  {et~al.}(2011{\natexlab{b}})\citenamefont {Andersen}, \citenamefont
  {Leganger}, \citenamefont {Strickland},\ and\ \citenamefont
  {Su}}]{Andersen:2011sf}%
  \BibitemOpen
  \bibfield  {author} {\bibinfo {author} {\bibfnamefont {J.~O.}\ \bibnamefont
  {Andersen}}, \bibinfo {author} {\bibfnamefont {L.~E.}\ \bibnamefont
  {Leganger}}, \bibinfo {author} {\bibfnamefont {M.}~\bibnamefont
  {Strickland}}, \ and\ \bibinfo {author} {\bibfnamefont {N.}~\bibnamefont
  {Su}},\ }\href {\doibase 10.1007/JHEP08(2011)053} {\bibfield  {journal}
  {\bibinfo  {journal} {JHEP}\ }\textbf {\bibinfo {volume} {08}},\ \bibinfo
  {pages} {053} (\bibinfo {year} {2011}{\natexlab{b}})},\ \Eprint
  {http://arxiv.org/abs/1103.2528} {arXiv:1103.2528 [hep-ph]} \BibitemShut
  {NoStop}%
\bibitem [{\citenamefont {Mogliacci}\ \emph {et~al.}(2013)\citenamefont
  {Mogliacci}, \citenamefont {Andersen}, \citenamefont {Strickland},
  \citenamefont {Su},\ and\ \citenamefont {Vuorinen}}]{Mogliacci:2013mca}%
  \BibitemOpen
  \bibfield  {author} {\bibinfo {author} {\bibfnamefont {S.}~\bibnamefont
  {Mogliacci}}, \bibinfo {author} {\bibfnamefont {J.~O.}\ \bibnamefont
  {Andersen}}, \bibinfo {author} {\bibfnamefont {M.}~\bibnamefont
  {Strickland}}, \bibinfo {author} {\bibfnamefont {N.}~\bibnamefont {Su}}, \
  and\ \bibinfo {author} {\bibfnamefont {A.}~\bibnamefont {Vuorinen}},\ }\href
  {\doibase 10.1007/JHEP12(2013)055} {\bibfield  {journal} {\bibinfo  {journal}
  {JHEP}\ }\textbf {\bibinfo {volume} {12}},\ \bibinfo {pages} {055} (\bibinfo
  {year} {2013})},\ \Eprint {http://arxiv.org/abs/1307.8098} {arXiv:1307.8098
  [hep-ph]} \BibitemShut {NoStop}%
\bibitem [{\citenamefont {Haque}\ \emph {et~al.}(2014)\citenamefont {Haque},
  \citenamefont {Bandyopadhyay}, \citenamefont {Andersen}, \citenamefont
  {Mustafa}, \citenamefont {Strickland},\ and\ \citenamefont
  {Su}}]{Haque:2014rua}%
  \BibitemOpen
  \bibfield  {author} {\bibinfo {author} {\bibfnamefont {N.}~\bibnamefont
  {Haque}}, \bibinfo {author} {\bibfnamefont {A.}~\bibnamefont
  {Bandyopadhyay}}, \bibinfo {author} {\bibfnamefont {J.~O.}\ \bibnamefont
  {Andersen}}, \bibinfo {author} {\bibfnamefont {M.~G.}\ \bibnamefont
  {Mustafa}}, \bibinfo {author} {\bibfnamefont {M.}~\bibnamefont {Strickland}},
  \ and\ \bibinfo {author} {\bibfnamefont {N.}~\bibnamefont {Su}},\ }\href
  {\doibase 10.1007/JHEP05(2014)027} {\bibfield  {journal} {\bibinfo  {journal}
  {JHEP}\ }\textbf {\bibinfo {volume} {05}},\ \bibinfo {pages} {027} (\bibinfo
  {year} {2014})},\ \Eprint {http://arxiv.org/abs/1402.6907} {arXiv:1402.6907
  [hep-ph]} \BibitemShut {NoStop}%
\bibitem [{\citenamefont {Andersen}\ \emph {et~al.}(2002)\citenamefont
  {Andersen}, \citenamefont {Braaten}, \citenamefont {Petitgirard},\ and\
  \citenamefont {Strickland}}]{Andersen:2002ey}%
  \BibitemOpen
  \bibfield  {author} {\bibinfo {author} {\bibfnamefont {J.~O.}\ \bibnamefont
  {Andersen}}, \bibinfo {author} {\bibfnamefont {E.}~\bibnamefont {Braaten}},
  \bibinfo {author} {\bibfnamefont {E.}~\bibnamefont {Petitgirard}}, \ and\
  \bibinfo {author} {\bibfnamefont {M.}~\bibnamefont {Strickland}},\ }\href
  {\doibase 10.1103/PhysRevD.66.085016} {\bibfield  {journal} {\bibinfo
  {journal} {Phys. Rev. D}\ }\textbf {\bibinfo {volume} {66}},\ \bibinfo
  {pages} {085016} (\bibinfo {year} {2002})},\ \Eprint
  {http://arxiv.org/abs/hep-ph/0205085} {arXiv:hep-ph/0205085} \BibitemShut
  {NoStop}%
\bibitem [{\citenamefont {Ee}\ \emph {et~al.}(2017)\citenamefont {Ee},
  \citenamefont {Jung}, \citenamefont {Kim},\ and\ \citenamefont
  {Lee}}]{Ee:2017}%
  \BibitemOpen
  \bibfield  {author} {\bibinfo {author} {\bibfnamefont {J.-H.}\ \bibnamefont
  {Ee}}, \bibinfo {author} {\bibfnamefont {D.-W.}\ \bibnamefont {Jung}},
  \bibinfo {author} {\bibfnamefont {U.-R.}\ \bibnamefont {Kim}}, \ and\
  \bibinfo {author} {\bibfnamefont {J.}~\bibnamefont {Lee}},\ }\href {\doibase
  10.1088/1361-6404/aa54ce} {\bibfield  {journal} {\bibinfo  {journal}
  {European Journal of Physics}\ }\textbf {\bibinfo {volume} {38}},\ \bibinfo
  {pages} {025801} (\bibinfo {year} {2017})}\BibitemShut {NoStop}%
\bibitem [{\citenamefont {Hahn}(2005)}]{Hahn:2004fe}%
  \BibitemOpen
  \bibfield  {author} {\bibinfo {author} {\bibfnamefont {T.}~\bibnamefont
  {Hahn}},\ }\href {\doibase 10.1016/j.cpc.2005.01.010} {\bibfield  {journal}
  {\bibinfo  {journal} {Comput. Phys. Commun.}\ }\textbf {\bibinfo {volume}
  {168}},\ \bibinfo {pages} {78} (\bibinfo {year} {2005})},\ \Eprint
  {http://arxiv.org/abs/hep-ph/0404043} {arXiv:hep-ph/0404043} \BibitemShut
  {NoStop}%
\bibitem [{\citenamefont {Shtabovenko}\ \emph {et~al.}(2016)\citenamefont
  {Shtabovenko}, \citenamefont {Mertig},\ and\ \citenamefont
  {Orellana}}]{Shtabovenko:2016sxi}%
  \BibitemOpen
  \bibfield  {author} {\bibinfo {author} {\bibfnamefont {V.}~\bibnamefont
  {Shtabovenko}}, \bibinfo {author} {\bibfnamefont {R.}~\bibnamefont {Mertig}},
  \ and\ \bibinfo {author} {\bibfnamefont {F.}~\bibnamefont {Orellana}},\
  }\href {\doibase 10.1016/j.cpc.2016.06.008} {\bibfield  {journal} {\bibinfo
  {journal} {Comput. Phys. Commun.}\ }\textbf {\bibinfo {volume} {207}},\
  \bibinfo {pages} {432} (\bibinfo {year} {2016})},\ \Eprint
  {http://arxiv.org/abs/1601.01167} {arXiv:1601.01167 [hep-ph]} \BibitemShut
  {NoStop}%
\bibitem [{\citenamefont {Vuorinen}(2003)}]{Vuorinen:2003fs}%
  \BibitemOpen
  \bibfield  {author} {\bibinfo {author} {\bibfnamefont {A.}~\bibnamefont
  {Vuorinen}},\ }\href {\doibase 10.1103/PhysRevD.68.054017} {\bibfield
  {journal} {\bibinfo  {journal} {Phys. Rev. D}\ }\textbf {\bibinfo {volume}
  {68}},\ \bibinfo {pages} {054017} (\bibinfo {year} {2003})},\ \Eprint
  {http://arxiv.org/abs/hep-ph/0305183} {arXiv:hep-ph/0305183} \BibitemShut
  {NoStop}%
\bibitem [{\citenamefont {Caron-Huot}\ and\ \citenamefont
  {Moore}(2008)}]{CaronHuot:2008uh}%
  \BibitemOpen
  \bibfield  {author} {\bibinfo {author} {\bibfnamefont {S.}~\bibnamefont
  {Caron-Huot}}\ and\ \bibinfo {author} {\bibfnamefont {G.~D.}\ \bibnamefont
  {Moore}},\ }\href {\doibase 10.1088/1126-6708/2008/02/081} {\bibfield
  {journal} {\bibinfo  {journal} {JHEP}\ }\textbf {\bibinfo {volume} {02}},\
  \bibinfo {pages} {081} (\bibinfo {year} {2008})},\ \Eprint
  {http://arxiv.org/abs/0801.2173} {arXiv:0801.2173 [hep-ph]} \BibitemShut
  {NoStop}%
\bibitem [{\citenamefont {York}\ \emph {et~al.}(2012)\citenamefont {York},
  \citenamefont {Moore},\ and\ \citenamefont {Tassler}}]{York:2012ib}%
  \BibitemOpen
  \bibfield  {author} {\bibinfo {author} {\bibfnamefont {M.~C.~A.}\
  \bibnamefont {York}}, \bibinfo {author} {\bibfnamefont {G.~D.}\ \bibnamefont
  {Moore}}, \ and\ \bibinfo {author} {\bibfnamefont {M.}~\bibnamefont
  {Tassler}},\ }\href {\doibase 10.1007/JHEP06(2012)077} {\bibfield  {journal}
  {\bibinfo  {journal} {JHEP}\ }\textbf {\bibinfo {volume} {06}},\ \bibinfo
  {pages} {077} (\bibinfo {year} {2012})},\ \Eprint
  {http://arxiv.org/abs/1202.4756} {arXiv:1202.4756 [hep-ph]} \BibitemShut
  {NoStop}%
\bibitem [{\citenamefont {Stevenson}(1984)}]{Stevenson:1982qw}%
  \BibitemOpen
  \bibfield  {author} {\bibinfo {author} {\bibfnamefont {P.~M.}\ \bibnamefont
  {Stevenson}},\ }\href {\doibase 10.1016/0550-3213(84)90307-9} {\bibfield
  {journal} {\bibinfo  {journal} {Nucl. Phys. B}\ }\textbf {\bibinfo {volume}
  {231}},\ \bibinfo {pages} {65} (\bibinfo {year} {1984})}\BibitemShut
  {NoStop}%
\bibitem [{\citenamefont {Fraga}\ and\ \citenamefont
  {Romatschke}(2005)}]{Fraga:2004gz}%
  \BibitemOpen
  \bibfield  {author} {\bibinfo {author} {\bibfnamefont {E.~S.}\ \bibnamefont
  {Fraga}}\ and\ \bibinfo {author} {\bibfnamefont {P.}~\bibnamefont
  {Romatschke}},\ }\href {\doibase 10.1103/PhysRevD.71.105014} {\bibfield
  {journal} {\bibinfo  {journal} {Phys. Rev. D}\ }\textbf {\bibinfo {volume}
  {71}},\ \bibinfo {pages} {105014} (\bibinfo {year} {2005})},\ \Eprint
  {http://arxiv.org/abs/hep-ph/0412298} {arXiv:hep-ph/0412298} \BibitemShut
  {NoStop}%
\bibitem [{\citenamefont {Kurkela}\ \emph {et~al.}(2010)\citenamefont
  {Kurkela}, \citenamefont {Romatschke},\ and\ \citenamefont
  {Vuorinen}}]{Kurkela:2009gj}%
  \BibitemOpen
  \bibfield  {author} {\bibinfo {author} {\bibfnamefont {A.}~\bibnamefont
  {Kurkela}}, \bibinfo {author} {\bibfnamefont {P.}~\bibnamefont {Romatschke}},
  \ and\ \bibinfo {author} {\bibfnamefont {A.}~\bibnamefont {Vuorinen}},\
  }\href {\doibase 10.1103/PhysRevD.81.105021} {\bibfield  {journal} {\bibinfo
  {journal} {Phys. Rev. D}\ }\textbf {\bibinfo {volume} {81}},\ \bibinfo
  {pages} {105021} (\bibinfo {year} {2010})},\ \Eprint
  {http://arxiv.org/abs/0912.1856} {arXiv:0912.1856 [hep-ph]} \BibitemShut
  {NoStop}%
\bibitem [{\citenamefont {Le~Bellac}(2011)}]{Bellac:2011kqa}%
  \BibitemOpen
  \bibfield  {author} {\bibinfo {author} {\bibfnamefont {M.}~\bibnamefont
  {Le~Bellac}},\ }\href {\doibase 10.1017/CBO9780511721700} {\emph {\bibinfo
  {title} {{Thermal Field Theory}}}},\ Cambridge Monographs on Mathematical
  Physics\ (\bibinfo  {publisher} {Cambridge University Press},\ \bibinfo
  {year} {2011})\BibitemShut {NoStop}%
\bibitem [{\citenamefont {Folland}(2008)}]{Folland:2008zz}%
  \BibitemOpen
  \bibfield  {author} {\bibinfo {author} {\bibfnamefont {G.~B.}\ \bibnamefont
  {Folland}},\ }\href {\doibase 10.1090/surv/149} {\emph {\bibinfo {title}
  {{Quantum field theory: A tourist guide for mathematicians}}}},\ Mathematical
  Surveys and Monographs\ (\bibinfo  {publisher} {American Mathematical
  Society},\ \bibinfo {year} {2008})\BibitemShut {NoStop}%
\end{thebibliography}%

\end{document}